# Plasmonic Doped Semiconductor Nanocrystals: Properties, Fabrication, Applications and Perspectives

Ilka Kriegel,[a, b,*] Francesco Scotognella,[c] Liberato Manna[a]

[a]Department of Nanochemistry, Istituto Italiano di Tecnologia (IIT), via Morego 30, I-16163 Genova, Italy, ilka.kriegel@iit.it, liberato.manna@iit.it

[b]Molecular Foundry Lawrence Berkeley National Laboratory, Berkeley, CA 94720, USA

[c]Dipartimento di Fisica, Politecnico di Milano, Piazza Leonardo da Vinci 32, I-20133 Milano, Italy, francesco.scotognella@polimi.it

[*]Corresponding author: ilka.kriegel@iit.it: tel. +39-010-71781502

**ABSTRACT:** Degenerately doped semiconductor nanocrystals (NCs) are of recent interest to the NC community due to their tunable localized surface plasmon resonances (LSPRs) in the near infrared (NIR). The high level of doping in such materials with carrier densities in the range of $10^{21}cm^{-3}$ leads to degeneracy of the doping levels and intense plasmonic absorption in the NIR. The lower carrier density in degenerately doped semiconductor NCs compared to noble metals enables LSPR tuning over a wide spectral range, since even a minor change of the carrier density strongly affects the spectral position of the LSPR. Two classes of degenerate semiconductors are most relevant in this respect: impurity doped semiconductors, such as metal oxides, and vacancy doped semiconductors, such as copper chalcogenides. In the latter it is the density of copper vacancies that controls the carrier concentration, while in the former the introduction of impurity atoms adds carriers to the system. LSPR tuning in vacancy doped semiconductor NCs such as copper chalcogenides occurs by chemically controlling the copper vacancy density. This goes in hand with complex structural modifications of the copper chalcogenide crystal lattice. In contrast the LSPR of degenerately doped metal oxide NCs is modified by varying the doping concentration or by the choice of host and dopant atoms, but also through the addition of capacitive charge carriers to the conduction band of the metal oxide upon post-synthetic treatments, such as by electrochemical- or photdoping. The NIR LSPRs and the option of their spectral fine-tuning make accessible important new features, such as the controlled coupling of the LSPR to other physical signatures or the enhancement of optical signals in the NIR, sensing application by LSPR tracking, energy production from the NIR plasmon resonance or bio-medical applications in the biological window. In this review we highlight the recent advances in the synthesis of various different plasmonic semiconductor NCs with LSPRs covering the entire spectral range, from the mid- to the NIR. We focus on copper chalcogenide NCs and impurity doped metal oxide NCs as the most investigated alternatives to noble metals. We shed light on the structural changes upon LSPR tuning in vacancy doped copper chalcogenide NCs and deliver a picture for the fundamentally different mechanism of LSPR modification of impurity doped metal oxide NCs. We review on the peculiar optical properties of plasmonic degenerately doped NCs by highlighting the variety of different optical measurements and optical modeling approaches. These findings are merged in an exhaustive section on new and exciting applications based on the special characteristics that plasmonic semiconductor NCs bring along.

# Contents

# 1. Introduction

Since the pioneering report by Zhao et al.[1] in 2009 demonstrating that copper sulfide ($Cu_{2-x}S$) nanocrystals (NCs) show localized surface plasmon resonances (LSPRs) in the near infrared (NIR), the field of degenerately doped semiconductor NCs has grown swiftly.[2–13] Shortly after, Luther et al.[11] have demonstrated that essentially any semiconductor NC with appreciable free carrier concentrations can sustain LSPRs.[11] Depending on the size and doping level of the nanostructure, LSPRs from the visible through the terahertz can be reached.[11] Strategies to intentionally add dopant atoms to the crystal of a semiconductor nanostructure have seen rapid progress and various doped semiconductor NCs were synthesized,[8] including phosphorous and boron-doped silicon NCs,[14–16] aluminum, germanium, or indium doped zinc oxide,[17] and indium doped tin oxide (ITO) NCs.[9] All these NCs have intense resonances in the mid to near IR due to plasmonic absorption. Additionally, NCs of other materials were found to show NIR plasmonic response, such as colloidal GeTe NCs,[18] $LaB_6$ NCs,[19,20] and Gd doped Prussian Blue.[21]

Degenerately doped semiconductor NCs form an exceptional case, as high levels of doping can be reached, leading to degeneracy of the doping levels. Carrier densities around $10^{21}$ $cm^{-3}$ place their LSPRs in the NIR,[2–13] and the high number of defect states leads to a widening of the bandgap in terms of the Moss-Burstein effect for heavily doped semiconductors.[12,22] Copper chalcogenides are typical members of degenerately doped semiconductors. The high density of free carriers is a result of vacancies in the structure, leading to strong LSPRs in the NIR.[1,11–13] Indeed, as shown by Zhao *et al.*[1] a variation of the copper to sulfide stoichiometry determines the exact position of the LSPR. Similarly, the nonstoichiometric composition in $Cu_{3-x}P$ NCs[23,24] and in $WO_{3-x}$ NCs[25] generates charge carriers that lead to plasmonic response in the NIR. Another prominent case of degenerately doped semiconductors are doped metal oxides.[9] By varying the numerous hosts and dopant atoms and playing with the level of doping, plasmon resonances in the entire spectral range from the mid to the near IR can be produced.[9] Notably, the lower order of magnitude of carrier density in degenerately doped semiconductors with respect to noble metals has a very attractive consequence: even a minor modification of the carrier density has a strong effect on the spectral position of the LSPR (**Figure 1**). This allows for a controlled plasmon tuning over a wide spectral range *via* chemical treatment, either during synthesis[1,8,26] or post-synthesis,[11–13,22] making these materials very appealing for numerous applications.[10,27–33] In vacancy doped semiconductors and in particular in copper chalcogenides, plasmon tuning goes in hand with redox-chemical reactions and a simultaneous removal/insertion of copper, as well as structural changes of the crystal lattice.[11–13] In metal oxide nanostructures instead the incor-poration of additional carriers can also be induced by capacitive charging. These extra carriers to the conduction band (CB) of the semiconductor increase the carrier density, enabling LSPR tunability over a wide frequency range.[10,22]

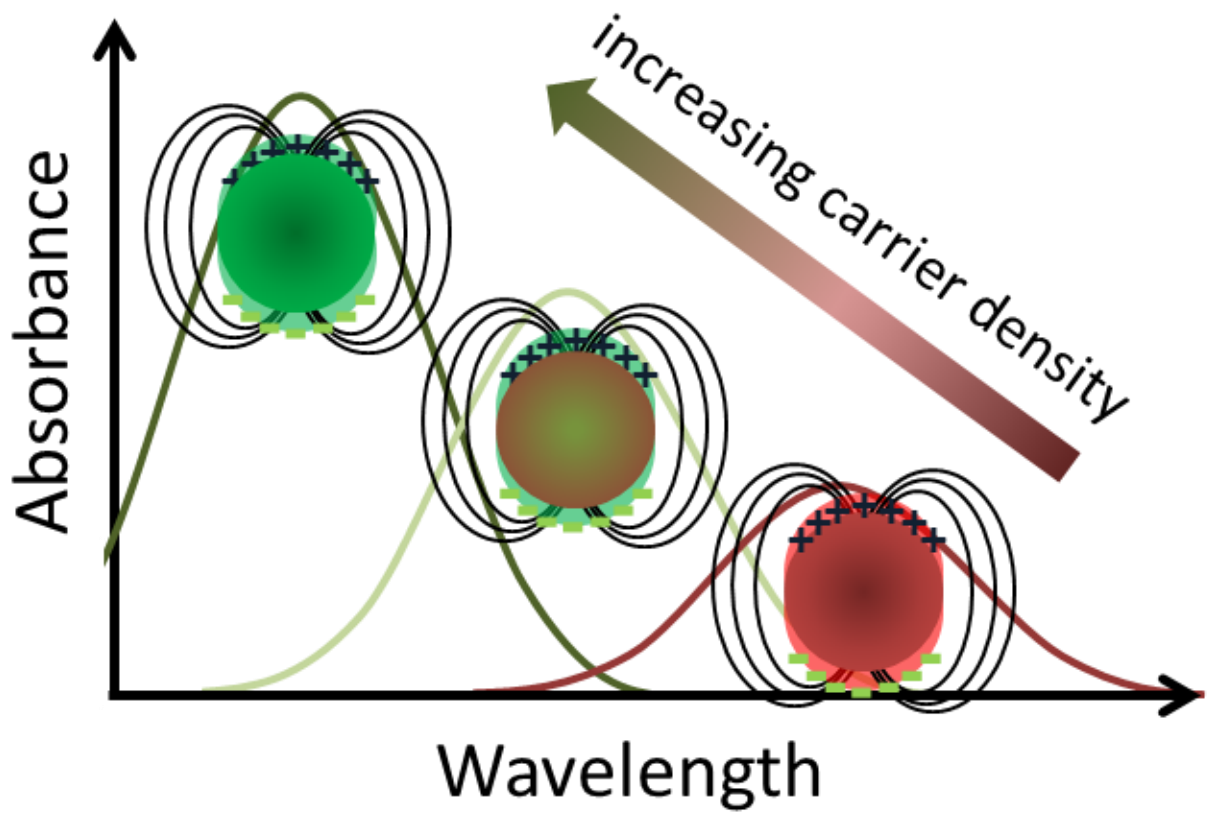

**Figure 1.** Illustration of LSPR tuning with increasing the carrier density in degenerately doped semiconductor NCs (pink to green curves).

The inherent tunability (by varying the carrier density) of degenerately doped semiconductor NCs and the LSPR characteristics stimulated intensive research in this area, ranging from synthesis to applications, based on existing and new emerging technologies.[1,3–11] To reach the precise control over dopant incorporation, new synthesis routes have been established together with techniques that help elucidating the structural characteristics.[9,34,35] Understanding the structure-properties relationship of plasmonic nanostructures requires a combined study of the local distribution of the dopant atom within the crystal, the role of vacancy formation, the presence of additional scattering centers or crystal anisotropies, and the extension of existing models (such as the Drude model) to simulate the plasmonic response of the nanostructure.[34–36] The possibility to tailor the plasmonic response of the degenerate nanostructure in the near to mid IR has been exploited for IR enhancing spectroscopies.[37–40] Photothermal heat release benefits from the strong light absorption in the NIR at wavelengths within the biological window.[41,27,42,43] The active control over the carrier density enables the fabrication of switchable electrochromic smart windows, where tuning of the LSPR allows filtering heat from the sunlight and at the same time ensures visible light transparency.[10,29] LSPR control in hybrid nanostructures also facilitates the study of plasmon-plasmon or exciton-plasmon coupling, delivering a tool to switch on and off the overlap between the resonances.[12,44,45] The regulation of the carrier density to ultralow levels in quantum confined systems enables the study of quantum plasmonics.[46–48] The combination of different functionalities in one material *via* the chemical incorporation of dopant atoms with specific physical properties, such as magnetism, enables the fabrication of switchable magnetic probes.[49] The chemical reactivity of some of these compound semiconductors, for example the ability of copper chalcogenide to lose or acquire copper ions, or yet to exchange them in part or in total with other ions, with concomitant changes in the plasmonic response, makes them interesting for various applications, including sensing.[50,51] Also, the sequestration of heavy metal cations, or their use as transport agents for radiolabels in biomedical applications become in reach.[32,52]

In this review, we will focus on the two most widely explored examples of degenerately doped semiconductor NCs, namely copper chalcogenide NCs[11–13] and doped metal oxide NCs.[9,22,26] We will first give a general introduction to basic concepts of plasmonics in nanoparticles. In past works, both types of degenerate semiconductors have been discussed next to each other due to their generally very similar properties.[2,53] Here instead we aim to give a separate picture of the most important characteristics of LSPRs in MO NCs and their means of LSPR manipulation, before introducing the characteristics of copper chalcogenide NCs. Thereby we will focus on the major differences of degenerate doping and plasmon tuning among these two types of heavily doped semiconductors. A section will be dedicated to the response of degenerately doped semiconductor NCs to light, including optical modelling, steady state properties and their ultrafast response. We will complete this review by providing an exhaustive overview over the newest directions for the application of degenerately doped semiconductor NCs in the biological, optical and energy fields, plus an outlook section at the end.

## 2. General introduction to plasmonics in nanoparticles

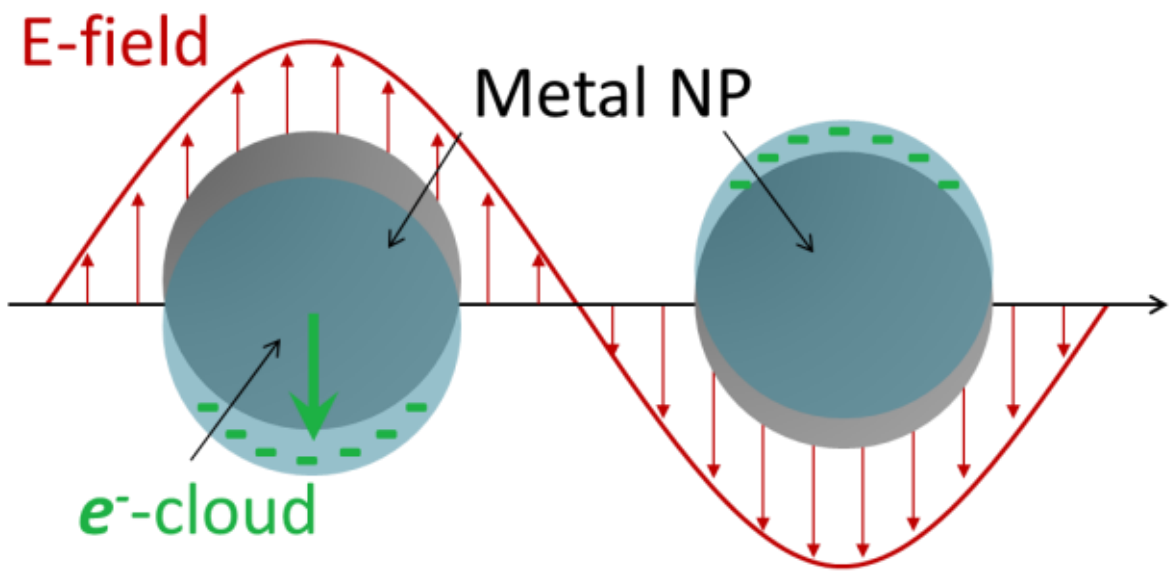

**Figure 2.** Schematic of plasmon oscillation for a spherical metallic NP, showing the displacement of the conduction electron charge cloud relative to the core in an oscillating electric field.

An LSPR can be described as a coherent oscillation of the free carrier gas in a metallic nanoparticle. The incoming electromagnetic wave drives the free carriers to oscillations, following the frequency of the incoming electric field of the light wave. The displacement of the carriers relative to the core leads to net charges at the nanoparticle surface, with the positive nuclei acting as restoring force. In the bulk, a classic harmonic oscillator model well describes the behavior of the free carriers. In nanoparticles, the free carriers are confined to a small volume localizing the surface plasmon to the nanoparticle. These so-called localized surface plasmon resonances (LSPRs) show pronounced absorption and arise when the resonance condition is fulfilled:

$$\boldsymbol{\omega_{LSPR} = \sqrt{\frac{\omega_p^2}{1+2\varepsilon_m} - \gamma^2}} \qquad (1)$$

where $\varepsilon_m$ is the dielectric constant of the surrounding medium and $\gamma$ is related to carrier damping. The resonance frequency for the collective oscillation of the carriers in the bulk metal is the so called plasma frequency $\omega_p$, defined as:

$$\boldsymbol{\omega_p = \sqrt{\frac{ne^2}{\varepsilon_0 m^*}}} \qquad (2)$$

where $e$ is the unit of elementary charge, and $\varepsilon_0$ is the permittivity of free space. It is related to the carrier density $n$ and the effective mass $m^*$ of the carriers. The given relation is valid for spherical particles in the quasi-static (dipole) limit, i.e. where the size of the nanoparticle is small enough that the electromagnetic interaction between different parts of the nanostructure is almost instantaneous and excitations are limited to dipolar modes.

Notably, changes in the carrier density directly affect the bulk plasma frequency $\omega_p$ and in turn also the resonance frequency $\omega_{LSPR}$ of the collective plasma oscillation. Thus, an increase in the carrier density $n$ will ultimately induce a spectral shift of the LSPR. Bulk metals like silver and gold have carrier densities in the range of $10^{23}cm^{-3}$ and show LSPRs in the UV-visible range. Instead, degenerately doped semiconductors have two orders of magnitude lower carrier densities, around $10^{21}cm^{-3}$ and show LSPRs in the NIR region. Importantly, this two orders of magnitude lower carrier density compared to metallic nanoparticles also induces a stronger effect on the LSPR, if small changes in the carrier density occur. This enables a precise control over the LSPR in a broad spectral range by controlling the density of carriers in the particle and in turn enables the detection of small carrier density changes by observing the LSPR.[11]

The damping constant $\gamma$ in equation 2 describes the damping of the collective oscillations of the carriers, which is a result of the scattering with electrons, phonons, lattice defects or impurities.[54,55] Surface scattering becomes important, when the effective path length of the electrons exceeds the nanoparticle size and results in a broadening of the LSPR line width.[54] The damping term can be written as the sum of all contributions:

$$\gamma = \gamma_{e-e} + \gamma_{e-phonon} + \gamma_{e-defect} + \gamma_{e-surface} \qquad (3)$$

The damping is indirectly related to the width of the plasmon resonance. The quality factor of an LSPR mode is defined by the ratio of the LSPR peak energy and the full width at half maximum (FWHM) of the resonance peak.[11,56] It generally describes the quality of a specific LSPR mode: the stronger the plasmonic response is, the higher the quality factor, i.e. the broader the resonance, the higher the damping. To calculate the exact value, the dielectric function needs to be taken into account.[56,57] Equation 2 also shows that the surrounding medium $\varepsilon_m$ is influencing the LSPR. The higher the medium dielectric constant is, the weaker the Coulombic restoring force of the electron cloud, in turn leading to a decrease of the LSPR.[58] The displacement of the charges in the NC can be seen as an oscillating dipole, which itself creates an oscillating field upon the excitation of the LSPR. This field is reaching into the surrounding of the nanoparticle. This so-called near-field is orders of magnitude enhanced with respect to the incoming electric field and decays exponentially with distance around the surface of the NC.[59] The near-field reaching into the surrounding is responsible for the high sensitivity of the plasmon resonance to the refractive index of the surrounding medium, and can be exploited in various applications, for example for sensing of single molecules or in signal enhancement.[60] The shape of the nanostructure has a very strong impact on the field distribution, and stronger localization of the field is found around tips or small radius apexes.[61] The absorption spectra of the differently shaped nanostructure are dominated by several resonances. For example in nanorods, two dominant resonances are usually found, one for the excitation along the rod diameter, the so called transverse resonance, and the other for the excitation along the length of the rod, the longitudinal LSPR, which is stronger in intensity and more red shifted with respect to the transverse one.[62] While an increase in size of the nanostructure usually results in a red shift of the LSPR, beyond a certain limit also higher order resonances, i.e. multipoles such as quadrupoles, are observed. However, in size ranges that are at least an order of magnitude lower than the wavelength of light, the quasi-static approximation (where only dipolar resonances play a role) is valid, as mentioned above.[59] Part of the energy transferred to the plasmon resonance is released upon scattering with electrons and phonons. This happens within several hundreds of femtoseconds and results in an efficient heating of the nanoparticle.[63] The energy of the LSPR is therefore released as heat, which then diffuses away from the NC and raises the temperature of the surrounding medium. The heating effect is stronger the more free carriers are present and is maximal at the LSPR.[64] This effect is exploited in many different applications, most notably in the treatment of cells *via* photothermal therapy.[65]

# 3. Impurity doped metal oxide NCs.

Oxide semiconductors such as zinc oxide,[66] cadmium oxide[67] or indium oxide[22] can be highly doped by adding n-type dopants. Thus, their optical response is governed by free electrons, whose density is controlled by the dopant concentration. Since these semiconductors have a large bandgap, they are transparent in the visible range. The free carrier concentration in MOs can be high enough for the MOs to exhibit metal-like behavior in the mid to near IR ranges. The properties of MOs depend strongly on the level of doping; hence their optical properties can be tailored for a specific device design. Differently from noble metals, the optical permittivity of MOs can be adjusted *via* post fabrication processes providing extraordinary tuning and modulation opportunities, because their carrier concentrations can be varied over several orders of magnitude.[68] Upon applying an electric field, either capacitive charge accumulation or depletion is induced, depending on the direction of the electric field. This causes a temporary change in the optical properties within a thin layer (~2–5nm) of the MO interface. Besides electrical tuning, also optical excitation allows dynamic modification of their optical properties. Light with energy greater than the bandgap is used to excite valence band (VB) electrons into the conduction band (CB), which in turn alters the properties of the material until they recombine and the material returns to equilibrium. This method has the benefit of generating free carriers throughout the bulk of the material (roughly hundreds of nanometers thick) instead of only a few nanometers.[68]

The carrier density in the range of $10^{21}cm^{-3}$ leads to pronounced absorption bands in the NIR ascribed to localized surface plasmon resonances (LSPRs), as explained above. Depending on the material, the type of synthesis, the dopant and the doping concentration, this resonance is tunable to cover the spectral range from the near-IR to the mid-IR. The tuning modalities as mentioned in the previous paragraph also hold for MO nanostructures. Due to their small volume, a switching from purely dielectric or semiconducting, i.e. with a bandgap, to metallic or plasmonic, i.e. governed by LSPRs, can be achieved upon posts-synthetic processing. A manifold of different materials have been investigated in the past and their plasmonic resonances explored. In the following section we aim to give an overview of the different MO NCs, their doping and plasmonic properties by highlighting several new recent approaches. Within this paragraph we will further review on recent advances in post-synthetic tuning of the LSPR achieved *via* photodoping and electrochemical charging.

### 3.1. Degenerate doping in metal oxide NCs

Degenerately doped metal oxide (MO) NCs can sustain extremely heavy doping without degradation of their morphological structure. Thus, high carrier densities (~$10^{21}$ $cm^{-3}$) can be achieved, which are intermediate between semiconductors and metals, placing their LSPR in the NIR, as described in the pre-vious section.[8,9] Due to such high level of doping, the donor levels are close enough to form bands rather than individual states. Subsequently, the optical properties of these materials display plasmonic response.[8,9,37,69] The incorporation of dopants leads to a steady increase of the doping level, i.e. carrier density with a continuous blue-shift of the LSPR.[9] Thus, a fine tuning of LSPRs is achieved by changing their dopant concentration.[9,26] Dopant atoms can be incorporated *via* interstitial doping, substitutional doping, oxygen-vacancies or a combination thereof.[8] Interstitial doping is observed in metal oxides having crystal structures that can incorporate interstitial atoms in their lattice. Tungsten bronzes ($M_xWO_3$) are a typical example in which alkali metal cations ($x$M) are accommodated in the structure, and this introduces shallow defect states in the oxide band structure.[9,70–72] $WO_6$ octahedra in the crystal structure of $WO_3$ create one-dimensional channels that are ideal for the incorporation of cations with specific ionic radii, such as $Cs^+$ (ionic radii of 169 pm), but also $Ce^+$, $Rb^+$ and $Na^+$ (Figure 2a and b).[70,71] Mattox et al. reported on the synthesis of three distinct shapes of $CsWO_3$ NCs—hexagonal prisms, truncated cubes, and pseudospheres—exhibiting strongly shape-dependent LSPRs in the NIR region by using varying ratios of oleylamine to oleic acid. With a similar synthetic procedure, Rb, Ce and Na tungstate NCs were synthesized with intense NIR LSPRs (Figure 3a).[70]

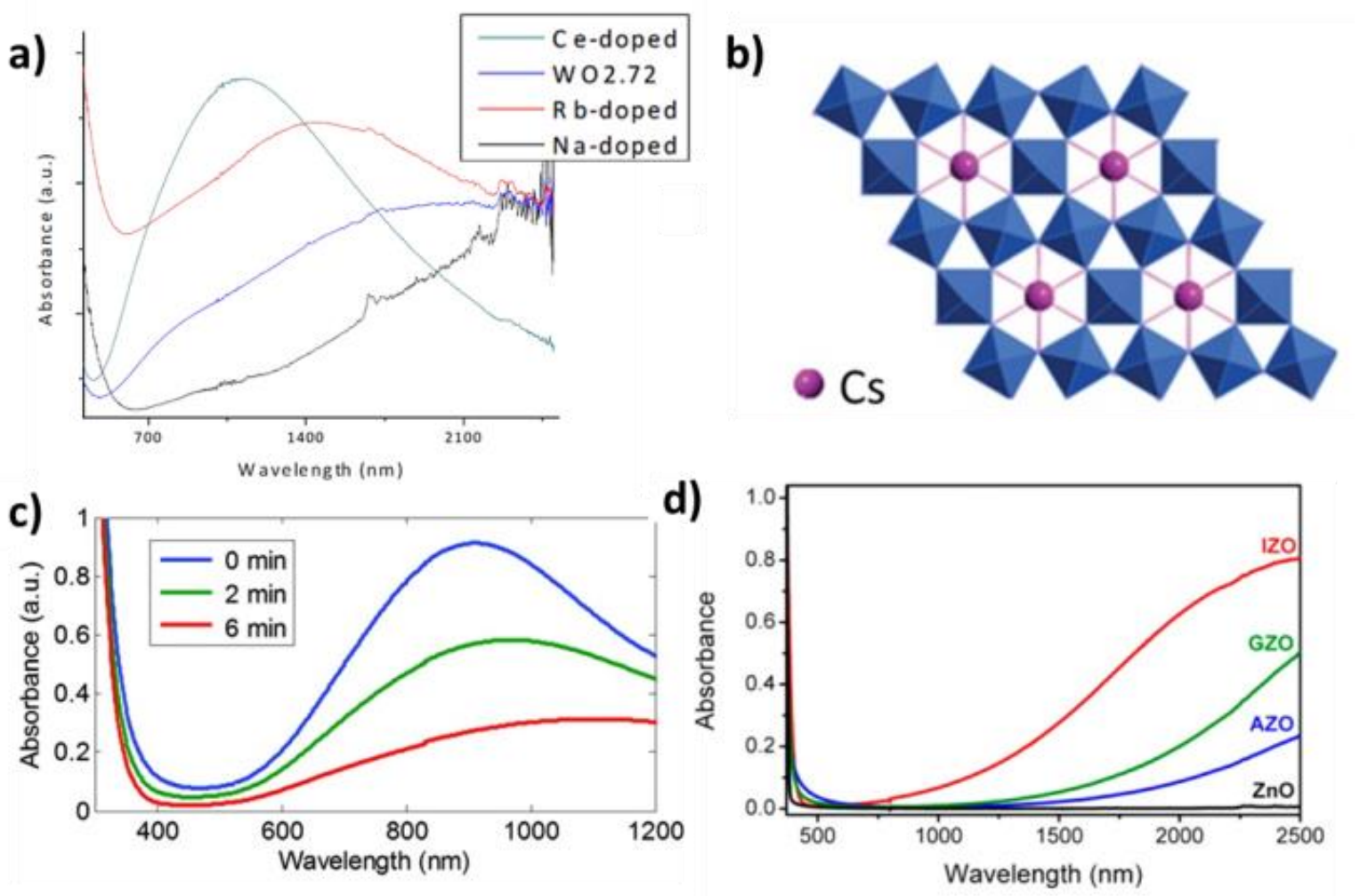

**Figure 3**. a) Absorbance spectra of $WO_{2.72}$ (structure based On XRD) doped with Rubidium, Sodium, and Cerium. b) Crystal structure of polyhedral model $Cs_xWO_3$. Adapted with permission from ref. [70] Copyright 2014 American Chemical Society. c) UV–vis–NIR absorption spectra of $WO_{2.83}$ nanorods in *N*-methylpyrrolidone upon heating at 175 °C in air for 0, 2, and 6 min. Reprinted with permission from ref [25] Copyright 2012 American Chemical Society. (d) Optical absorption spectra of equimolar solutions of ZnO-based NCs in TCE. Adapted with permission from ref [17] Copyright 2014 American Chemical Society.

In doped metal oxides, the role of oxygen vacancies and of other ionic defects crucially influences the electron or hole donor activity. In tungsten oxides ($WO_{3-x}$) oxygen vacancy formation is sufficient to increase the carrier concentration up to the $10^{21}$ $cm^{-3}$ range with LSPRs in the NIR, similar to copper chalcogenides. The more vacancies are formed, the more the LSPR blue shifts (**Figure 3**c).[25,73] In other metal oxides as well, for example undoped ZnO, NIR resonances are observed due to vacancy doping.[35] In the case of aliovalent substitutional doping, a lattice ion (usually the metal cation) is substituted with an ion of different valence. This replacement introduces shallow electron or hole states (depending on the valence of the dopant) in the band structure and leads to n- or p-doping of the system.[9] Examples are Aluminium doped Zinc Oxide (AZO),[26,35] Gallium doped Zinc Oxide (GZO)[17] and Germanium doped Zinc Oxide (GeZO)[74] with LSPRs up to around 3000 nm. Indium doped Tin Oxide (ITO)[22,75–77] and Antimony doped Tin Oxide (ATO)[78] support strong LSPRs up to about 1600 nm,[9,10] and In-doped CdO (ICO) NCs up to 2000 nm.[67] Della Gaspera et al.[17] developed a non-injection synthesis of ZnO NCs doped with three different metal cations, namely Al, Ga, and In, by mixing zinc stearate with the metal acetylacetonate as dopant source, and 1-dodecanol as activating agent. A comparison of undoped and doped ZnO NCs, prepared using a 20% molar ratio of dopant in solution with respect to Zn, showed that pronounced absorption was found in the NIR for all doped samples. The undoped ZnO NCs instead remained fairly transparent in the investigated spectral range. For doped ZnO NCs a LSPR shift to the blue was observed from about 7.5, to 5, to 3 μm for AZO, GZO, and IZO NCs, respectively (**Figure 3**d and e). This blue-shift can be correlated to an increased carrier density. For a detailed understanding of this finding the authors evaluated the actual dopant incorporation in the ZnO NCs *via* elemental analysis (by inductively coupled plasma optical emission spectroscopy). Notably, a lower doping efficiency for aluminum (Al/Zn = 3.1%) was found with respect to gallium (Ga/Zn = 13.2%) and indium (In/Zn = 15.4%), confirming the observed trend in the optical absorption.[17] An explanation for this trend in doping efficiency was given by comparing the ionic radii of the trivalent dopants with that of $Zn^{2+}$ (60 pm). The ionic radii of $Al^{3+}$, $Ga^{3+}$, and $In^{3+}$ in tetrahedral coordination are 39, 47, and 62 pm, respectively. Among the three, the radius of $In^{3+}$ has the best match with that of $Zn^{2+}$, and explains the more efficient lattice substitution of $In^{3+}$ compared to the other cations.[17] Many other materials, including vanadium oxide [79–81], rhenium oxide [82], or hydrogen doped semiconductors such as $TiO_2$, ZnO, $WO_3$ or $MoO_3$ [83,84], and Nb doped $TiO_2$ [85–88] display tunable plasmonic response by controlling the doping concentration [89].

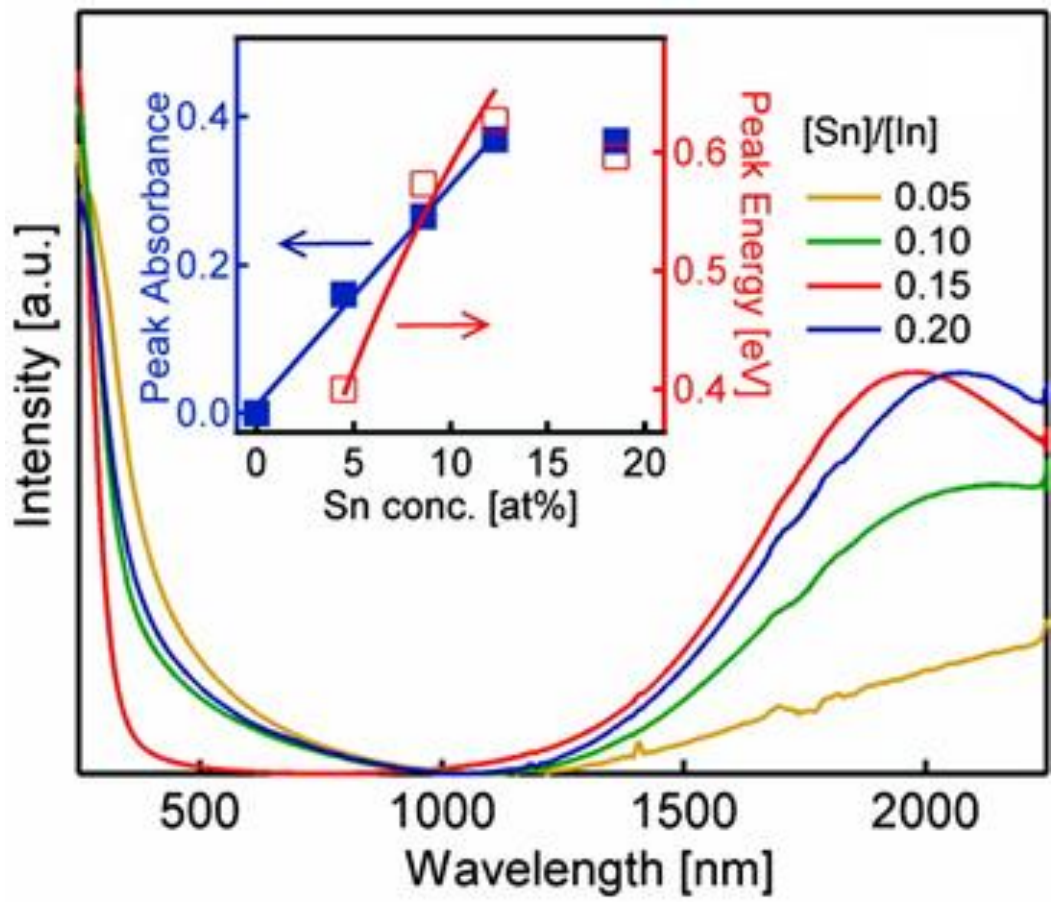

**Figure 4.** Band-gap-normalized absorption spectra of ITO synthesized with different precursor concentration ratios ([Sn]/[In]), as indicated in the graph. Inset: maximum plasmon band absorbance (blue squares) and peak energy position (open red squares) as a function of $Sn^{4+}$ doping concentration. Adapted with permission from ref [77] Copyright 2011 American Chemical Society.

A variety of other parameters, besides the density of charge carriers and the doping level, determine the location of the LSPR: these are NC size, shape, defects, and segregation of dopants. Defect cluster formation or structural defect concentration largely influence the free carrier density.[9] As a result, it was observed that increasing the doping beyond a certain level will not result in a further blue shift of the LSPR.[34,66] This suggests that not all of the introduced dopants successfully add excess degenerate charge carriers to the crystal, but instead are compensated elsewhere and excess doping leads to a saturation of the carrier density and sets an upper limit to the accessible intensity and spectral position of the LSPR.[34,66] In **Figure 4** are given the absorption spectra of ITO NCs with various doping levels, as reported by Wang and Radovanovic.[77] The inset depicts the plasmon band maxima (solid blue squares) and positions (open red squares) as a function of the $Sn^{4+}$ doping concentration. The absorption intensity increases linearly with increasing $Sn^{4+}$ doping owing to the corresponding increase in the free electron concentration, up to 12 % doping, above which a reverse behavior is observed. This was explained by electron trapping around $Sn^{4+}$ doping sites, which decreases the free electron density.[34,75,77]

A way to greatly expand the spectral tunability of LSPRs in doped MO NCs is *via* cation–anion cooping. As an attractive example, cation dopants in cooperation with the anion dopant fluorine is advantageous for increasing the effective carrier concentration and, hence expanding the wavelength range of LSPRs in doped MO NCs. Anion doping with fluorine, which has one valence electron more than oxygen, acts as electron donor when replacing oxygen atoms. This occurs readily as the ionic radius of $F^-$ (133 pm) is slightly smaller than that of $O^{2-}$ (140 pm) and fluorine is more electronegative than oxygen.[37] Ye et al.[37] demonstrated a generalized cation-anion cooping methodology for the synthesis of monodisperse, plasmonic, doped MOs, using fluorine doping in CdO NCs: Fluorine- and indium-codoped CdO (FICO) NCs and fluorine- and tin-codoped CdO (FSCO) NCs. Such NCs display LSPRs tunable across a significantly broader wavelength range (from 1.5 and 3.3 μm, see **Figure 5**a) than by changing only the concentrations of substitutional cation dopants. In addition, the LSPR in FICO NCs exhibit high plasmon quality factors due to low impurity scattering, making them ideal for applications based on local-field enhancement.[37]

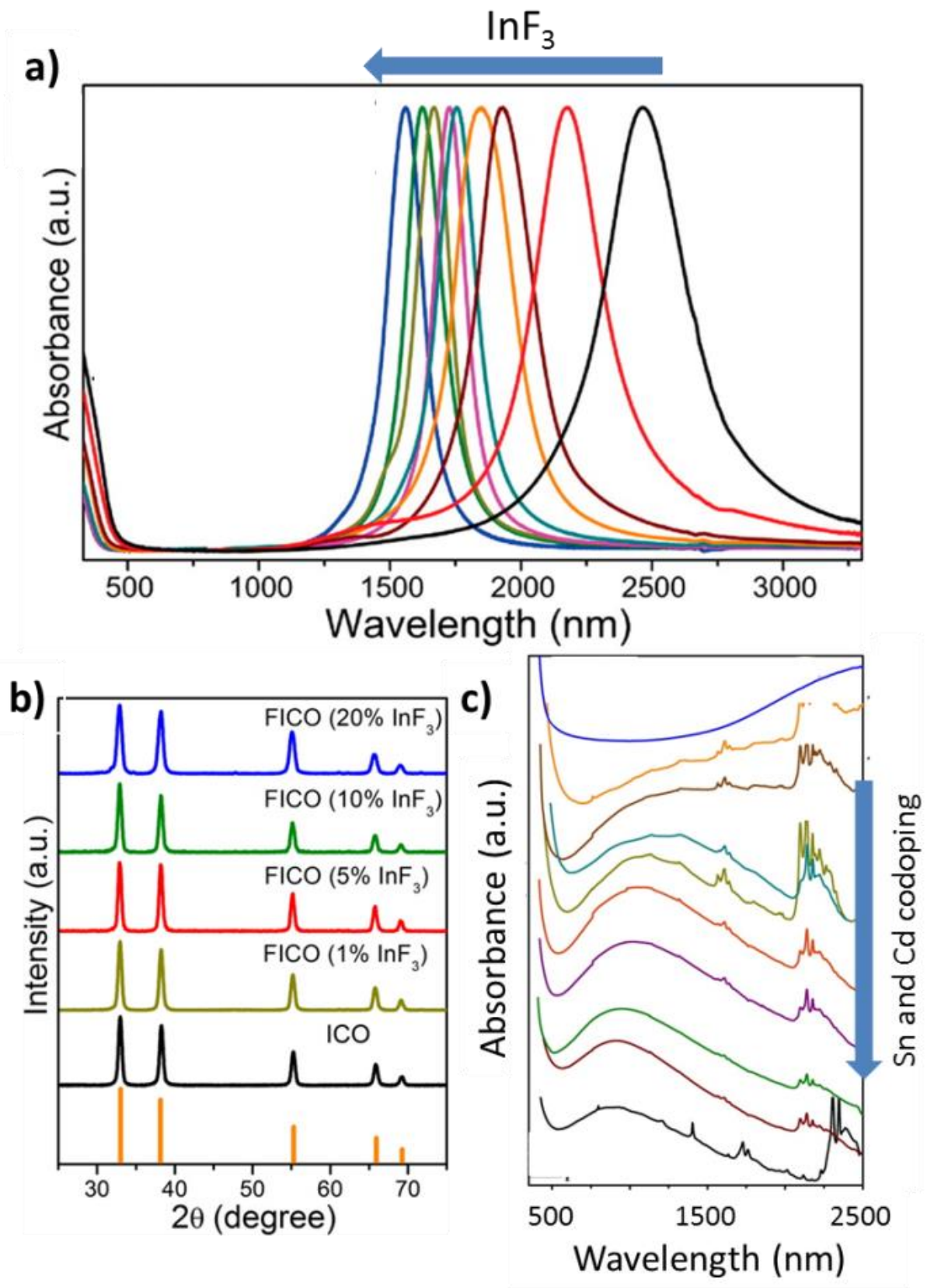

**Figure 5.** (a) UV–vis-NIR spectra of spherical fluorine- and indium-codoped CdO (FICO) NCs dispersed in $CCl_4$ with increasing amount of $InF_3$ used for NC synthesis. b) The powder X-ray diffraction (XRD) patterns reveal that the FICO NCs adopt the cubic CdO-type crystal structure. For FICO NCs synthesized using different percentages of the $InF_3$ precursor (up to 25%), negligible shifts in the XRD peak positions are observed. Adapted with permission from ref. [37] Copyright 2014 American Chemical Society. c) Optical absorbance spectra of equimolar dispersions of Sn and Cd codoped ZnO NCs. Adapted with permission from ref [69] Copyright 2014 American Chemical Society.

Alternatively, Sn and Cd codoped ZnO NCs, with simultaneous Sn doping and Cd alloying were reported by Ghosh *et al.*[69] through a one-pot reaction. Controlled Sn doping and a precise degree of Cd alloying were demonstrated to enable tuning of the LSPR band in a wide spectral region between 890 and 2600 nm as the Cd% increases, while the amount of Sn% stays fixed (**Figure 5**b). A further blue shift of the plasmon resonance were reported for Sn doping of ZnO NCs reaching values up to the visible spectral region.[69] Multifunctional Fe–Sn codoped $In_2O_3$ NCs that simultaneously exhibit LSPRs and high electrical conductivity were presented by Shanker et al.[90] who demonstrated the possibility of charge mediated magnetic coupling. Interesting magneto-optic and magneto-electric properties are expected from such multifunctional NCs.[90]

The incorporation of dopants is not always straightforward and faces several challenges. These include electronic passivation of surface defects (as discussed above), control over the activation of dopants, i.e. those dopants that successfully contribute to the carrier density, and achieving a uniform distribution of the dopant atoms within the entire NC volume.[35] Such factors influence the LSPR wavelength but also the line shape of the plasmon resonance. Lounis *et al.*[34] showed that the synthesis procedure has a high impact on the dopant distribution within the NC volume and in turn their ability to donate an electron to the NC. Tin-rich surfaces exhibit reduced dopant activation due to compensation of the dopant ions by electronic defects and oxygen interstitials, i.e. the formation of electronic trap states at the NC surface.[34] To account for such local changes of the dopant distribution in the volume of the NC, Mehra *et al.*[91] suggest a synthetic strategy in which undoped zinc oxide nanorods serve as templates. The addition of dopant salts to the ongoing reaction leads to dopant incorporation, and subsequent overgrowth with an undoped shell ensures that surface segregated dopants are integrated into the core.[91] In an alternative approach Jansons *et al.*[92] presented a method of growth for $In_2O_3$ NCs that relies on the slow, continuous addition of the precursor. Linear volume increase with added monomer is observed and the reaction can be stopped and restarted repeatedly. The number of NC remains constant, resulting in a tight size control and enabling the deposition of metal oxide shells to synthesize ITO/$In_2O_3$ core/shell NCs as well as $In_2O_3$/ITO/$In_2O_3$ core/shell/shell NCs. Greenberg *et al.*[35] recently suggested nonthermal plasma synthesis as an alternative strategy for the production of doped metal oxide NCs.[35,93] This strategy is successful even for the doping throughout the entire NC, also reaching the central cores and for heavy doping of small NCs. In a typical synthesis, the Zn and Al precursors are injected into a low-pressure radio frequency Ar/$O_2$ plasma. Dopants are irreversibly incorporated into the NC *via* nucleation and growth in the highly reactive gas-phase. This results in a radially uniform dopant distribution, allowing high doping levels even if the growth is stopped early. Finally, the AZO NCs are deposited onto a substrate. Atomic layer deposition is then used to infill the films with $Al_2O_3$, forming a $ZnO/Al_2O_3$ composite. This process also removes surface OH groups and seals the NCs, resulting in air-stable conductive films with LSPR around 3000 nm.[35] The successful doping of small NCs and the absence of undoped cores are explained by the different environments that the reactants experience in the nonthermal plasma synthesis with respect to the solvent based synthesis and which influence their chemical potentials. In solution phase synthesis, the solvated species have much lower chemical potential than the gas-phase radicals and ions in plasma synthesis (**Figure 6**a) and, as a consequence, the NC nuclei can form and dissolve multiple times before growing past a critical size. In this formation and re-dissolution process, the nuclei can easily expel the dopant ions. On the other hand, in the plasma synthesis the nucleation process is practically irreversible, such that once a dopant ion is added to the growing nucleus, it will stay there. It follows that in a solution phase synthesis dopants are more easily added at a later stage during NC growth, where the concentration of the dopant ions relative to the ions used to grow the NCs increases to a point that they start competing for incorporation in the NCs (**Figure 6**b). This results in a dopant distribution that is peaked in regions closer to the surface of the NCs (**Figure 6**b, sketch on the left). In the plasma synthesis instead the distribution of dopant ions in the NCs is more homogeneous (**Figure 6**b, sketch on the right). By exploiting the non-equilibrium synthesis conditions of plasma synthesis, uniformly doped AZO NCs with high dopant activation fractions and high local doping levels throughout the NC volume can be synthesized.[35]

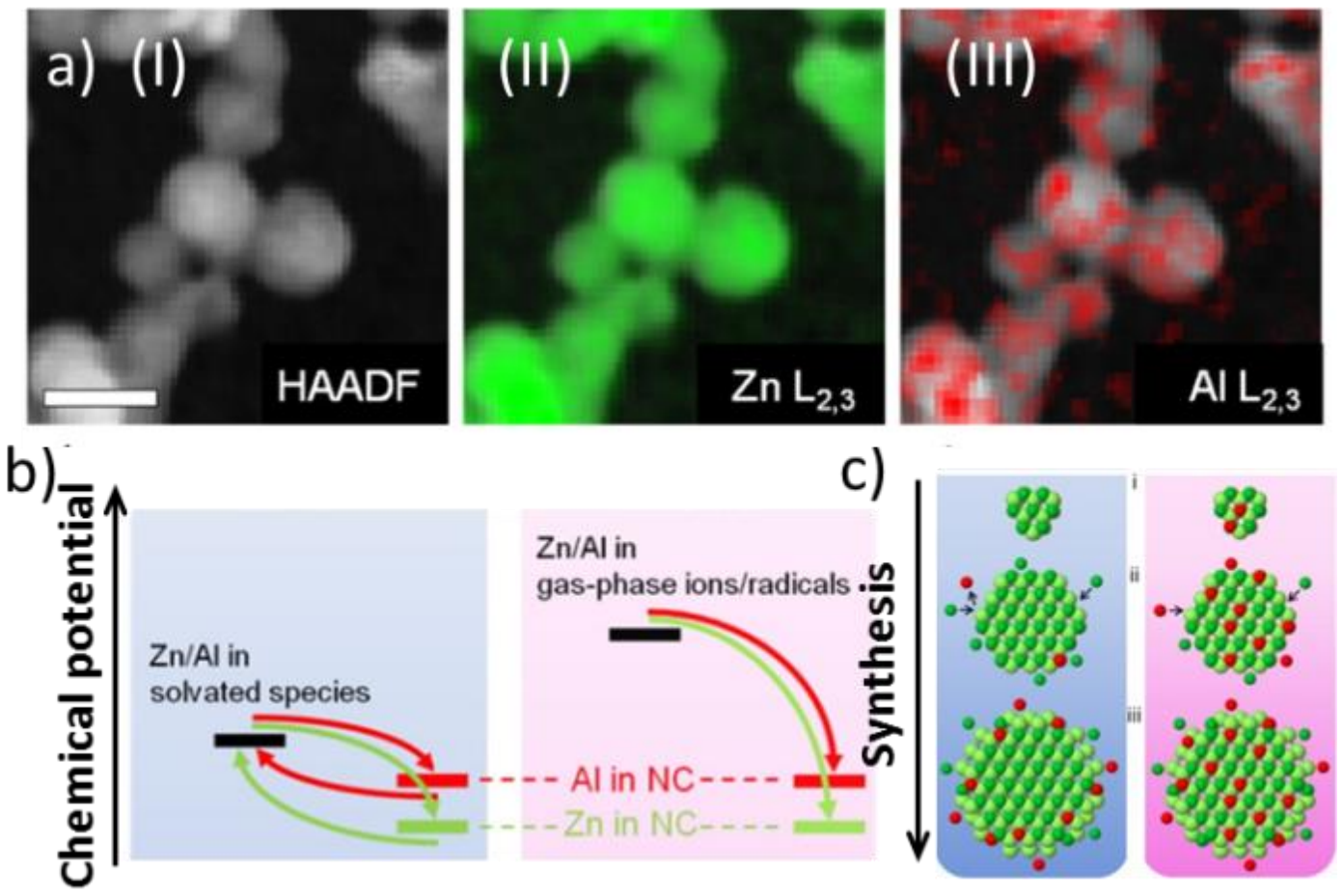

**Figure 6.** a) HAADF-STEM image of AZO NCs (I–III) Zn and Al L2,3 edges overlaid on the HAADF-STEM image, showing that the Al dopants are distributed throughout the NC volume. The scale bar is 10 nm. (b) and c) Schematic representations of AZO NC doping in a solvent (shaded blue area) and in the plasma (shaded red area). Zn, O, and Al atoms are dark green, light green, and red, respectively. a) In solution phase synthesis, the solvated species have much lower chemical potential than the gas-phase radicals and ions in plasma synthesis leading to the for-mation and dissolution of the nanocrystal nuclei multiple times before growing past a critical size. b) This might lead to the expulsion of dopant ions to the NC surface, whereas in the plasma synthesis the nucleation process is practically irreversible, such that dopant atoms, once placed in the nuclei remain there and doping throughout the core of the NC is achieved. Adapted with permission from ref [35] Copyright 2015 American Chemical Society.

Ionized impurity scattering arising from dopant ions or ionized vacancies in metal oxides is a reason for enhanced carrier damping and broad, asymmetric LSPRs in metal oxide NCs.[34] The impurities in metal oxides are the most important scattering centers and strongly influence carrier motion and carrier mobility. Deliberate defect engineering and rational dopant selection is a viable strategy to avoid such additional damping. Indeed, by choosing a dopant that minimizes ionized impurity scattering high electronic mobilities, very narrow LSPR peaks, and strong near-field enhancement can be reached, as found by Runnerstrom *et al.*.[94] For this purpose the authors chose cerium as a promising candidate for indium oxide doping due to several reasons: as cerium is more electropositive than other cations (for example tin), the Ce states are expected to lie within the conduction band (CB) of $In_2O_3$. This limits hybridization of the impurity states with the CB maximum and thus results in screening of the ionized impurities. Moreover, Ce is suggested to be more energetically stable as electron donor than oxygen vacancies, thus inhibiting their formation. Finally, as the ionic radius of $Ce^{4+}$ (101 pm) is closer to that of $In^{3+}$ (94 pm) than to the more common dopant $Sn^{4+}$ (83 pm), the introduction of $Ce^{4+}$ ions leads to a lower strain in the crystal lattice than for the $Sn^{4+}$ doping case. Indeed, the resulting $Ce:In_2O_3$ NCs show resonances with line widths as narrow as 479 $cm^{-1}$ (59 meV), corresponding to a quality factor of 5.5 for a single NC and field enhancement factors of 215 relative to the incident intensity at a frequency of around 3800 nm. This highlights the use of defect engineered $Ce:In_2O_3$ NCs for field-enhancement applications such as surface enhanced infrared spectroscopy.[94]

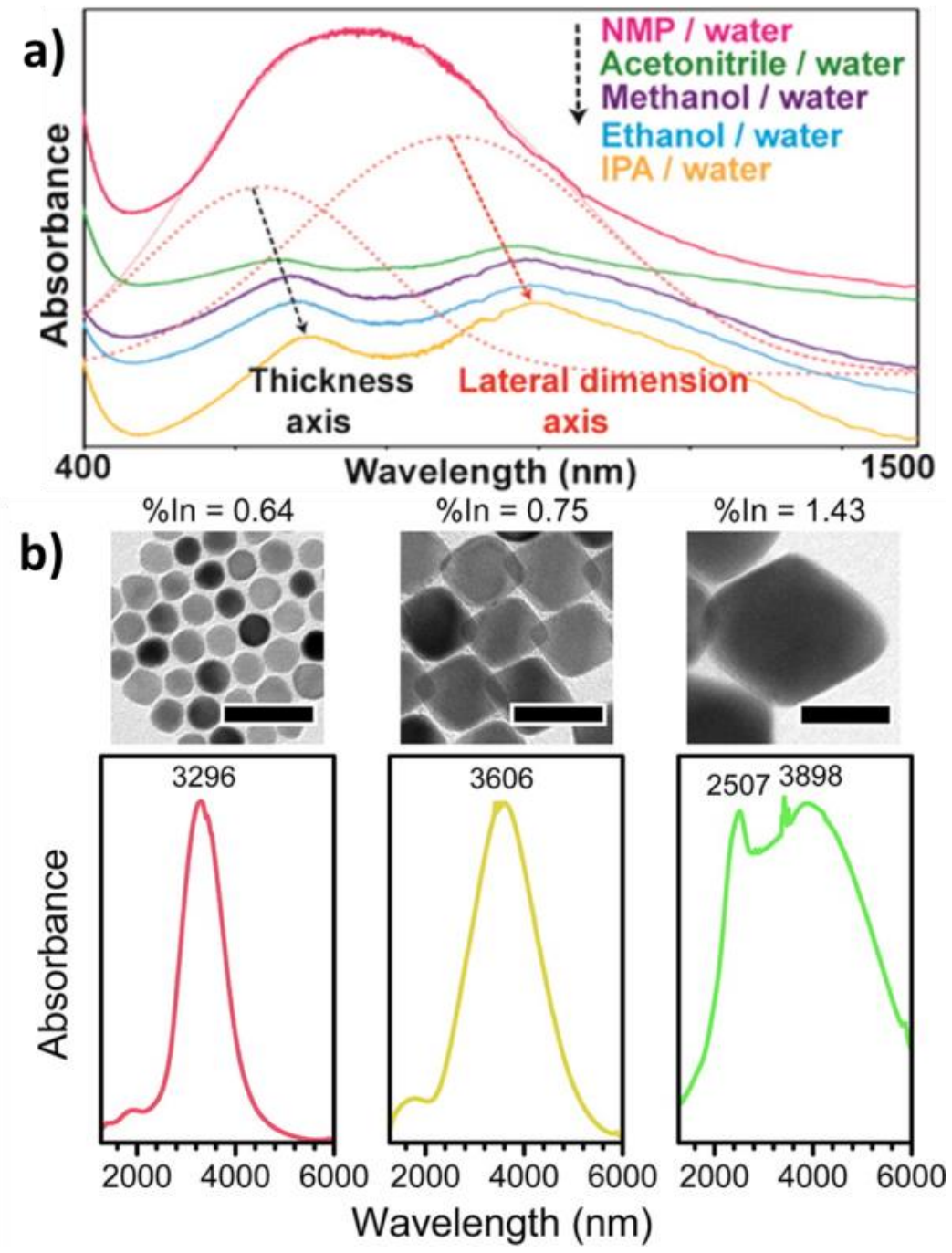

**Figure 7.** a) Absorbance spectra of the $MoO_3$ nanoflakes in different solvent/water mixtures. Adapted with permission from ref. [95] Copyright 2016 American Chemical Society. b) Solution phase spectra of three shapes/sizes (spherical, small octahedral, and large octahedral) of NCs with similar levels of In doping in $CCl_4$, The TEM images above corresponds to the sample measured optically (scale bar = 50 nm). Adapted with permission from ref [67]Copyright 2013 American Chemical Society

Notably, doped metal oxide NCs also show shape dependent plasmonic response in the NIR, which provides an additional means of control over the LSPR.[36,38,67,70,91,95–99] For example Mattox *et al.* showed strong shape dependent behavior of their Cs interstitial doped tungsten bronze $Cs_xWO_3$ NCs.[70] By varying the ratio of oleylamine to oleic acid, three distinct shapes of these NCs—hexagonal prisms, truncated cubes, and pseudospheres—were synthesized exhibiting strongly shape-dependent absorption features in the NIR. Two distinct resonances are observed for the hexagonal prism, which largely overlap in the truncated cubes, eventually becoming one peak in the pseudosphere. This indicates that the increasing number of facets on a nanocrystalline tungsten oxide surface causes the LSPRs to shift toward one another.[70] Two plasmon resonances were also presented in two-dimensional molybdenum oxide ($MoO_{3-x}$) nanoflakes.[95,96] The plasmon resonance in this material arise from the oxygen vacancies, where the thickness and the lateral dimension of the $MoO_{3-x}$ flakes (**Figure 7**a) lead to a shape dependent plasmonic response.[95,96] Also in indium-doped cadmium oxide (ICO) NCs a shape dependent response was observed (**Figure 7**b).[38,67] The authors presented a synthetic method which enables the control over shape with high level of monodispersity together with a control over the carrier concentration, thus allowing for a shape dependent investigation of the nanostructures. Modification of the reaction temperature and oleic acid concentration allowed the tuning of the ICO NCs' morphology and size, where the adjustment of the oleic acid concentration lead to size control, while an increase in temperature enabled morphology control. By changing the amount of In the level of doping can be directed and the LSPR tuned over a broad wavelength range.[67] Interestingly, a saturation of the LSPR peak shift upon doping is observed at high doping concentrations, which might well be due to trapping or scattering of the free carriers by In centers, as suggested for the case of ITO NCs.[34] Spherical NCs and two different sizes of octahedral NCs (in the dipolar, quasi-static limit) were compared with similar doping concentrations. The highly symmetrical resonance for the spherical ICO NC is observed at 3296 nm, while the octahedral NCs displayed a significant red-shift and broader peaks. More importantly, the larger octahedral NC sample had two distinct peaks, one at 2507 nm and one at 3898 nm, an effect of the shape of the ICO NCs on the LSPR (**Figure 7**b).[67] Indeed, the emergence of a single resonance for the smaller octahedral NC might be a result of the rounding of the edges and corners of the NC, as shown in simulations of ICO nanocubes by Agrawal *et al.*,[38] where blue shift and narrowing of the peaks was observed upon rounding, and lower- and higher-order modes moved more close to each other, until only the most intense dipolar mode dominated the spectrum.[38]

## 3.2. Plasmon tuning in impurity doped metal oxide NCs by capacitive charging: photodoping and electrochemical doping

### 3.2.1. Capacitive charging of metal oxide NCs

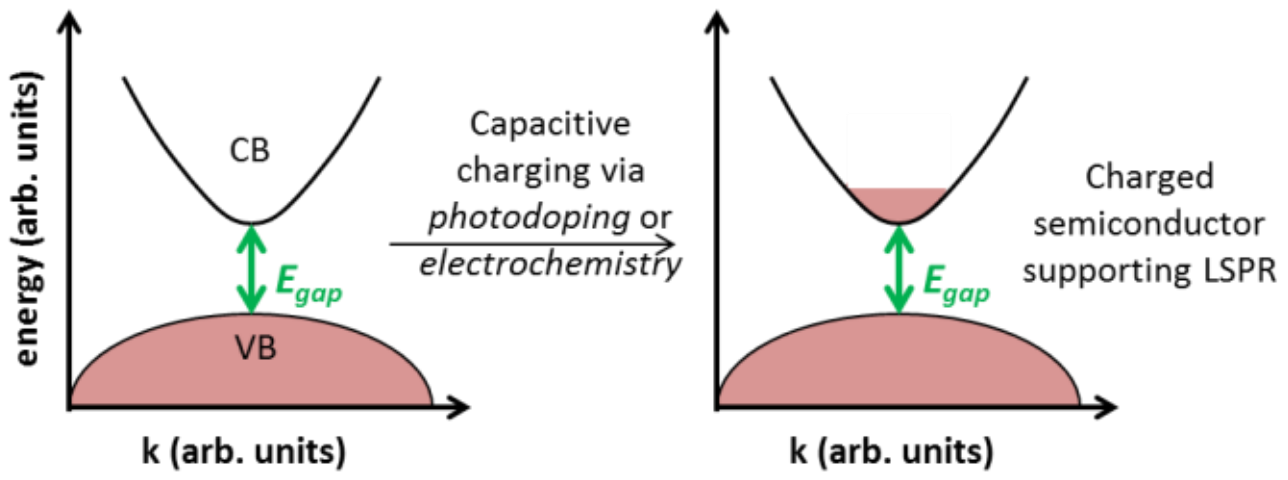

**Figure 8.** Illustration of capacitive charging of metal oxides. Extra carriers are added to the conduction band (CB) of the semiconductor either *via* photodoping or *via* electrochemical charging.

A notable characteristic of metal oxides is their capability to uptake additional charge carriers into acceptor states in the conduction band (CB) upon capacitive charging (**Figure 8**). The introduction of excess delocalized charge carriers via post-synthetic processes ultimately results in a blue-shift of the plasmon resonance, which can be modified over a large spectral range by controlling the level of doping.[9,22,100] In early works, excess band-like carriers were introduced into colloidal semiconductor NCs such as ZnO, CdSe, and PbSe by electron transfer from solvated reductants, leading to excitonic bleach in the visible and new intense absorption bands in the infrared, due to delocalized electrons in the quantized CB of such NCs.[101–106] In contrast, metal oxides present a manifold of states in the CB that can be filled and depleted by interaction with their environment. The order of magnitude lower carrier density of MOs relative to noble metals makes their LSPR very amenable to manipulation. The slight variation of the carrier density induces a large shift of the LSPR. Thus, a post-synthetic charging or discharging of MO NCs is directly related to a modulation of the LSPR frequency and intensity and delivers a tool for the modification of the dielectric properties in the region of the free carrier density. This high tunability offered by MO NCs was addressed by two different post-synthesis directions: photodoping and electrochemical doping. The first is based on the absorption of light in the region of the interband transition, which promotes an electron from the VB to the CB. The scavenging of the hole allows the photoexcited electron to remain in the CB, and this is tracked as a more intense and blue shifted LSPR due to the increased carrier density.[102,105,107,108] In another approach, charges are injected and extracted from the NC by electrochemical doping. In this case the modification of the LSPR can be tracked as a function of the applied bias.[22,94] In the following, we present these two approaches of carrier introduction into the CB of a metal oxide in a capacitive manner and give an outlook on the challenges and the issues connected with the manipulation of the LSPR properties of MO NCs.

### 3.2.2. Photodoping of MO NCs

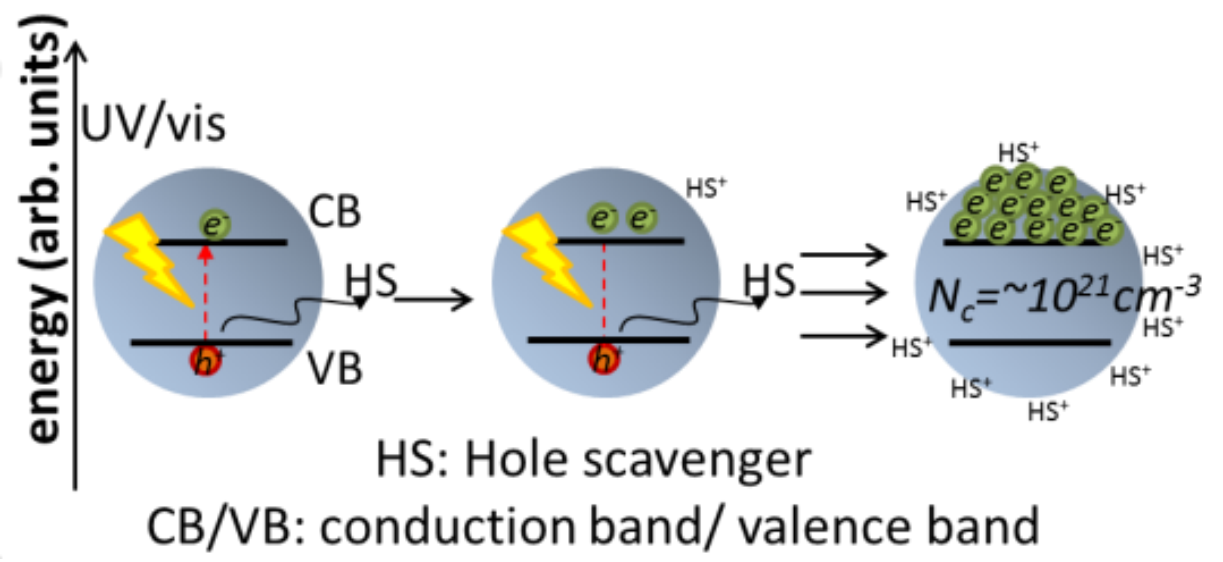

**Figure 9.** a) Illustration of the capacitive process: light with energy larger than the optical bandgap (usually in the UV/vis) excites an electron from the valence band (VB) to the conduction band (CB). The hole reacts with a hole scavenger (HS) leaving the electron behind, charge balanced by a positive cation at the surface ($H^+$). The additional electron remains in the NC tracked as a higher carrier density, i.e. stronger and blue-shifted LSPR. The amount of stored charges can be controlled by adjusting the pump intensity.

Anaerobic photoreduction in the presence of a hole scavenger represents an attractive option for LSPR modification *via* photodoping of MO NCs dispersed in a solvent (**Figure 9**a). Light with energy higher than the optical bandgap (typically in the UV-vis) excites an electron from the valence band (VB) to the CB. In the bare system, the excited electron in the CB adds to the inherent carrier density, leading to a temporary higher number of charges with a lifetime of several picoseconds, before it recombines with the hole in the VB.[109,110] This process can be detected by a blue shift of the LSPR with respect to the initial one due to the temporarily increased carrier density. In time resolved experiments (such as transient absorption spectroscopy) this is measured on ultrafast time scales as a photoinduced absorption (PIA) band in the NIR close to the frequency of the LSPR.[109,111,112] By increasing the pump power, more electrons accumulate in the CB and a higher intensity PIA signal is detected, which blue shifts with increasing pump power.[109,112] By adding a hole scavenger (HS) to the solution the extra charges can be stored in the NC under anaerobic conditions (**Figure 9**a).[102,105,107,108] This chemical reactant captures the hole from the NC leaving only the electron behind and charge balanced by a positive cation (such as protons) at the NC surface.[66,102,113,114] In this way, the additional charges remain stable in the NC and can be tracked as an increased and blue shifted LSPR due to the increased carrier density (**Figure 9**b).[114,115] The amount of additional charges in the NC is regulated by adjusting the pump intensity.[100,108,114,115]

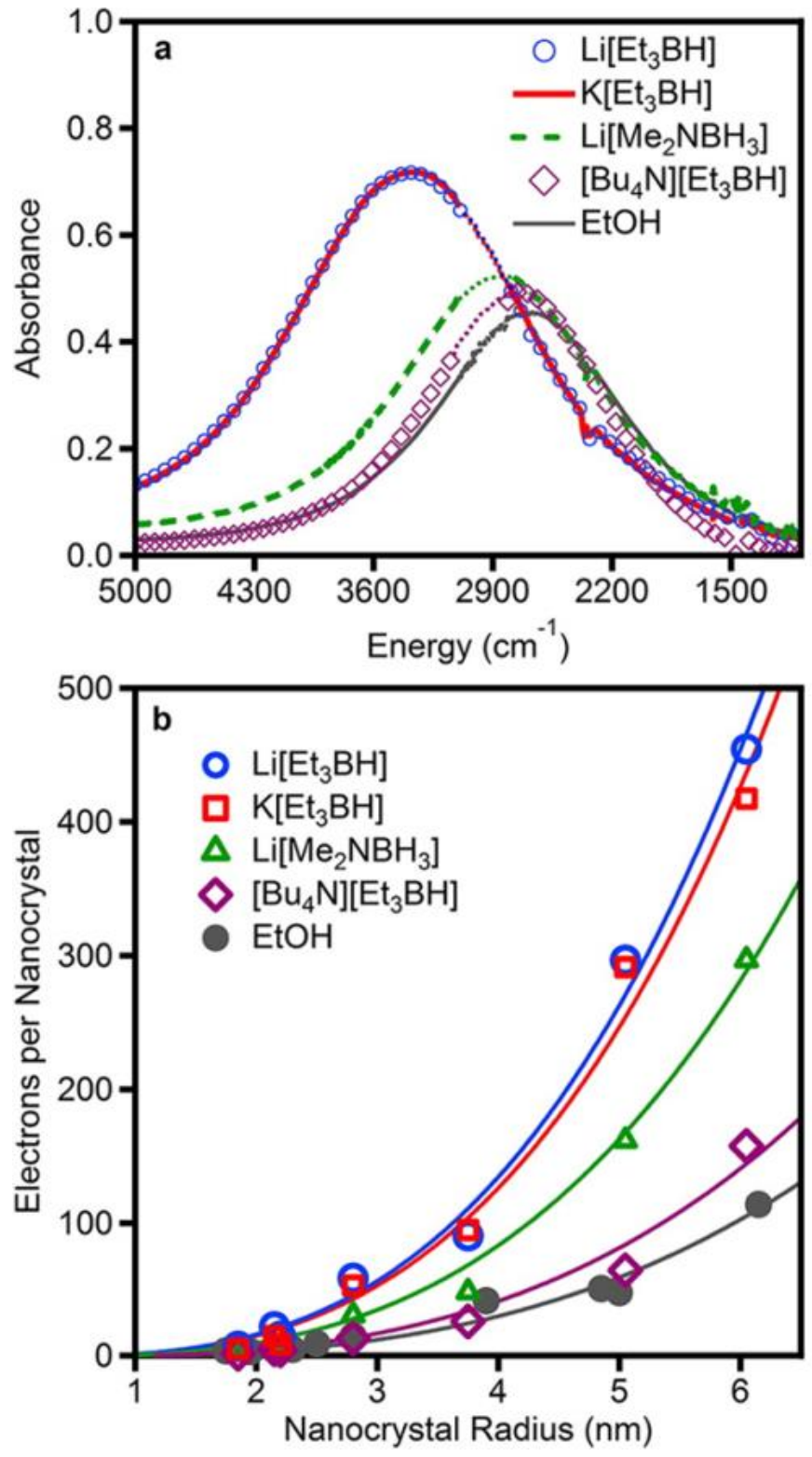

**Figure 10.** (a) IR absorption of colloidal $r$ = 2.8 nm ZnO NCs (TOPO-capped in 1:1 toluene/THF) photodoped in the presence of EtOH (solid gray line), $[Bu_4N][Et_3BH]$ (purple diamonds), $Li[Me_2NBH_3]$ (dashed green line), $K[Et_3BH]$ (solid red line) and $Li[Et_3BH]$ (blue circles) hole quenchers. The region of intense ligand and solvent C–H stretches (~2800–3100 cm–

1) was interpolated for clarity (dotted lines). (b) Size dependence of $\langle n_{max} \rangle$ of ZnO nanocrystals photodoped using the hole quenchers from (a). The solid lines show the relation of $\langle n_{max} \rangle \propto r^p$ in which $p$ is fixed to 3.0. Reprinted with permission from ref [100] Copyright 2013 American Chemical Society.

A prominent model system for photodoping of colloidal MO nanocrystals is ZnO.[100,115] Schimpf *et al.*[100] demonstrated that photodoping of ZnO NCs leads to an average maximum number of excess electrons per NC, $\langle n_{max} \rangle$, which increases in proportion to the NC volume for a specific hole scavenger, such as ethanol. Thus, the maximum carrier density remains constant for all NC sizes, solvent, surface-capping ligands, or photoexcitation rates.[100] However, a strong influence was ascribed to the charge compensating cation at the NC surface.[116] Therefore, Schimpf et al.[100] investigated a variety of different hole quenchers, as depicted in **Figure 10**. Indeed, depending on the reductant chosen as sacrificial hole scavenger, a different number of CB electrons can be accumulated by photochemical oxidation, and thus is not an intrinsic property of the multiply reduced ZnO NCs themselves. $\langle n_{max} \rangle$ can be markedly increased by changing the hole quencher (see LSPR intensity for different hole quenchers in **Figure 10**a). Ethanol is typically employed as a hole quencher, although in comparison up to 4-fold carrier densities were reached when using substituted borohydrides as quenchers.[100] For all investigated hole quenchers the relationship of $\langle n_{max} \rangle$ values as a function of NC radius shows the same increasing trend in proportion with the radius, as explained above (**Figure 10**). Notably, all hydrides examined yielded substantially faster photochemical electron accumulation than EtOH, despite their lower concentration.[100]

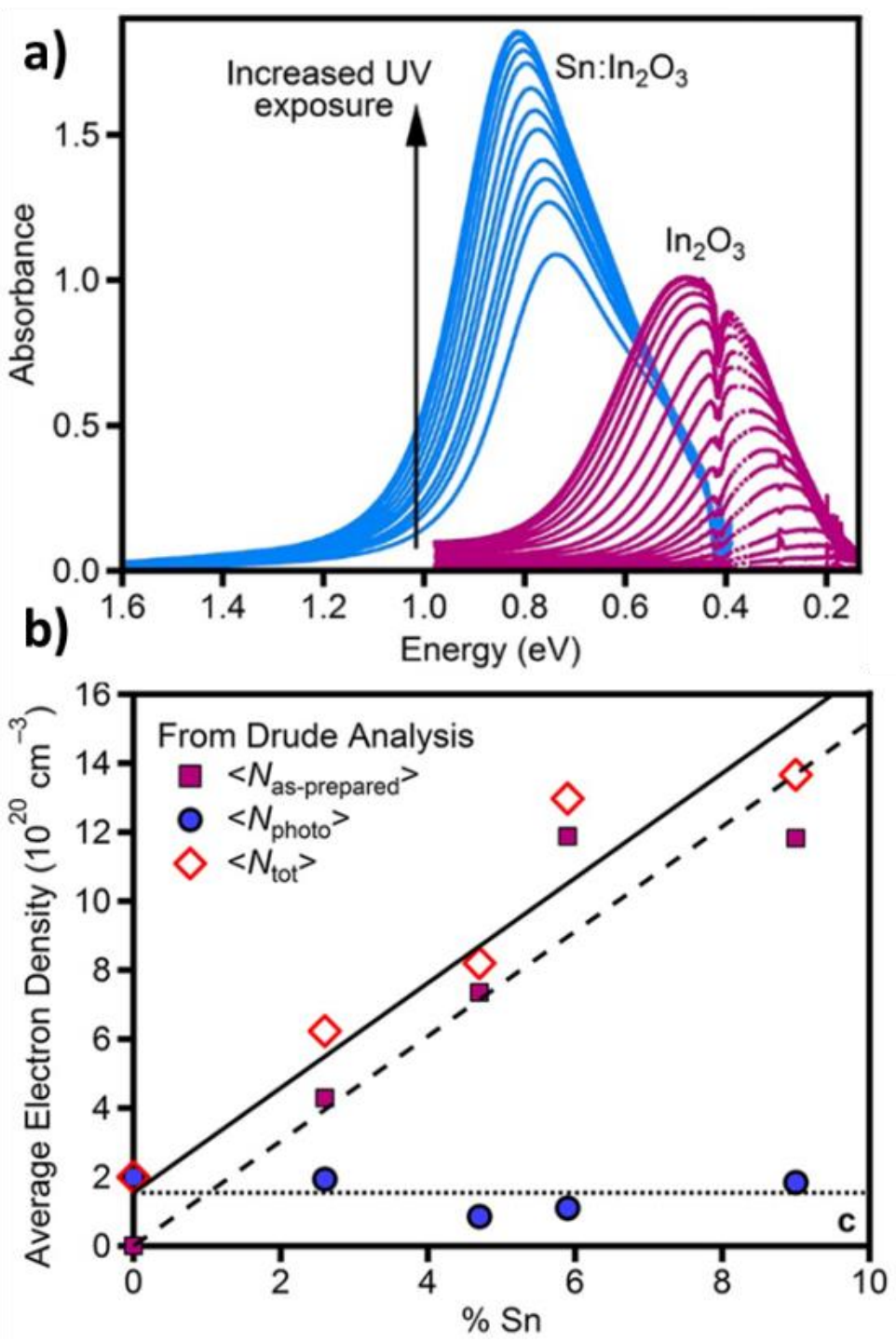

**Figure 11.** Absorption spectra before and following various extents of photodoping of $In_2O_3$ (right, ∼0.4 eV) and 9.0% Sn-doped $In_2O_3$ (left, ∼0.8 eV) NCs. The arrows show the direction of increasing photodoping. (b) Electron densities, $\langle N_{\text{as-prepared}} \rangle$ (squares), $\langle N_{\text{photo}} \rangle$ (circles), and $\langle N_{\text{tot}} \rangle$ (diamonds) obtained from Drude analysis plotted versus nanocrystal Sn content. These results were confirmed from titrations. Adapted with permission from ref [117] Copyright 2015 American Chemical Society.

While the hole scavenger plays an important role in determining the maximum number of extra electrons that can be added photochemically, such number is largely unaffected by the starting number of electrons introduced by aliovalent doping. In transfer between the two species in solution was observed, while in a mixture with aliovalently doped ITO NCs with undoped $In_2O_3$ NCs no electron transfer was detected.[117,118] This stabilization of the CB electrons in the aliovalently doped ITO is due to strong $Sn^{4+}$–electron Coulomb interactions.[119] Indeed, also in Aluminium doped ZO NCs (AZO) NCs the electrostatic interaction between lattice-bound $Al^{3+}$ and electrons in the CB appears to confer oxidative stability that is not found in photochemically reduced ZnO NCs, where charge compensation involves positive carriers. This leads to qualitatively different chemical reactivities.[66,116] This characteristic can be beneficially exploited for a controlled, fast and efficient extraction of the previously photochemically introduced charges. The addition of an electron acceptor to the reaction solution leads to a complete reversion of the photodoping, with a recovery of the spectroscopic properties to their original values.[100] In this manner, the photochemically added electrons can be directly quantified by titration against mild oxidants such as $[FeCp^*_2]^+$.[66,114,115,117] The extra electrons are removed, while no effect is observed on those deriving from chemical doping,[117] a result of the greater chemical stability of the free carriers introduced via aliovalent doping.[66,117] On the other hand, the LSPR energies can be affected separately by both the number of conduction band electrons and the number of impurity ions in the NC: greater Sn incorporation leads to higher LSPR energies, even without introducing more conduction-band electrons. This effect suggests that the LSPR frequency is not purely dependent on the carriers contributing to the plasmon resonance and that there is a marked influence from the doping mechanism.[117]

Figure **11**a photodoping experiments are shown for two different samples: undoped $In_2O_3$ (right, ~0.4 eV) and 9.0% Sn-doped $In_2O_3$ (left, ~0.8 eV) NCs.[117] The maximum number of added charges was shown to be approximately the same in the two samples; therefore it was not related to the initial doping level (

Figure **11**b). This is rationalized by considering that each conduction band electron within the doped ITO NC is compensated by a $Sn^{4+}$ ion. The electrons that are then added by photodoping are instead stabilized by charge-compensating protons at the NC surface, and thus their chemical potentials are essentially the same in all NC samples, irrespective of the Sn content or the initial electron density.[117] Along these lines, in an experiment with photodoped NCs mixed with undoped $In_2O_3$ NCs fast electron

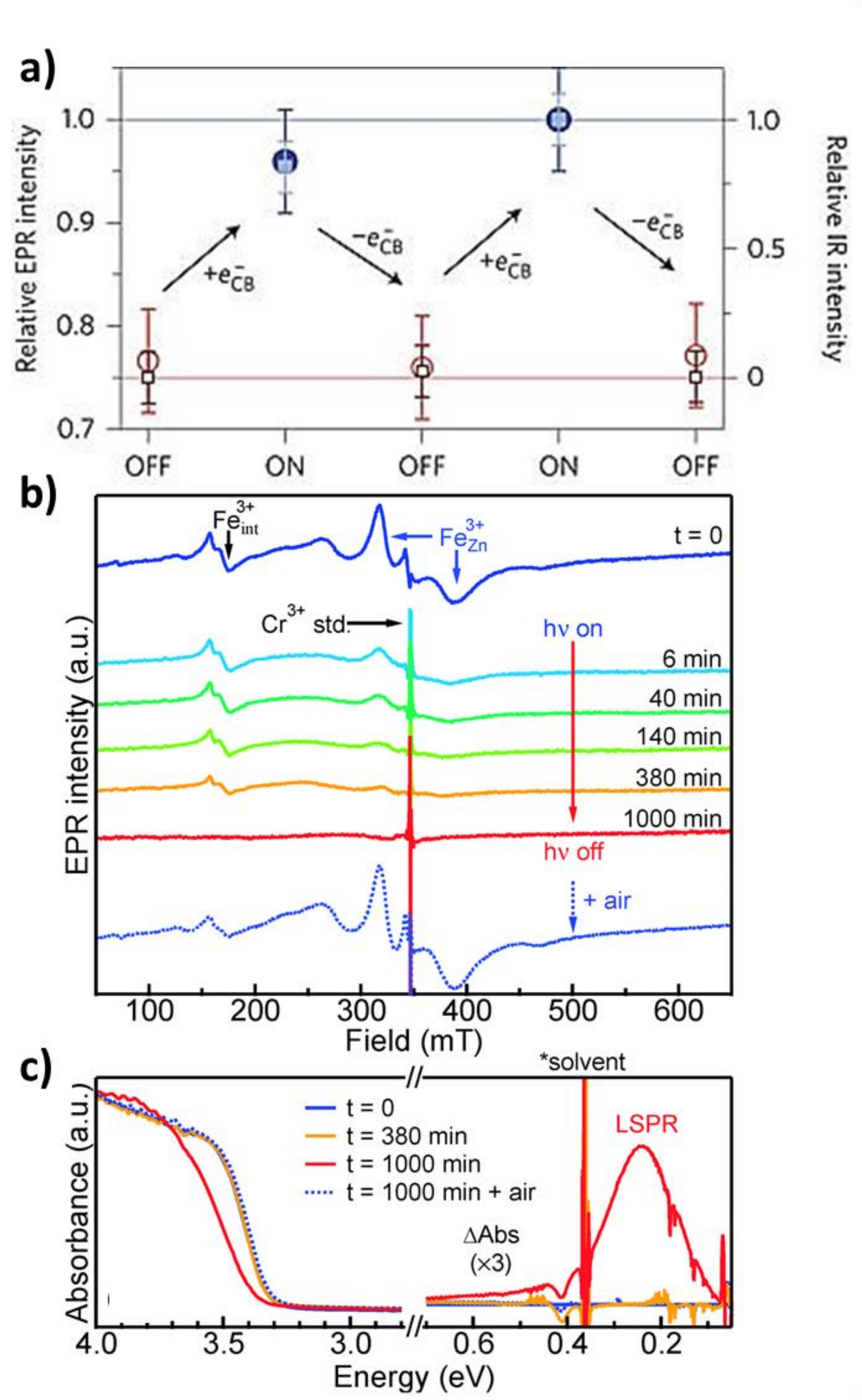

**Figure 12.** a) Relative intensity of the EPR signal of $Mn^{2+}$:ZnO NCs (open red circles) and photodoped $Mn^{2+}$:ZnO (filled blue circles) and IR absorption intensity of $Mn^{2+}$:ZnO NCs (open brown squares) and photodoped $Mn^{2+}$:ZnO NCs (filled blue squares) normalized at the highest observed intensity with repeated stepwise anaerobic UV irradiation (charging) and re-oxidation. Reprinted by permission from Macmillan Publishers Ltd: Nature Nanotechnology [49], copyright 2009. Evolution of the b) EPR and c) UV-Vis and FTIR spectra of 1% Fe-doped ZnO nanocrystals collected at various photodoping times. The sharp transition in b) at 350 mT is from $Cr^{3+}$-doped $SrTiO_3$ as a frequency standard (g = 1.978). Adapted from Ref [120] with permission of The Royal Society of Chemistry.

An attractive property of photodoping experiments and the accumulation of excess charge carriers in the CB is that several detectable signatures are involved: bleaching of the excitonic transition and photoluminescence,[121,122] the appearance/blue-shift of the LSPR (as discussed above), but also the emergence of a signal in the electron paramagnetic resonance (EPR) spectrum of the free electrons generated in the CB.[108] Thus, photodoping is a tool for active switching of a manifold of different signatures, among which the manipulation of magnetic moment is particularly interesting. The addition of unpaired electrons into colloidal MO NCs *via* photodoping reveals size-dependent g values in the range of $1.960 < g < 1.968$, as detected by EPR spectroscopy.[108,123] The appearance of the EPR signal confirms that these extra electrons are delocalized in the conduction band.[106] By choosing multiple dopants as reported recently in ($Fe^{3+}$,$Sn^{4+}$) co-doped $In_2O_3$-based NCs, free charge carriers are created by $Sn^{4+}$-doping,[90,124] while a magnetic moment is introduced by $Fe^{3+}$ doping. Charge mediated magnetic coupling between the free carriers in the system and the $Fe^{3+}$ ions can be useful for spin-based applications.[90] Active switching of the magnetic moment in MO NCs is achieved when exploiting photodoping instead of aliovalent doping. For example it was observed that photochemical reduction can induce a large, reversible, room-temperature magnetic response in colloidal ZnO NCs doped with $Mn^{2+}$ ions. The $Mn^{2+}$ ions in doped NCs – without photodoping - couple to form antiferromagnetic dimers that show basically no magnetization. The introduction of extra charges helps to reduce the pairwise $Mn^{2+}$–$Mn^{2+}$ magnetic exchange interactions via dopant–carrier exchange coupling that induces magnetic ordering in its vicinity and facilitates the observation of the magnetization of all $Mn^{2+}$ ions in the crystal.[49] **Figure 12**a depicts the relative EPR intensity and IR absorption intensity due to the introduced CB electrons with repeated stepwise anaerobic UV irradiation (charging) and re-oxidation (decharging). With photodoping (ON), charges accumulate in the CB, resulting in the evolution of the IR band and the disappearance of the EPR signal. This confirms the role of extra electrons in the CB of the $Mn^{2+}$ doped MO NC on the manipulation of spins and demonstrating the reversible switching of the magnetic properties of the $Mn^{2+}$:ZnO NCs.[49]

Another way to switch magnetization was demonstrated by Zhou *et al.*[120] in $Fe^{3+}$ doped ZnO NCs. Initially two different $Fe^{3+}$ sites in the crystal show a signal detected *via* EPR. In this case, photodoping leads to a depletion of the previously observed EPR signal of $Fe^{3+}$ with time of the photochemical reaction in the presence of a hole scavenger. This is explained by the reduction of $Fe^{3+}$ to the EPR-silent $Fe^{2+}$. As the $Fe^{3+/2+}$ redox potentials of the Fe sites in the ZnO NCs are placed at energies lower than the ZnO NC CB, the excited electron in the CB will first relax into these lower lying ‘trap’ states and thereby reduce $Fe^{3+}$. The photodoped electrons start to occupy the CB and induce LSPR absorption only after all $Fe^{3+}$ in the lattice were reduced to $Fe^{2+}$. Reaction with oxygen leads to the recovery of the EPR signal, along with the depletion of the LSPR. This new feature of combining magnetic properties with optical charging in doped semiconductor NCs reveals attractive new opportunities for the study of spin effects relevant to spintronics technologies, where the change in magnetic properties controlled by light is of interest for data storage and information technologies.[49,120]

### 3.2.3.Electrochemical charging of MO NCs

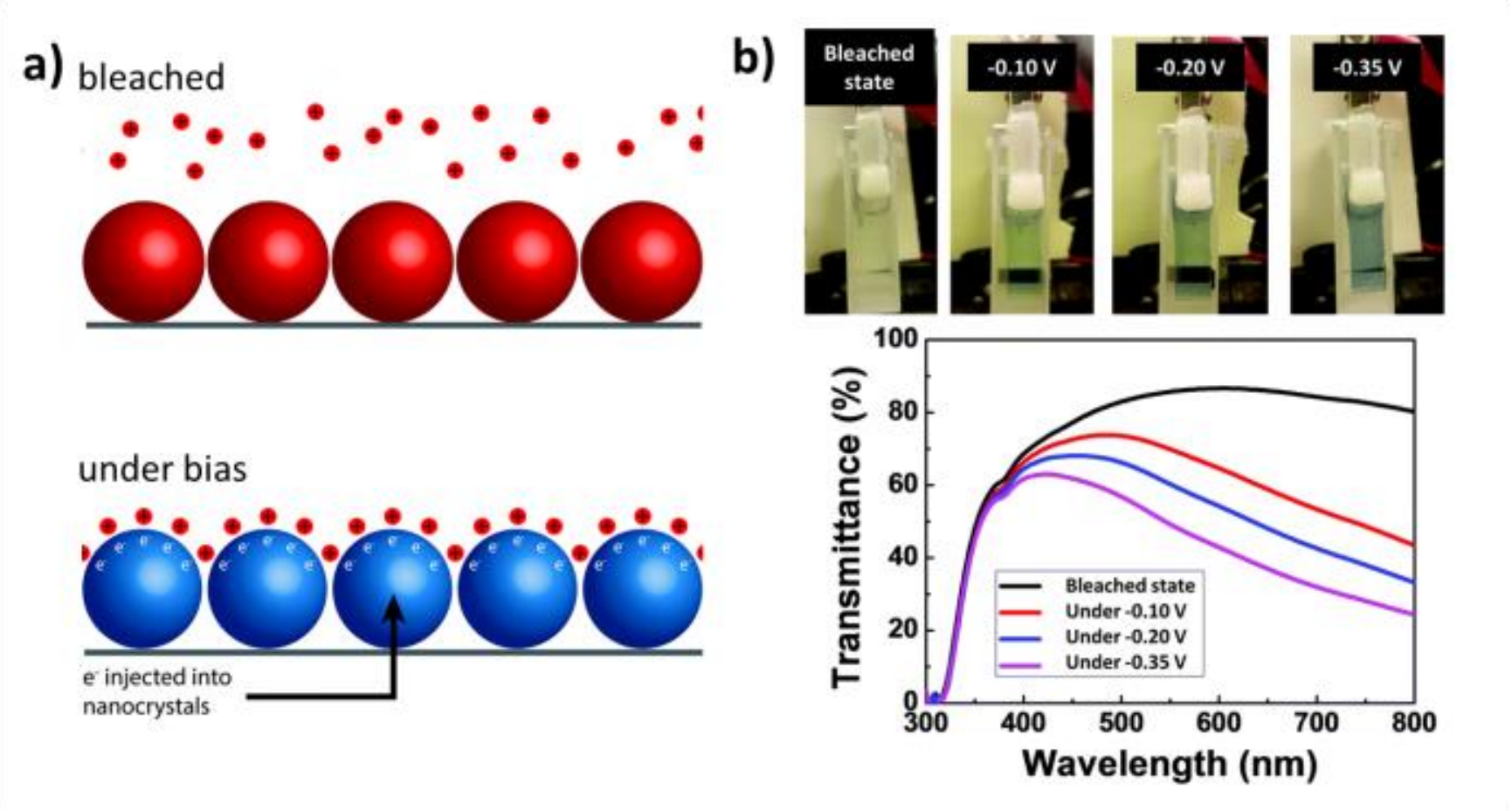

**Figure 13.** a) Depiction of the microscopic charging process of the plasmonic MO NC electrochromic film, along with the associated optical changes, where a positive potential leads to the depletion of electrons and the optical bleached state. The electron injection by applying a negative bias leads to the optical transmission changes. The positive counter ions are attached to the surface compensating the injected negative charge carriers capacitively.[10] - Published by The Royal Society of Chemistry. b) Color changes and the corresponding transmission spectra of a thin film of hexagonal $WO_3$ nanowires grown directly onto a bare fluorine-doped tin oxide glass. In the bleached state, the film was highly transparent and the intensity of the blue color can be controlled by adjusting the applied potential. Reproduced from Ref. [125] with permission from The Royal Society of Chemistry.

Another way of inducing a shift in the plasma frequency of metal oxides is by the accumulation of carriers under an applied field, as already mentioned in the introduction to this chapter. By applying a field across an ITO layer in a metal-oxide-semiconductor heterostructure, the carrier concentrations at the dielectric/conducting oxide interface can be increased from $1 \times 10^{21}/cm^3$ to $2.8 \times 10^{22}/cm^3$.[126] Accumulation of carriers is induced up to 5 nm into the 300 nm thick ITO layer. The changes in refractive index and carrier density are limited to this region, which is commonly called the accumulation layer.[126] By applying an electric field it is relatively straightforward to accumulate carriers across the entire volume of a nanostructure. Such kinds of modifications were induced *via* electrochemical charging of conductive films of plasmonic NCs (**Figure 13**), such as ITO, but also in noble metal nanoparticles in solution. The modification of the carrier density of polymer-stabilized silver nanoparticles immersed in an electrolyte solution in an electrochemical cell was studied by Ung et al.[127] The changes in the LSPR were monitored directly as a potential was applied. The observed shift of the LSPR was ascribed entirely to electron transfer between the working electrode and the colloid particles, where charges are exchanged directly through the collision with the electrode.[127] Equilibration of the particles with the electrode upon applying an electrode potential was found to occur on a time scale as short as 150-200s, with the number of transferred carriers of the order of 1600 (electrons). This transfer of charges corresponds to a shift of the LSPR from 400 nm to about 392 nm.[127]

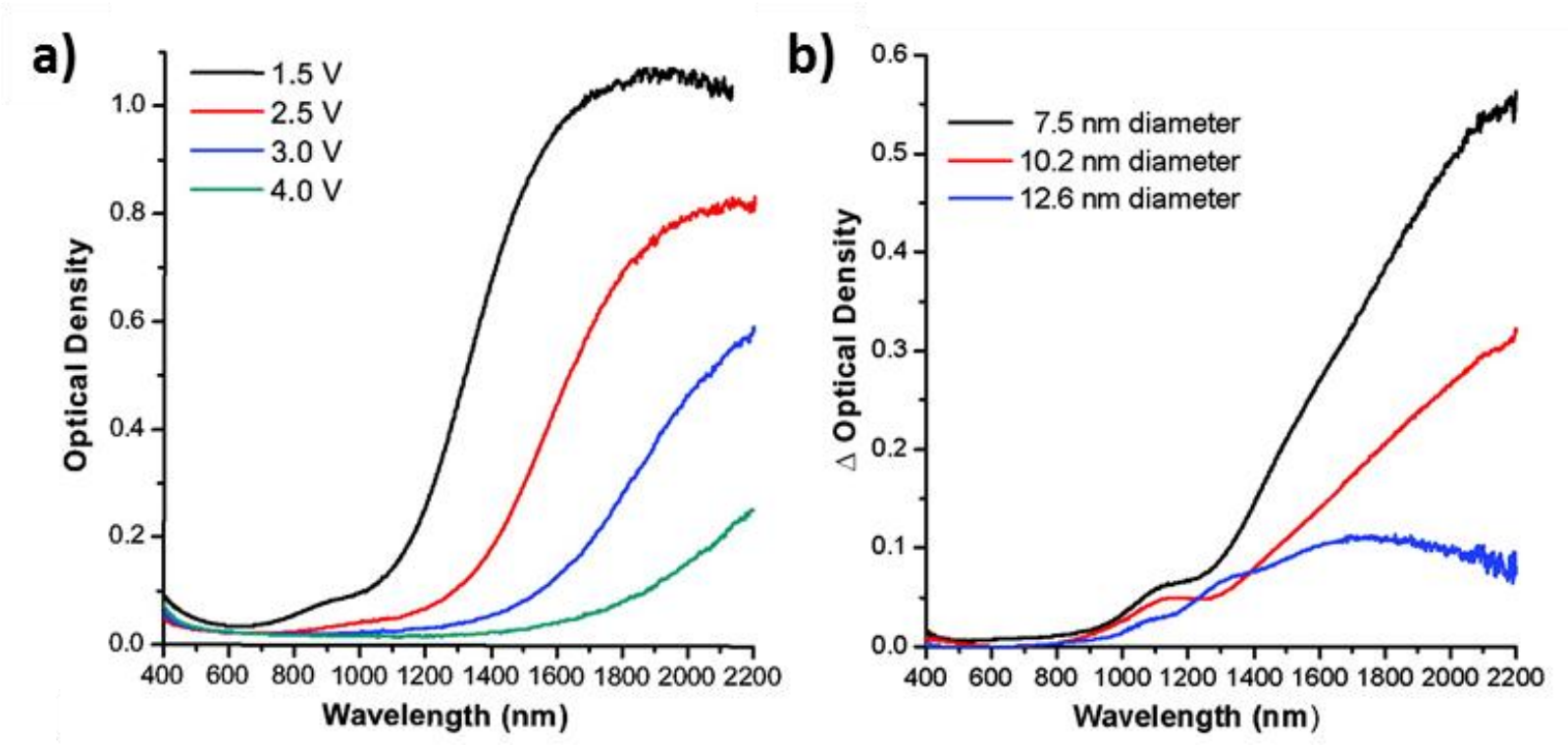

**Figure 14.** Optical properties of ITO NC films during electrochemical modulation. (a) Bias-dependent optical density of a film composed of 4.1 nm diameter, 16.8% Sn NCs. (b) Change in optical density between 1.5 and 4 V for various NC sizes, each with 4.65 ± 0.25% Sn. All potentials are referenced to a $Li/Li^+$ electrode in 0.1 M $LiClO_4$ in propylene carbonate. Adapted with permission from ref [22] Copyright 2011 American Chemical Society.

Considering these results, electrochemical charging in an electrochemical cell appears as an attractive alternative to introduce capacitive charge carriers also into MO NCs.[22,128] Indeed, the order of magnitude lower carrier density in MOs with respect to noble metals makes these materials more accessible to LSPR modification over a larger spectral range.[10,22] For example, when immersing an ITO NC film in a $LiClO_4$ electrolyte solution and applying a negative voltage to the electrochemical cell, the LSPR blue shifts down to 1200 nm, as demonstrated by Garcia et al.[22] **Figure 14**a shows the absorption of an ITO NC film during electrochemical modification for various potentials between 1.5 and 4V. As a negative bias is applied (1.5V) the LSPR shifts to higher energies and becomes more intense (black line). Such peak modifications are in accordance with the modulation of the free carrier concentration with electrochemical potential, due to carrier injection into the accumulation region of the material.[10,22] A positive potential reverses the optical changes with an additional red shift from its equilibrium position. No change in the electrochemical properties of the NCs was observed over multiple charge–discharge cycles. In this way the carrier density is modified between 4.5 and 14.5 × $10^{20}/cm^3$. For ITO NCs of 4.1 nm diameter this corresponds to a transfer of approximately 40 carriers. Notably, this is a markedly lower number of charges transferred compared to that found by Ung *et al.*[127] for silver nanoparticles, yet it induces a much larger spectral shift in the LSPR in the ITO NC film com-pared to the silver particles case.[22] This effect, which is due to the lower initial carrier density of the doped MO with respect to noble metals, underlines the potential of electrochemical charging as an efficient tool to manipulate the carrier density of MO NCs, hence their LSPR.

The use of NC films in which the NCs are bound into continuous, electrically connected films represents an appealing alternative to a planar ITO film. In MOs charge accumulation occurs only at a thin layer at the surface. The greater surface area of the NC film made of NCs that possess sizes within the accumulation region of the MO enables the injection of carriers throughout the entire volume of the NC film and thus results in greater tunability of the LSPR.[22] Indeed, as shown in **Figure 14b**, a greater LSPR modulation is achieved when smaller NCs are used instead of bigger particles.[10,22] The introduced charges are balanced by a charged layer at the NC surface (**Figure 13**a).[10] To ensure charge balancing, films with pores and high surface areas that are easily accessed by the electrolyte are important.[10,22] As such, the proper design of a nanocomposite architecture can contribute largely to its switching efficiency. As shown by Kim *et al.*,[129] $WO_{3-x}$ NCs were arranged into a mesoporous architecture to achieve rapid and reversible plasmonic switching. By contrast, a NC film made without mesopores exhibited a broad, red-shifted optical response due to strong LSPR coupling of the densely packed NCs.[129]

Despite very similar results for the electrochemical charging of ITO NCs when using tetrabutylammonium ($TBA^+$) electrolyte instead of LiClO4,[22] it is conceivable that the charge balancing ions limit the maximum number of electrons that can be injected, similar to the discussion on photodoping (section 3.2.2). Also the size of the counter ion might play a dominating role. The presence of dielectric layers and the potential dependent chemisorption of stabilizer molecules are particularly important when reconciling peak positions exactly with theory, as highlighted by Ung *et al.* in the case of silver nanoparticles.[127] Similarly, Brown *et al.*[130] developed a theoretical model that describes the optical properties of electrochemically modified gold nanoparticles in terms of changes in plasmon damping and changes in the index of refraction of the surrounding electrolyte medium, induced by the formation of an electric double layer at the NC surface.[130] In order to predict the optoelectronic behavior of the LSPR upon charging, models are required that assess the relative spectral contributions of several different effects separately, such as changes in carrier density, changes in damping, and of the index of refraction of the surrounding medium. This is particularly important for applications where knowing the exact carrier density is of importance, such as for sensing with single electron sensitivity as suggested by Mendelsberg *et al.*[31] A better quantification of the various contributions that affect the shape and position of the LSPR are crucial to improve applications based on tunable plasmonics.

# 4. Copper chalcogenide ($Cu_{2-x}E$; E=S, Se, and Te) NCs

Colloidal binary copper chalcogenide ($Cu_{2-x}E$; E=S, Se, and Te) NCs have attracted increasing attention over the last decade, because of their notable optical and electronic properties.[2,3,5,7,53,131,132] Differently from the impurity doped metal oxides previously discussed, and similarly to oxygen vacancy doped MOs (i.e. for $WO_{3-x}$ NCs[25]), the formation of copper vacancies in copper chalcogenides leads to self-doping of the material with carrier densities that can be as high as $10^{21}$ $cm^{-3}$. These intrinsically doped semiconductors can exist in a manifold of different stoichiometries, with very rich phase diagrams and a variety of equilibrium crystal structures.[133–135] Copper sulfide for example exists in the near stoichiometric chalcocite ($Cu_{2.00-1.995}S$), with a copper to sulfur ratio close to 2:1, in the sub-stoichiometric compounds such as djurleite ($Cu_{1.96}S$), digenite ($Cu_{1.8}S$), roxbyite ($Cu_{1.78}S$), or anilite ($Cu_{1.75}S$), and in the limiting case, stoichiometric covellite (CuS).[136] $Cu_{2-x}Se$ and $Cu_{2-x}Te$ can exist in several different structures too. Interestingly, in all these materials the optical properties are closely related to the structural and compositional parameters, since the ratio of copper to chalcogen determines the vacancy density, hence the level of doping. The stoichiometric chalcocite ($Cu_2S$) exhibits purely semiconducting behavior with a bandgap around 1.2 eV.[137] This feature, together with the low cost of $Cu_2S$, suggested its use for photovoltaic applications, with some interesting results.[6,138–141] On the other hand, the optical and electronic properties of chalcocite $Cu_2S$ are altered in oxidative environments due to tendency of the material to become sub-stoichiometric in Cu, with the missing $Cu^+$ ions compensated by a partial oxidation of the sulfide sublattice. A change from moderate doping ($10^{18}$ $cm^{-3}$) in $Cu_2S$ to carrier densities in the range of $10^{21}$ $cm^{-3}$ goes hand in hand with the formation of copper vacancies and the solid state transformation to copper-deficient crystal structures.[136,142] The increase in carrier density in copper chalcogenides is accompanied by changes in the optical constants in the NIR, where intraband absorption is observed. In nanoparticles, this increased carrier density results in strong LSPRs in the NIR.[1,11–13] A notable feature connected with the formation of vacancies in sub-stoichiometric copper chalcogenide NCs is the tunability of the LSPR. By controlling the copper stoichiometry in the nanostructure, the plasmon resonance can be manipulated over a broad range of frequencies, a feature that makes $Cu_{2-x}E$ (E=S, Se, and Te) NCs[1,11–13] interesting candidates for several applications in photocatalysis,[42,143,144] photothermal therapy,[27,28] and biomedical sensing.[6,131,132,145]

In the following, will we give a brief overview of the structural-electronic correlations in copper chalcogenides, with a focus on copper sulfide and its many crystallographic phases. We will describe the most popular synthesis methods for NCs of copper chalcogenides, and the means of controlling their LSPR. We will then give a brief summary of other copper and chalcogenides based NCs that present LSPRs, and hybrid structures thereof.

### 4.1. Structural characteristics of copper chalcogenides

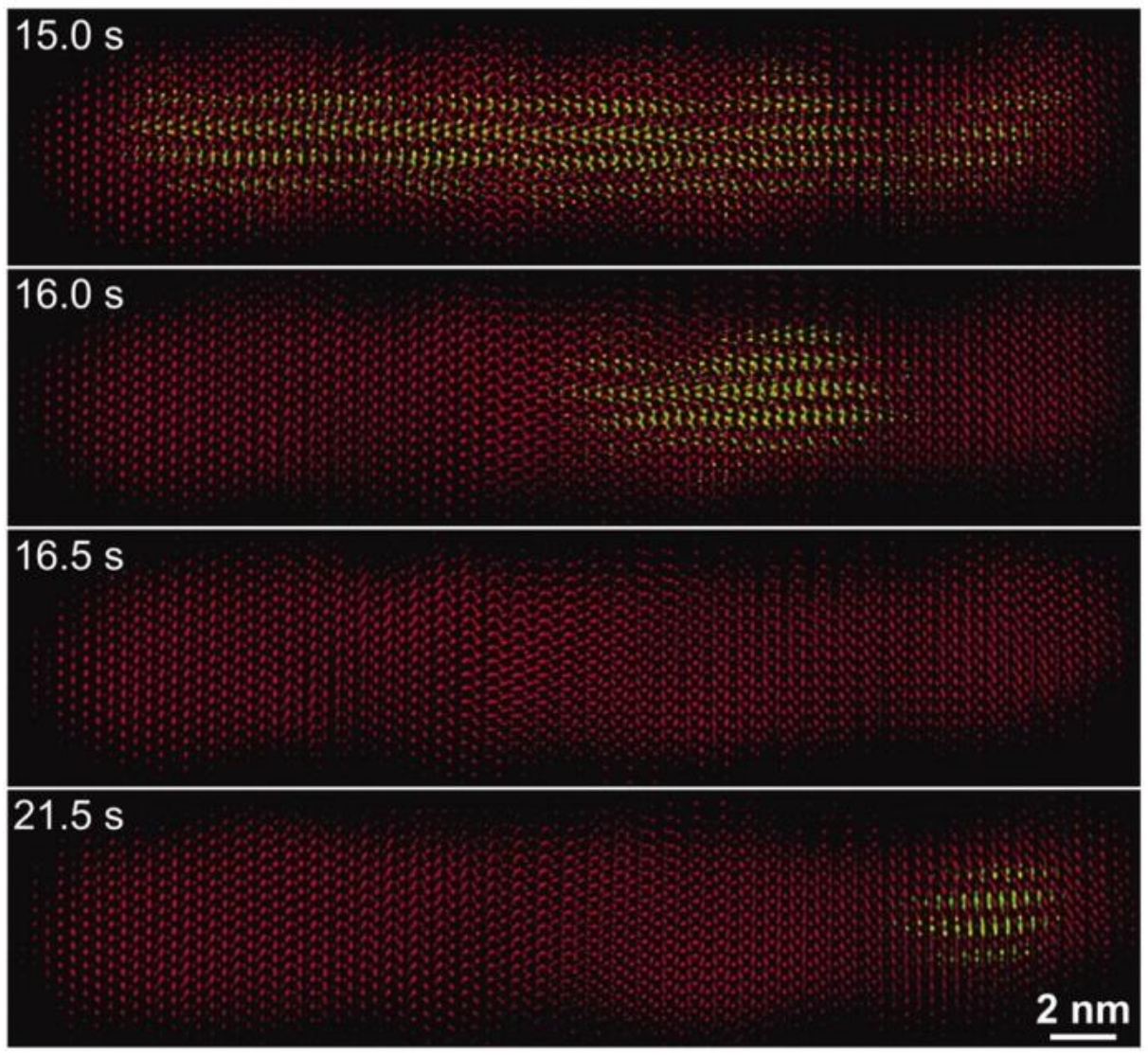

**Figure 15.** Trajectory of the structural transformations in a single $Cu_2S$ nanorod upon electron beam induced heating. Sequential images showing the α-chalcocite structure (green), mixed structure (mixed green and red domains), and the pure β-chalcocite structure (red). From ref [146]. Reprinted with permission from AAAS.

As mentioned before, copper chalogenides exist in a variety of crystallographic phases with varying copper to chalcogenide ratios.[136] Notably, chalcocite with the copper to sulfur ratio close to 2:1 exists in several different phases, such as the monoclinic (α-$Cu_2S$) and the hexagonal chalcocite (β-$Cu_2S$) structure. The α-chalcocite has a space group of $P2_1/c$ or Pc. The copper atoms partly occupy the lattice sites and the hexagonal sulfur lattice frame is distorted.[146] The β-chalcocite structure has a space group of $P6_3/mmc$, with a similar hexagonal sulfur sublattice and copper atoms occupying different lattice sites.[146] The energy involved in a structural transformation is close to room temperature, and thus thermal energy is sufficient to induce structural transformation. This was exploited by Zheng *et al.*[146] who report the direct observation of a structural transition between the two phases within a single copper sulfide nanorod *via* electron beam induced heating in the transmission electron microscope with atomic resolution (**Figure 15**).[146]

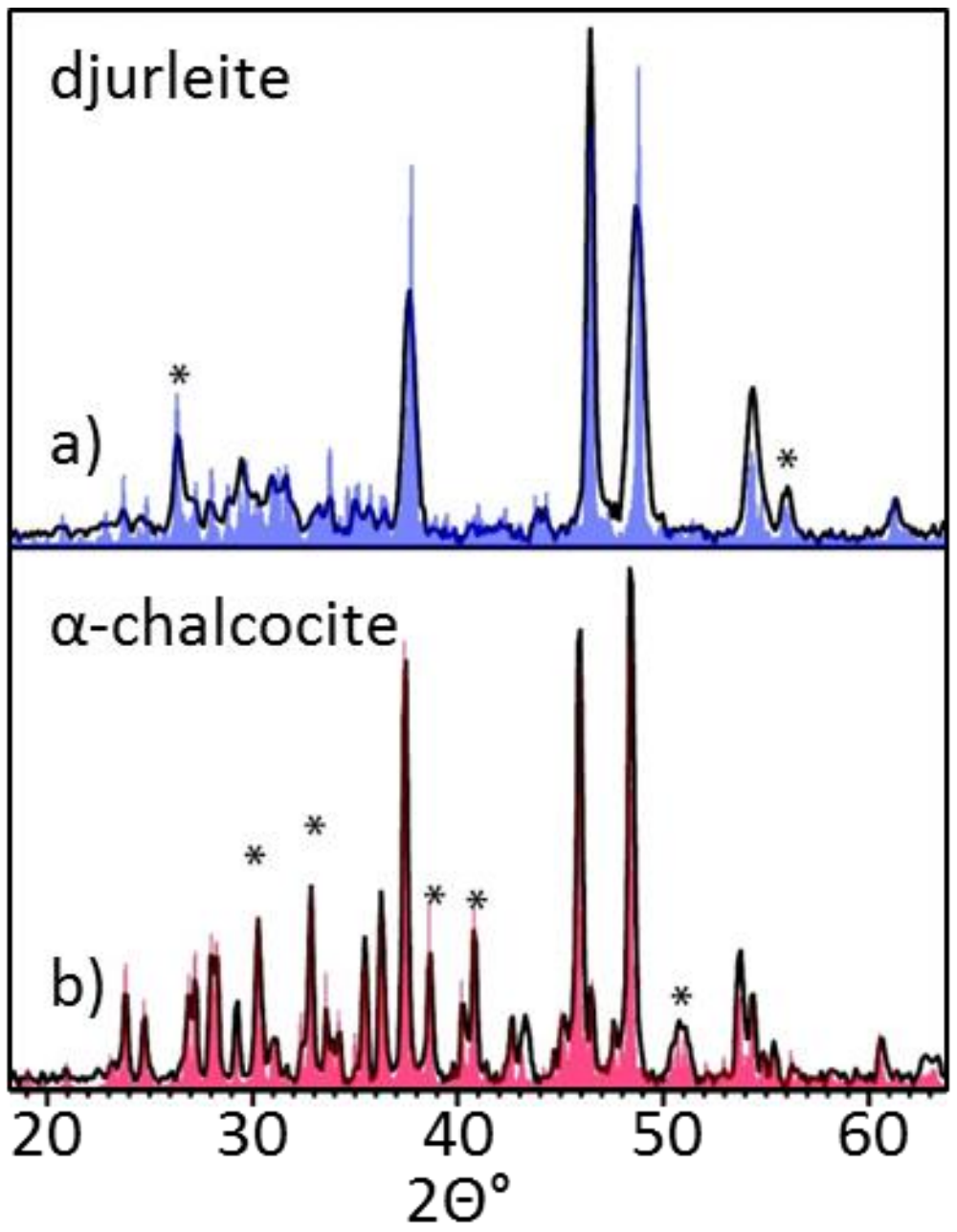

**Figure 16.** XRD patterns of copper sulfide NCs (black curves) with a) djurleite crystal structure and b) α-chalcocite. Matching patterns are overlaid calculated from the Inorganic Crystal Structure Database (a) 100334-ICSD and b) 100333-ICSD.) Noted with asterisks are some prominent peaks that distinguish α-chalcocite and djurleite. Adapted with permission from ref [147] Copyright 2015 American Chemical Society.

Chalcocite can additionally exist in a tetragonal phase, which represents the high-temperature structure of $Cu_2S$, and was successfully synthesized at the nanoscale by iron-induced stabilization.[148] In this way either monoclinic or tetragonal chalcocite NCs were selectively synthesized and stabilized with respect to the tendency to transform to djurleite, the copper deficient copper sulfide phase, as discussed later.[148] In tetragonal chalcocite, copper ions occupy trigonal sites in a body-centered tetragonal sulfide lattice, transforming to body-centered packing in tetragonal chalcocite.[148] In monoclinic and hexagonal chalcocite (α- and β-$Cu_2S$) and in substoichiometric djurleite ($Cu_{1.96}S$) the XRD patterns are very similar, with the main peaks nearly identical, and shifted only slightly towards higher 2θ angles in djurleite compared to chalcocite (**Figure 16**).[147–149] These structures are closely related through a nearly identical sulfide sublattice, with the main structural differences given by the mobile copper cation positions.[149] Djurleite is the thermodynamically favored phase at room temperature. Indeed, the formation of pure chalcocite with a Cu:S ratio of 2:1 is very unlikely at ambient and oxidizing environments, and its transformation to djurleite is kinetically favored.[148]A lower Cu:S stoichiometric ratio (1.8:1) is found in the digenite crystal structure. Digenite forms a highly symmetric face-centered cubic anion structure. Bulk digenite has three polymorphs. The cubic phase of digenite has a face-centered cubic structure while the rhombohedral phase resembles the hexagonal chalcocite, with one copper atom distributed over four equivalent positions in each sulfur tetrahedron.[150] The limiting case of covellite with the Cu/S ratio of 1:1 represents a special case in terms of its electronic properties, as will be discussed in detail in the following chapter. The XRD pattern of covellite is characterized by four strong characteristic diffraction peaks that represent a unique fingerprint of the hexagonal CuS phase.[151]

### 4.2. LSPR tuning: Oxidation and reduction affect the electronic properties of copper chalcogenide NCs.

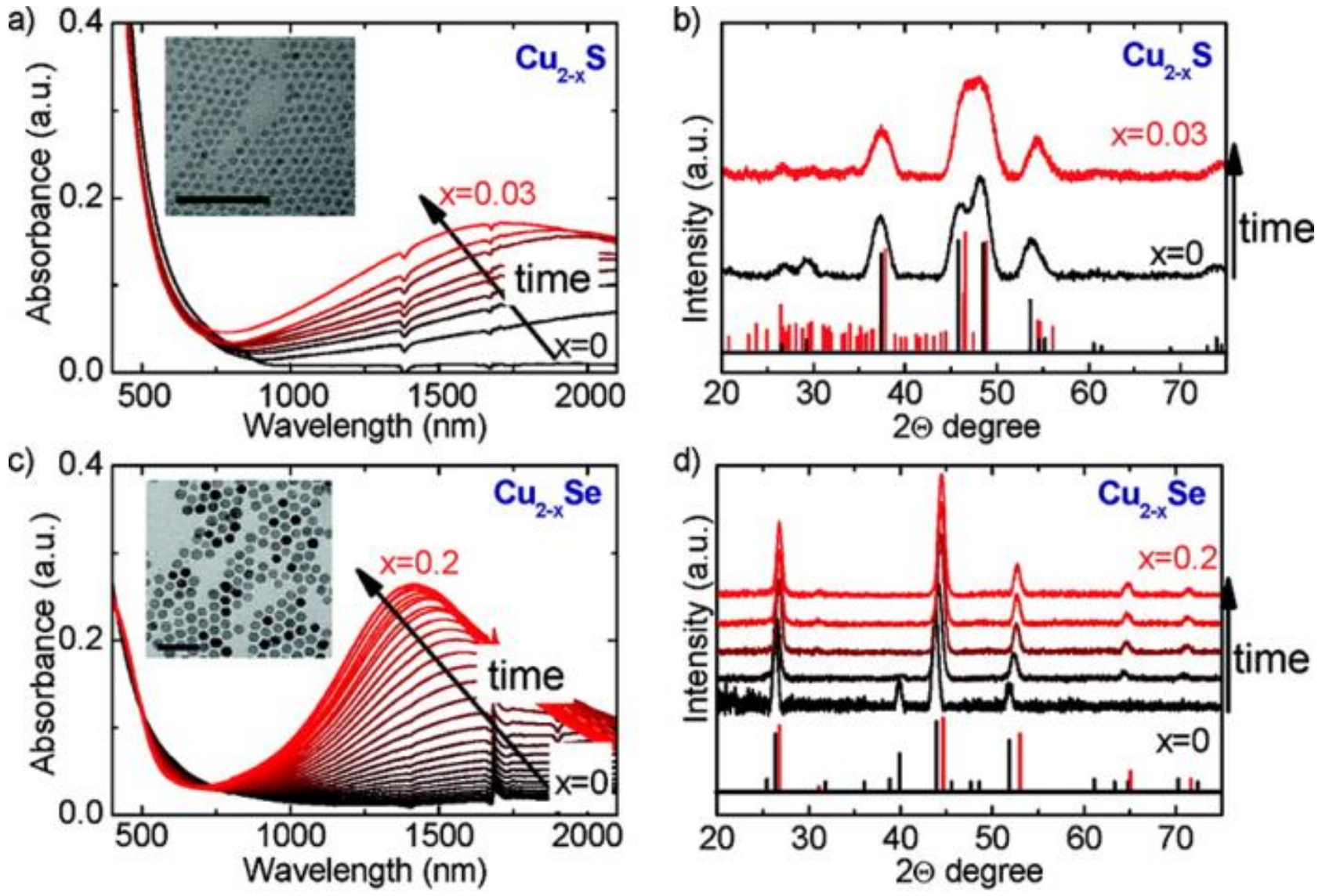

**Figure 17.** Time evolution of the absorbance spectra of stoichiometric (a) $Cu_{2-x}S$ (x = 0, black curve) and (c) $Cu_{2-x}Se$ (x = 0, black curve) NCs in toluene during oxidation (from black to red) with a gradual increase of copper vacancies (x increases). Insets: TEM micrographs of the non-oxidized samples (the scale bar is 50 nm). (b) XRD patterns of the $Cu_{2-x}S$ NCs shown in (a) before (x = 0, black curve) and after (x = 0.03, red curve) exposure to oxygen. (d) Time evolution of the XRD patterns of the $Cu_{2-x}Se$ NCs during oxidation (from the black to the red curve). Adapted with permission from ref [12] Copyright 2015 American Chemical Society.

In copper chalcogenides ($Cu_{2-x}S$, $Cu_{2-x}Se$, $Cu_{2-x}Te$), the top of the valence band is essentially built from *p* orbitals of the chalcogenide anions, while the bottom of the conduction band has strong contribution from the Cu *4s* orbitals. In the fully stoichiometric materials, the copper ions in the lattice have a +1 valency, while the chalcogenide ions have a -2 valency. When these materials are exposed to an oxidizing environment, electrons are extracted from them along with $Cu^+$ ions in order to ensure electrostatic neutrality.[13] In the case of copper chalcogenide NCs stabilized in a solution environment, the extracted $Cu^+$ ions will generally form complexes with the solvent or with ligands molecules present in solu-tion.[12,152] The removal of electrons from these materials will occur from the electronic levels at the top of the valence band, with their predominantly chalcogen *p* type character. It turns out then that such oxidation process affects mainly the anion framework, which will have oxidation state between -2 and -1 in the substoichiometric compounds, while the Cu ions in the lattice will continue to have oxidation state around +1.[152]

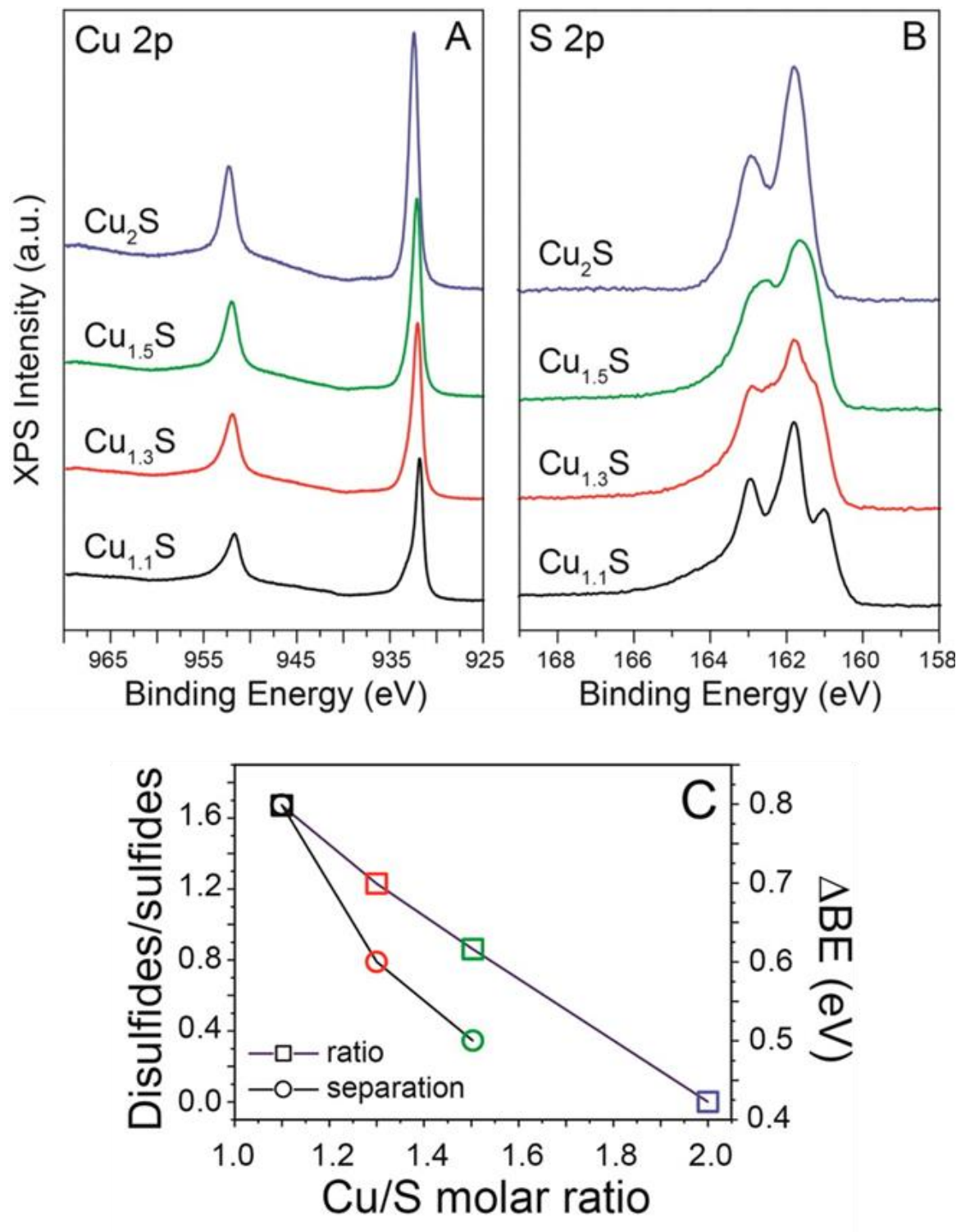

**Figure 18.** High-resolution XPS characterization of covellite CuS NCs and of three samples reduced by the addition of increasing amounts of Cu(I). (A) Cu 2p region. (B) S 2p region. (C) Ratio and energy separation of the disulfide and sulfide components as functions of the Cu/S ratio. With reduction the relative amount of disulfides decreased, until its absence in the stoichiometric $Cu_2S$ NCs. Adapted with permission from ref [152] Copyright 2015 American Chemical Society.

As a result of the copper removal, a high density of free holes characterizes the deficient copper chalcogenide NCs, and is responsible for the intense LSPR in the NIR.[1,12,136,142,152,153] In comparison, the chalcocite phase with Cu:S ratio in the narrow range from 1.99 to 2.00 exhibits only a faint absorption or no absorption at all in the NIR, indicating a much lower density of free carriers as a result of the close to $Cu_2S$ stoichiometry.[147] Djurleite has instead Cu:S molar ratios between 1.94 and 1.97,[147] which results in an intense LSPR absorption in the NIR due to the high density of free carriers.[147–149] Indeed, the oxidation of the fully stoichiometric $Cu_2S$ or $Cu_2Se$ NCs leads to the gradual evolution of a LSPR in the NIR as shown in **Figure 17**, due to the concomitant formation of copper va-cancies.[11–13] This ultimately also results in a change of the energy band structure of the copper chalcogenide. Upon their oxidation from the initial $Cu_2S$ in-gap states near the VB are induced accompanied by a small shift of the Fermi level towards the VB edge, i.e. *p*-type doping.[154] In parallel to this, there are progressive changes in the crystal structure towards lower Cu stoichiometries (**Figure 17**).[12] A more controlled way of vacancy formation is given by the post synthetic, stepwise oxidation with a Ce(IV) complex, for example $(NH_4)_2Ce(NO_3)_6$.[13] The redox potential of Ce(IV) is slightly higher than that of oxygen, inducing a quasi-instantaneous reaction with the addition of a controlled amount of the oxidizing agent. This leads to a decrease of the copper to chalcogenide stoichiometry and an evolution of the LSPR band.[13] Notably, reduction of the resulting NCs and the restoration of the optical response is achieved by the addition of $Cu^+$ ions to the sample or the addition of a strong reducing agent such as diisobutylaluminium hydride.[12,13,152,155] This property was used to switch between the different crystal structures, upon copper extraction in oxidizing environments and copper re-insertion in reducing environments.[11–13,150,152]

An interesting case is represented by covellite (CuS), which has a metallic behavior, and NCs of this material exhibit LSPR absorption in the NIR[156]. Here, the metallic and plasmonic behavior do not arise from the presence of a large number of Cu vacancies in the lattice, but rather from its intrinsic electronic structure.[156] The covellite crystal structure is built by stacks of trilayers of Cu and S atoms with both planar and tetrahedral coordination, each trilayer bound to a top and a bottom trilayer by S-S covalent bonds. In covellite, the Cu cation framework has valence close to +1, while the S anion framework has valency close to -1.[152,156] The covellite structure (and consequently NCs of this material) is very stable in oxidizing conditions, but tends to be unstable in the combined presence of reducing conditions and of extra cations in solution. Under these conditions, electrons from the reducing environment tend to break the S-S covalent bonds[152] and at the same time covellite sequesters cations from the external environment to maintain charge neutrality, a process that overall leads to the red-shift and damping of the LSPR.[152] This behavior was recently exploited by Xie *et al.* to introduce in covellite NCs either $Cu^+$ ions or divalent metal cations such as $Cd^{2+}$ and $Hg^{2+}$ (see 7. Applications of degenerately doped semiconductor NCs).[32,152] In all these cases, the covellite structure was progressively lost, along with partial loss of crystallinity of the particles and a damping and red-shift of their plasmon resonance.[32] The reaction of covellite CuS NCs with Cu(I) species as sacrificial electron donors leads to its complete reduction to chalcocite $Cu_2S$.[152] It is accompanied by the evolution of the framework of S anions from a −1 valency to −2 in the final $Cu_2S$ NCs, while the valency of Cu cations stays close to +1 (**Figure 18**).[152] A similar study were performed on copper selenide nanosheets, where Cu(I) treatment of klockmannite CuSe nanosheets causes the incorporation of Cu ions and leads to a solid state transformation to antifluorite $Cu_{2-x}Se$.[157]

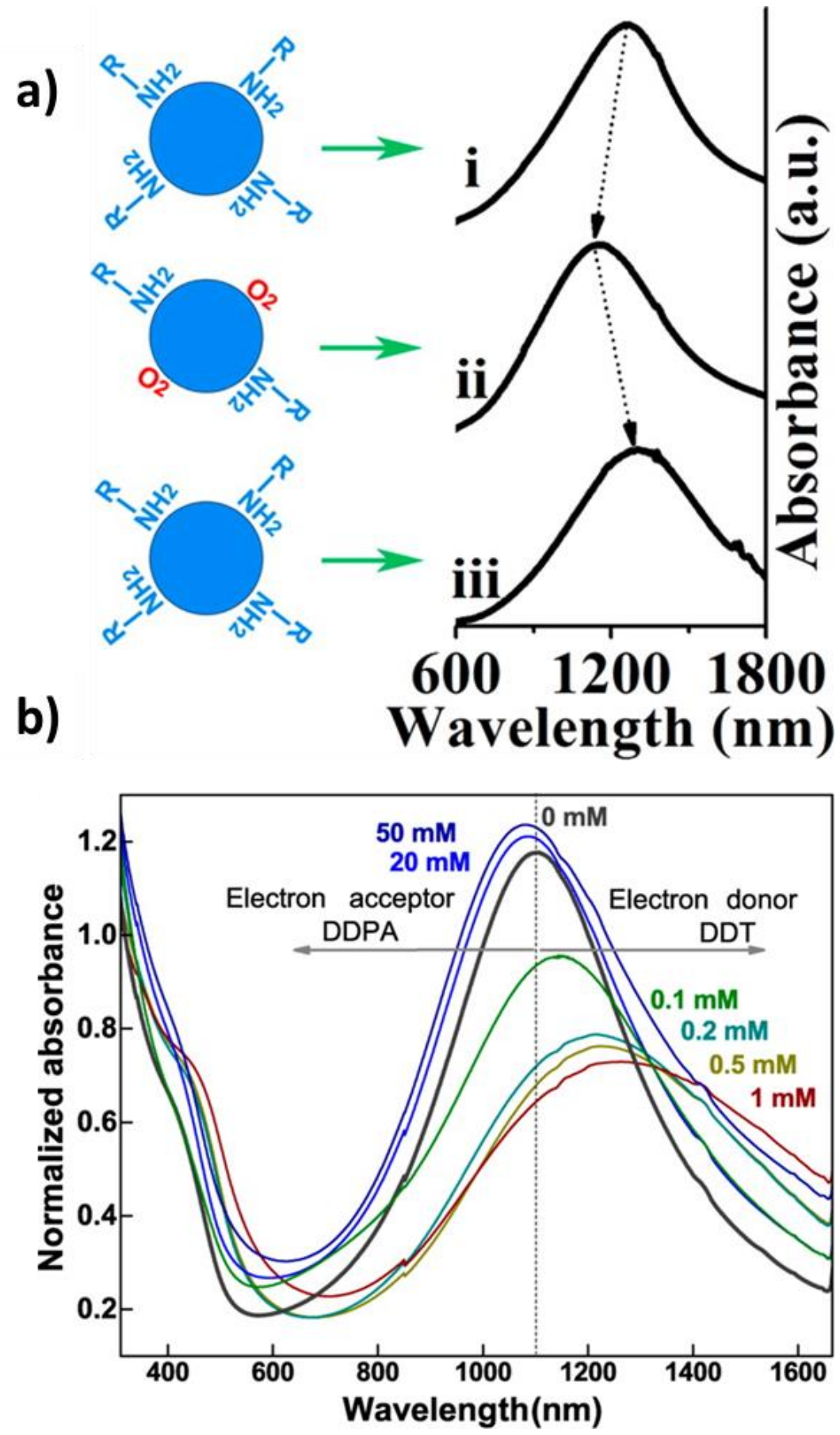

**Figure 19.** a) Schematic illustration of the process of LSPR shifts. LSPR blue-shift (curve i to ii) when the surface ligand is unloaded and oxygen is adsorbed on the surface of CuS nanodisk. The LSPRs reverts (curve ii to iii) when the CuS nanodisks are repassivated. Adapted with permission from ref [158] Copyright 2015 American Chemical Society. b) Tuning the plasmon resonance of $Cu_{2-x}Se$ NCs by ligand attachment with various concentrations of 1-dodecylphosphonic acid (DDPA) and 1-dodecanethiol (DDT) ligands. Adapted with permission from ref [158] Copyright 2015 American Chemical Society.

Various studies have shown that the adsorption of different ligands and various other molecules to the NC surface causes substantial shift to the LSPR of copper chalcogenide NCs.[151,158–163] In a study by Wei *et al.*[158] the effect of oxygen exposure on CuS nanodisks was investigated. CuS should be stable against oxidation, as the sulfide sublattice is already in an “oxidized” state. Nevertheless, an oxygen induced blue-shift of the sample was observed that could be recovered by the repassivation with surface ligands (**Figure 19**a). This is explained by the fact that oxygen adsorbs to the NC surface acting as an electron acceptor, which withdraws electrons from the NC and thereby increases the hole density. Notably, this effect was completely reversed when reloading the NC surface with ligands.[158] Additionally, the effect of surface ligands with approximately the same refractive index likewise caused a shift in the LSPR position due to the different electron-withdrawing abilities of the molecules in terms of Lewis acids and bases.[158] In a report by Balitskii *et al.*[160] the carrier density and the LSPR in $Cu_{2-x}Se$ NCs was controlled by passivation with electron withdrawing or donating ligands, which could induce a wavelength shift of up to 200 nm (**Figure 19**b). This corresponds to a variation of the carrier density by +7% and −17%, respectively.[160] A shift of the LSPR is detected (**Figure 19**b) even through the attachment of single electron donating/accepting ligands, making such plasmonic semiconductor NCs suitable for sensing of analytes with functional groups.[160] Rabkin *et al.*[159] demonstrated that not only the type of anchoring group plays a role on the LSPR position, but also the reorganization of the surface ligand arrangement on the NC surface.[159] This discussion highlights the importance to understand the surface chemistry when addressing the exact LSPR position and carrier density of this material.

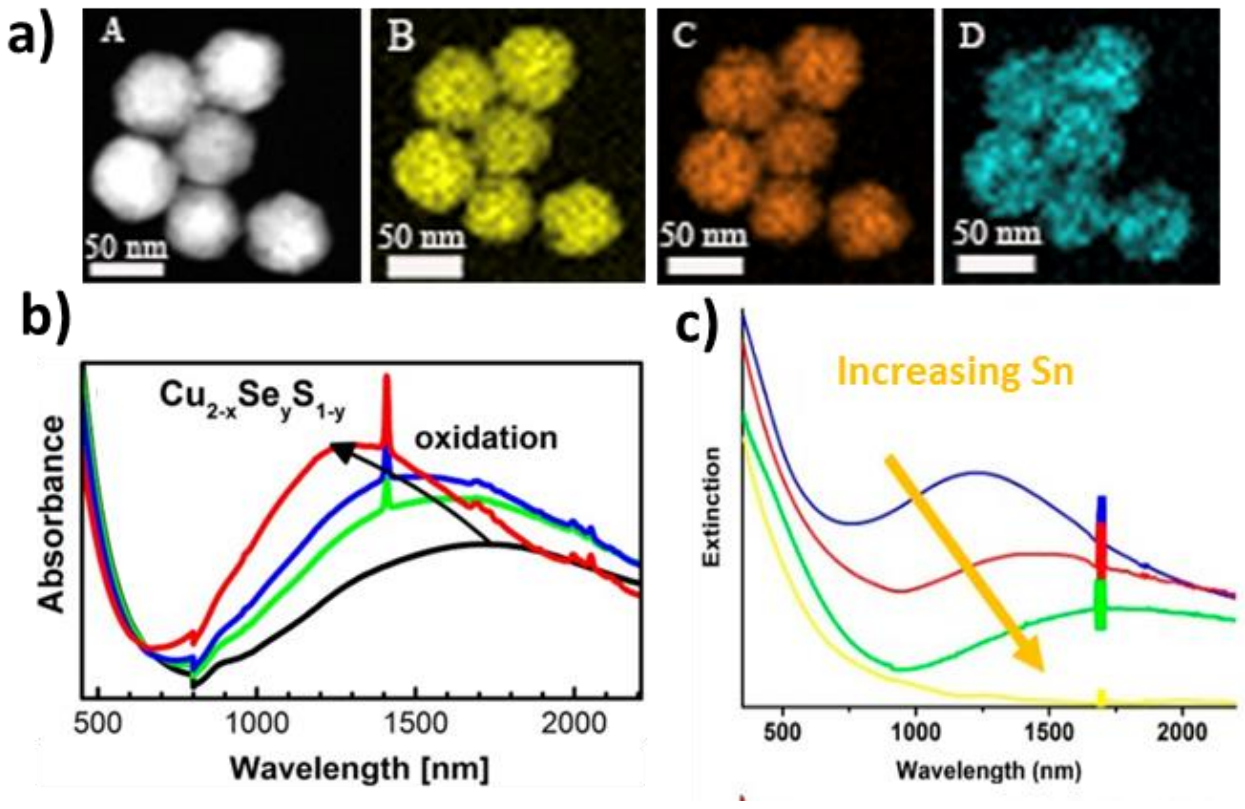

**Figure 20.** a) Chemical elemental mapping via scanning transmission electron microscopy (STEM) (A) STEM image, (b) Cu elemental mapping, (c) S elemental mapping, and (d) Se elemental mapping of hexagonal porous $CuS_ySe_{1-y}$ nanoparticles with the S/Se molar ratio of 2.5:2.5. Reprinted from Ref. [164], with permission from Elsevier. b) Absorption spectra of $Cu_{2-x}Se_yS_{1-y}$ NCs and after partial oxidation by stepwise addition of the oxidizing agent cerium(IV) ammonium nitrate $(NH_4)_2Ce(NO_3)_6$ in methanol indicated by arrows. Adapted with permission from ref [165] Copyright 2014 American Chemical Society. c) Absorbance spectra of CTSe alloy NCs prepared with varying chalcogenide ratio. Adapted with permission from ref [166] Copyright 2015 American Chemical Society.

Beyond the addition of Cu(I) also the post-synthetic addition of other metal cations was demonstrated to result in LSPR modification. While for example the addition of Ag(I) salt results in a reduction of the LSPR due to the incorporation of Ag ions into the lattice of the copper chalcogenide, the reaction with other metal cations induces the formation of domains of the metal chalcogenide.[32,155,167] In general, these processes are irreversible. An alternative means of tuning LSPR in copper chalcogenide is by mixing the chalcogenides and varying the chalcogenide ratio in alloyed self-doped copper chalcogenides, such as $Cu_{2-x}S_ySe_{1-y}$ or $Cu_{2-x}Te_yS_{1-y}$.[164,165,168–171] As shown in **Figure 20**a both chalcogenide ions are equally distributed within the NCs.[164] The LSPRs are modified by tuning the molar ratios of Se and S (**Figure 20**b).[164,165,168] Also the controlled oxidation generating additional copper vacancies allows LSPR control. Additionally, alloys of copper selenide with tin (copper-tin-selenide, CTSe) [166,172] or alloys of copper sulfide with indium NCs (copper-indium-sulfide, CIS)[173] provide tunable LSPRs through the variation of the Cu:Sn[166] or Cu:In[173] ratio.

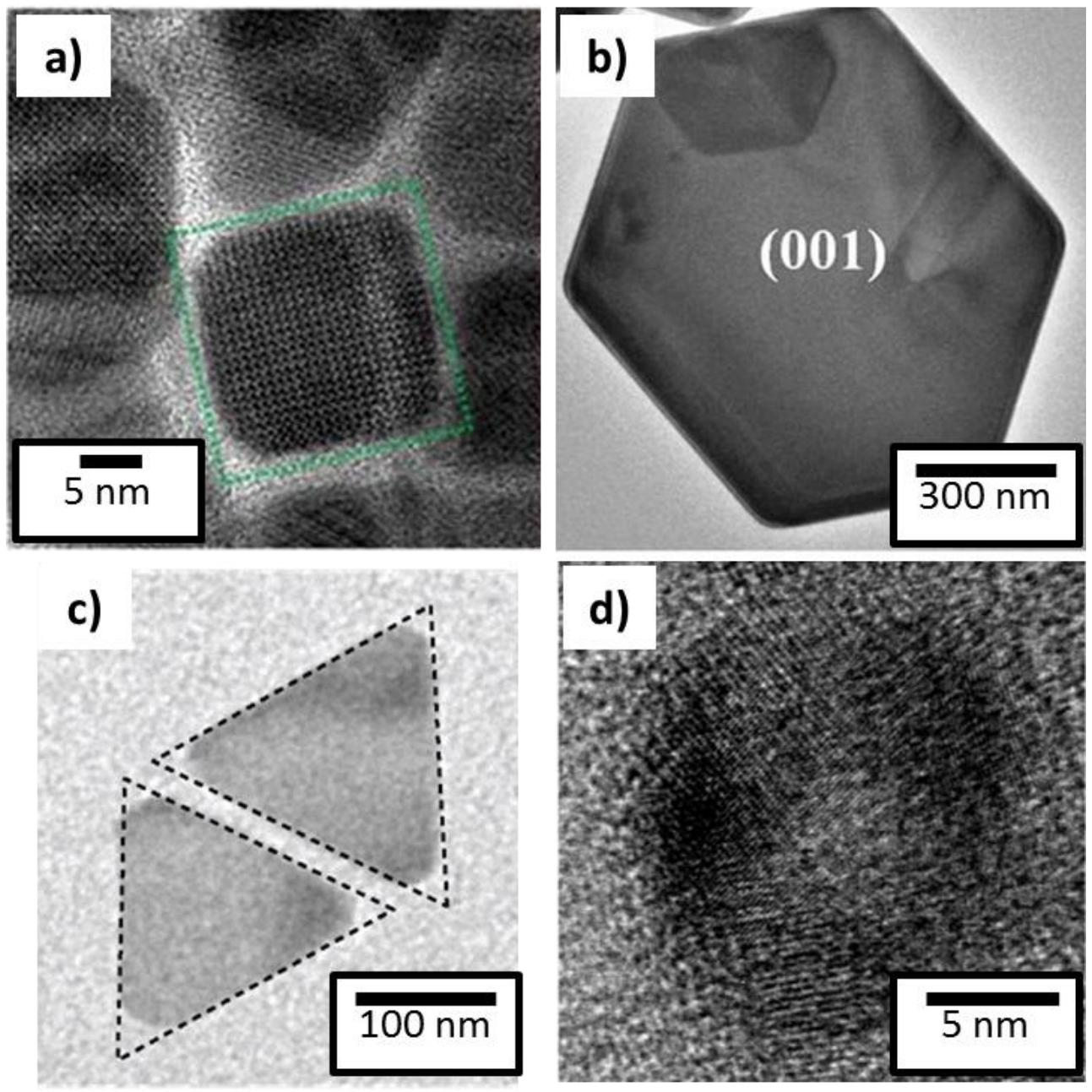

**Figure 21.** a) HRTEM micrograph of $Cu_3Se_2$ nanocubes. Adapted with permission from ref. [174] Copyright 2013 American Chemical Society. b) TEM image of a single as-prepared CuSe nanosheet with the indexed (001) facet. Reprinted by permission from Macmillan Publishers Ltd: Scientific Reports (ref. [175]), copyright 2014. c) TEM image of triangle CuS nanoparticles indicating that the nanoprisms are close to equilateral triangles. Adapted with permission from ref [176] Copyright 2015 American Chemical Society. d) TEM image of Cu2–xTe hollow NCs. Adapted with permission from ref [177] Copyright 2013 American Chemical Society.

Besides tuning of the LSPR in copper chalcogenide NCs, there is interest in finding ways to stabilize the plasmon resonance. Indeed, the affinity of copper chalcogenides to the reaction in oxidative environments often results in uncontrolled copper vacancy formation and unwanted LSPR shifts. This is of essential importance, not only for sensing applications, but also for photothermal therapy and for energy related applications (see later),[155,178,179] and, at a more fundamental level, to better correlate the NC shape and size to the plasmonic response. Surface stabilization through the right choice of surface ligand is a way to approach this problem. For this purpose Georgieva *et al.*[178] chose tetrathiomolybdate ($MoS_4^{2-}$) as the replacement ligand because of its strong chelating ability to copper ions. A remarkable stability of the LSPR in oxidizing conditions was demonstrated upon $(NH_4)_2MoS_4$-treatment, while at the same time the $Cu_{2-x}S$ NCs remained crystalline. The authors explained this by an initial removal of copper from the surface before passivation. This cleaning step might be the reason for the more efficient passivation of $MoS_4^{2-}$ ligands compared to the various organic ligands that had been previously tested.[178] Notably, the process of LSPR stabilization upon $(NH_4)_2MoS_4$ treatment is concentration dependent. The increase of the $(NH_4)_2MoS_4$ concentration leads to the dissolution of $Cu_{2-x}S$ particles and the reformation of amorphous $Cu_2MoS_4$.[178] In a recent work by Wolf *et al.*[180] the LSPR in copper selenide NCs was stabilized by growing a ZnS shell around them. Indeed, it was observed that the oxidation of the copper chalcogenide core under ambient conditions is slowed down dramatically compared to the bare $Cu_{2-x}Se$ NCs. Instead, the core was effectively protected from reduction, even in the presence of reducing agents such as borane tert-butyamine complex and diisobutylaluminum hydride, giving rise to a stable particle LSPR, also under strongly reducing conditions.[180]

The synthesis of copper sulfide NCs in the presence of metal cations such as iron can lead to the stabilization of specific crystallographic phases and thereby provides a means of regulating the copper vacancy density, as discussed above.[148,181] However, the presence of cations can also influence the shape of the NC.[182] For example Chen *et al.*[183] demonstrated a morphological evolution of the $Cu_{31}S_{16}$ nanodisks to tetradecahedra due to the $Sn^{4+}$-directed modification of the vertical crystal planes of the copper sulfide disk seeds. This leads to the preferential deposition of the copper and sulfur source on the vertical crystal planes, favoring the formation of the tetradecahedra. Along a similar line, Li *et al.*[174] synthesized $Cu_{2-x}Se$ nanocubes in the presence of $Al^{3+}$ ions (**Figure 21**a). The $Al^{3+}$ ions were not incorporated in the $Cu_3Se_2$ particles, nor were they found on their surface, indicating that their role in dictating shape control was probably not that of preferentially passivating a particular facet, but is yet to be clarified.[174] In another report, CuTe nanocubes were prepared through the reaction of a copper salt with trioctylphosphine telluride in the presence of lithium bis(trimethylsilyl)amide and oleylamine. This salt enabled the growth of highly uniform copper telluride nanocubes, nanoplates, and nanorods.

Notably, only a slight dependence of the LSPR on the NC shape was observed.[42] This peculiar feature is a general characteristic of copper chalcogenide nanostructures and suggests that most likely the free carriers are not entirely delocalized over the whole nanostructure.[184] This aspect will be discussed in greater detail later in this review (section 5 Optical modelling of degenerately doped semiconductor nanostructures.).

Far more often, organic molecules that selectively bind to different NC facets are used to control NC morphology, as for example in the synthesis of CuSe nanosheets (**Figure 21**b).[157,175,185–190] After the initial crystalline nucleation, the oriented growth of CuSe crystal seeds, effectively stabilized by octylamine, leads to the preferential growth of CuSe nanosheets.[175] Hsu *et al.*[176] showed that the formation of faceted CuS triangular nanoprisms is strongly mediated by the presence of halide ions (**Figure 21**c). Shape focusing occurs, favoring faceted nanoprisms larger in edge length at the expense of unfaceted nanodisks, upon Ostwald ripening. Such behavior was ascribed to the presence of halide ions, which are capable of disrupting the binding interactions between oleylamine capping ligands and the CuS surface.[176] In another approach, the synergistic interaction between halides (Br or Cl) and copper-thiolate metal–organic frameworks was exploited to synthesize 2 nm thick colloidal $Cu_{2-x}S$ nanosheets with well-defined shape (triangular or hexagonal) and sizes up to 3 μm.[185] Nethravathi *et al.*[186]demonstrated that a solvothermal reaction between a colloidal dispersion of dodecylsulfate intercalated copper hydroxide layers in ethylene glycol with a solution of $TeO_2$ yielded single crystalline 2D nanosheets of copper telluride. The solvated copper hydroxide layers act as templates to facilitate the formation of nanosheets of a few nanometers in thickness and micron lateral dimensions.[186] Thermoelectric measurements evidenced a marked decrease in the lattice thermal conductivity while preserving the electrical conductivity, due to the different length scales of phonons (through the thickness of the nanosheets) and charge carriers.[186] Cation exchange too was used to synthesize ultrathin and thickness-controlled $Cu_{2-x}Se$ nanosheets from CdSe nanosheets preserving the starting morphology and retaining their cubic crystal structure.[187] The resulting $Cu_{2-x}Se$ nanosheets had a cubic structure because the parent anion sublattice was maintained during cation exchange. Notably, the obtained copper selenide nanosheets were stable against oxidation even after exposure to air for two weeks. Sheets with dimensions as low as 1.6, 1.8 and 2.2 nm were prepared that showed evident NIR LSPRs despite their very low lateral dimensions. The optical red shift with decreasing size was assigned to decreased carrier damping, when the sheets become thicker, while a contribution due to a possible variation in the density of copper vacancies cannot be excluded.[187] An interesting synthesis approach was demonstrated in the report given by Cheung *et al. via* a continuous-flow millifluidic synthesis.[191] Indeed, the authors demonstrated a "surfactant-free" approach to produce various different sizes and shapes of $Cu_{2-x}S$ NCs.

By adjusting the flow rate and precursor concentrations, spherical, triangular prism and rod shaped NCs could be prepared. Uniform heating was ensured by the continuous-flow on the millifluidic chip and facile variation of the precursor ratios was given, representing a simple platform for nanoparticles' synthesis with high reproducibility.[191] A variety of other interesting shapes were reported, such as nanodisks, nanoplatelets, as well as various polyhedral shapes.[131,177,192–195] Hollow copper chalcogenide nanostructures were also prepared: for example, hollow $Cu_{2-x}Te$ NCs were formed by implementing the nanoscale Kirkendall effect, in which first Cu nanoparticles were nucleated. The addition of the tellurium precursor then led to the formation of thin layers of $Cu_{2-x}Te$ onto the Cu core surface, followed by the nanoscale Kirkendall process occurring at the interface between the Cu core and shell, leading to the observed hollow structure (**Figure 21**d).[177] While both, hollow and solid $Cu_{2-x}Te$ NCs exhibit a single plasmon resonance, it was shown that hollow particles are more sensitive to the refractive index of the environment, which makes them interesting for sensing.[177]

# 5. Optical modelling of degenerately doped semiconductor nanostructures.

Important characteristics of the relationships between the structure and composition, and the optical properties of degenerately doped semiconductor NCs remain poorly understood. Optical modeling using theoretical methods to predict and understand the optical response of degenerately doped nanoparticles is a way to access some of these features. For noble metals, the most prominent theory approximating the dielectric properties in the region of the free carriers is the Drude model given by the complex function:

$$\varepsilon(\omega) = \varepsilon_1(\omega) + i\varepsilon_2(\omega) \qquad (4)$$

where

$$\varepsilon_1 = \varepsilon_\infty - \frac{{\omega_p}^2}{(\omega^2+\Gamma^2)} \qquad (5)$$

and

$$\varepsilon_2 = \frac{{\omega_p}^2 \Gamma}{\omega(\omega^2+\Gamma^2)} \qquad (6)$$

are the real and imaginary part of the dielectric function. The parameter $\Gamma$ is the free carrier damping constant, which describes the damping of the collective oscillations due to scattering of electrons at electrons, phonons, lattice defects or impurities. The plasma frequency of the free carriers is given by equation (2):

$$\omega_p = \sqrt{\frac{ne^2}{\varepsilon_0 m^*}} \qquad (2 \qquad (2)$$

with $e$ the electron charge, $m^*$ the effective mass, $\varepsilon_0$ the vacuum dielectric permittivity, and $n$ the carrier density. This underlines that an increase in $n$ ultimately results in an increase of the plasma frequency $\omega_p$ and a blue shift of the LSPR. In the quasi-static regime for particles much smaller than the incoming light wave, *i.e.*, $R << \lambda$ an analytical solution for the response of a small metallic particle to the electromagnetic radiation is given within the framework of the Mie theory. For small particles, where only dipolar contributions play a dominant role, the extinction coefficient $C_{ext}$ for a spherical particle is given by:

$$\boldsymbol{C_{ext} = \frac{18\pi V \varepsilon_m^{3/2}}{\lambda} \frac{\varepsilon''(\omega)}{(\varepsilon'(\omega)+2\varepsilon_m)^2+\varepsilon''^2(\omega)}} \qquad \textbf{(7)}$$

In the expression above, $\varepsilon_m$ is the surrounding dielectric constant, $\varepsilon(\omega)$ is the dielectric function of the particle, $V$ is the particle volume. For ellipsoidal particles the Mie theory can be extended according to R. Gans, by adding depolarization factors for the three axes of the particle to account for the contribution of the different axes to the nanoparticle spectrum.[62] Also the effective medium approximation is a valid tool to calculate the plasmonic response of nanoparticles dispersed in a medium as solution or film.[196] Nanoparticles of arbitrary geometry require computational methods solving the Maxwell's equations by using numerical tools,[61] such as the finite difference in the time domain method (FDTD),[197,198] the boundary element method (BEM),[61] and the discrete dipole approximation (DDA).[199,200] The DDA is a numerical solution to calculate the extinction cross section ($C_{ext}$). A continuous material is approximated as discrete elements that are polarized by the superposition of the external incoming electric field and the induced dipoles in the surrounding elements.[199–201] Calculations with the DDA require intensive numerical efforts because of the large number of coupling terms, but have advantages when addressing very complex geometries.

The implementation of a bulk dielectric function for the theoretical modelling by using any theoretical method, such as the Mie-theory, DDA, BEM, FDTD etc.[62] is a common way to calculate the optical response of plasmonic nanomaterials and has also found application in plasmonic semiconductor nanostructures. Dorfs *et al.*[13] for example demonstrated a qualitatively good agreement between the spectra of $Cu_{2-x}Se$ NCs with different Cu:Se ratios and the theoretically calculated extinction spec-tra.[13] For the latter, they implemented the dielectric function of thin films of $Cu_{2-x}Se$ with different values of x (x = 0.1, 0.13, 0.15, and 0.2) extracted from bulk reflectivity measurements together with the Mie scattering theory in the quasi static approximation (**Figure 22**a).[13] Also Kriegel *et al.*[184] ($Cu_{2-x}Te$ NCs) and Agrawal *et al.*[38] (ICO NCs, **Figure 22**b) demonstrated a reasonable agreement between the absorption spectra of NCs of different shapes with the theoretically calculated ones by using the DDA and thin film dielectric data.[38,184]

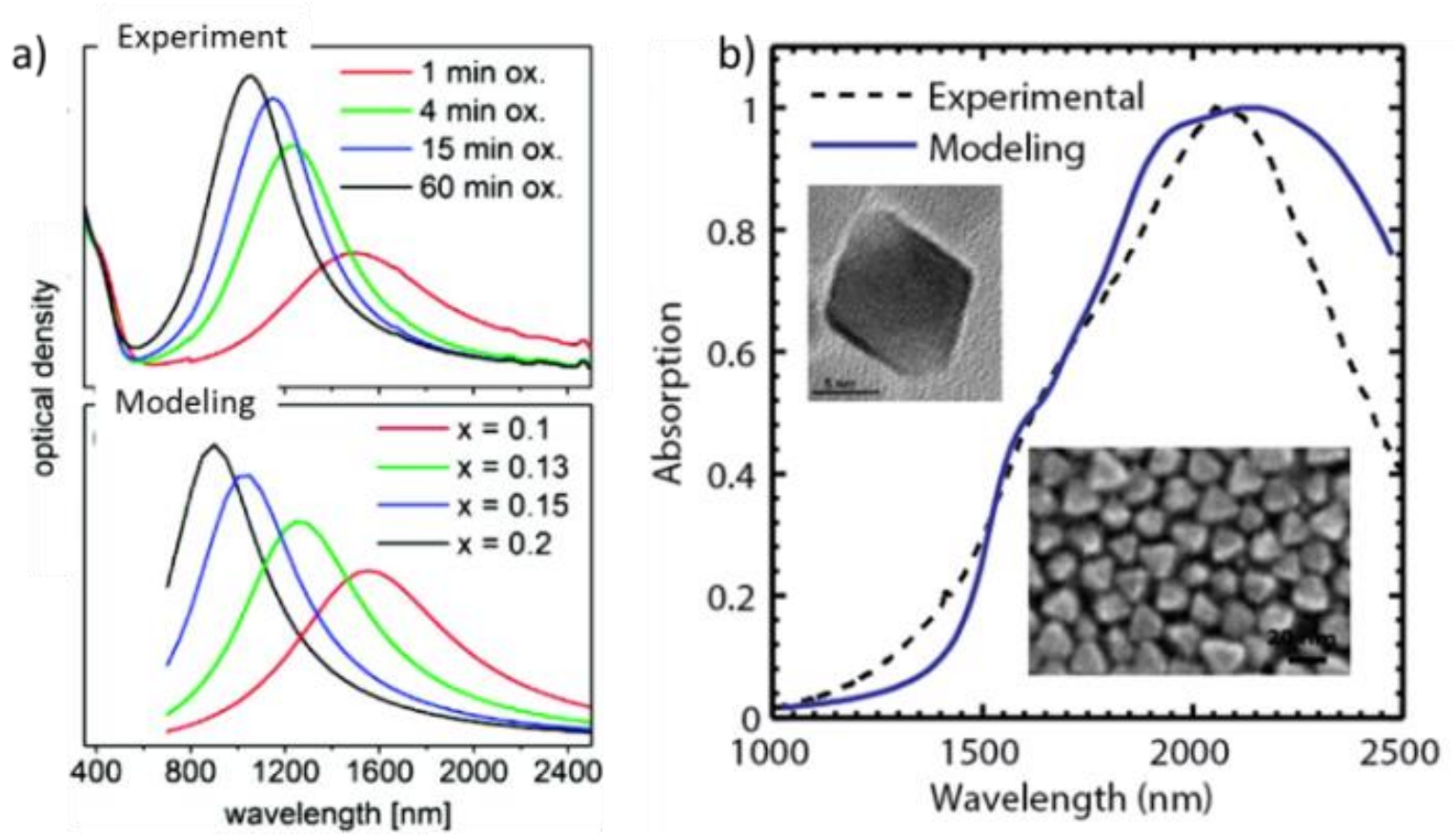

**Figure 22.** (a) Top: Evolution of the extinction spectrum of the "as-synthesized" $Cu_{1.96}Se$ nanocrystals (dissolved in tetrachloroethylene) over time, when they are gradually oxidized, under ambient conditions, to $Cu_{1.81}Se$ (as found by elemental analysis). Bottom: Calculated extinction spectra of single $Cu_{2-x}Se$ nanocrystals for different *x* values using dielectric functions of $Cu_{2-x}Se$ for x=0.1, 0.13, 0.15 and 0.2 as calculated from bulk reflectivity measurements reported in ref.[202] Reprinted with permission from ref [13] Copyright 2011 American Chemical Society. (b) Experimental and theoretical absorption spectra of around 10 nm ICO octahedral particles. Insets: TEM and SEM images of synthesized octahedral particles show a deviation from the perfect octahedron shape due to rounding of the edges and corners. Reprinted with permission from ref [38] Copyright 2015 American Chemical Society.

### 5.1. The case of copper chalcogenide NCs

As shown in eq. 2, $\omega_p$ directly relates to the carrier density $N_c$, and, thus, provides a tool to determine the carrier concentration of a specific sample by applying theoretical models, such as the Drude model to fit the LSPR. Unfortunately, in a compound such ad $Cu_{2-x}Se$, due to the non-parabolicity of the valence band, the carrier effective mass is highly dependent on the vacancy density, meaning that one cannot extract the carrier density directly from eq. 2.[196] To account for this, Dorfs et al.[13] determined the vacancy concentration in $Cu_{2-x}Se$ NCs by elemental analysis and extracted a carrier density of $3.07 \times 10^{21} cm^{-3}$ by assuming that each copper vacancy contributed one free carrier. The hole effective mass thus extracted from the plasma frequency $\omega_p$, was in agreement with $m^*= 0.2-0.25m_e$ expected for copper selenides with free hole concentrations $>10^{21}cm^{-3}$.[13] This report demonstrates the importance of structural and compositional analysis to support the extraction of the optically derived carrier density and other parameters in copper selenide NCs via optical modeling.

In copper sulfide, at least five different crystal phases have been documented (see previous sections)[152,203] and the control over structure and stoichiometry can be a tool for achieving dynamic, reversible LSPR tuning.[12,136,142,152,153,204] The various crystal phases are correlated to the LSPR and the deep understanding of the influences of phase transformation on the plasmonic behavior are of major importance. An estimate of the charge carrier density can be given by applying eq. 2 to the LSPR of NCs of several different copper sulfide phases, taking into account the different effective hole masses of the various copper sulfide phases, as explained above.[205] However, the different factors that influence the plasmon resonance frequency can only be extracted by comparing carefully the experimental spectra during phase transformation with theoretical extinction spectra.[150,205] In a work by Liu *et al.*[150] three related parameters including the plasmon frequency $\omega_p$, the high frequency dielectric constant $\varepsilon_\infty$ and the carrier density $n$ were extracted by investigating the phase transition from rhombohedral to hexagonal CuS clusters. The best fitted parameters indicate that the transformation process from $Cu_{1.8}S$ to CuS includes a transition of the dielectric constant from around 2.0–2.6 for rhombohedral NCs to 3.8–4.4 for hexagonal CuS clusters. It was demonstrated that both the change in dielectric constant upon phase transition and the composition induced variation in the carrier concentration influence the LSPR.[150] This study highlights that it is of major importance to correlate carefully structural and optical properties in copper chalcogenide NCs, as the important parameters for the Drude analysis, such as the plasmon frequency $\omega_p$, the high frequency dielectric constant $\varepsilon_\infty$, the carrier damping $\Gamma$, and the effective mass $m^*$ are all changing when a phase transition takes place.

As mentioned before, covellite is the crystal phase with an exact Cu:S ratio of 1:1 and represents a special case. In a careful work, Xie *et al.*[156] investigated covellite CuS NCs of various sizes in an approach using the discrete dipole approximation (DDA) to model their shape and size dependent optical spectra. They extracted the Drude parameters that allowed the reproduction of the extinction spectra of several NCs of various sizes, showing that the optical properties of stoichiometric covellite CuS are consistent with the assumptions made in the Drude model and the carriers in the system can be treated as essentially free. This is likely due to the fact that the carriers in covellite CuS arise from the inherent metallic-like character due to a significant density of lattice-constitutional valence-band free holes, as mentioned above.[156] This characteristic is different from the other copper sulfide phases, where the copper vacancies are responsible for the free hole density, making it difficult to distinguish between contributions from shape effects and from carrier density to the LSPR.[1,11,12] Phase purity in covellite NCs therefore renders possible to distinguish the influence that the NC shape has on the LSPR. Liu *et al.*[203] investigated CuS nanoplatelets at fixed composition and carrier density with a large range of aspect ratios, exploring quantitatively the optical extinction spectra. The LSPR showed a size dependent red-shift with increasing diameter, which was attributed to the changing aspect ratio of the nanoplatelets with a constant thickness. Optical modeling with the full-wave time-harmonic field theory and the Drude model allowed separating the in-plane and out-of-plane LSPR modes of the anisotropic NC (**Figure 23**a) and it was found that the out-of-plane mode strongly overlaps with the in-plane mode, as also shown by Xie *et al.* (**Figure 23**b). From quantitative modeling consistent values of free carrier concentration and background polarizability were extracted that apply over a wide range of NC sizes. The corresponding value of $\Gamma$, the damping constant for the in-plane mode, decreased with increasing size. This is reasonable, being in a regime of diameters where surface scattering of charge carriers is dominant.[203]

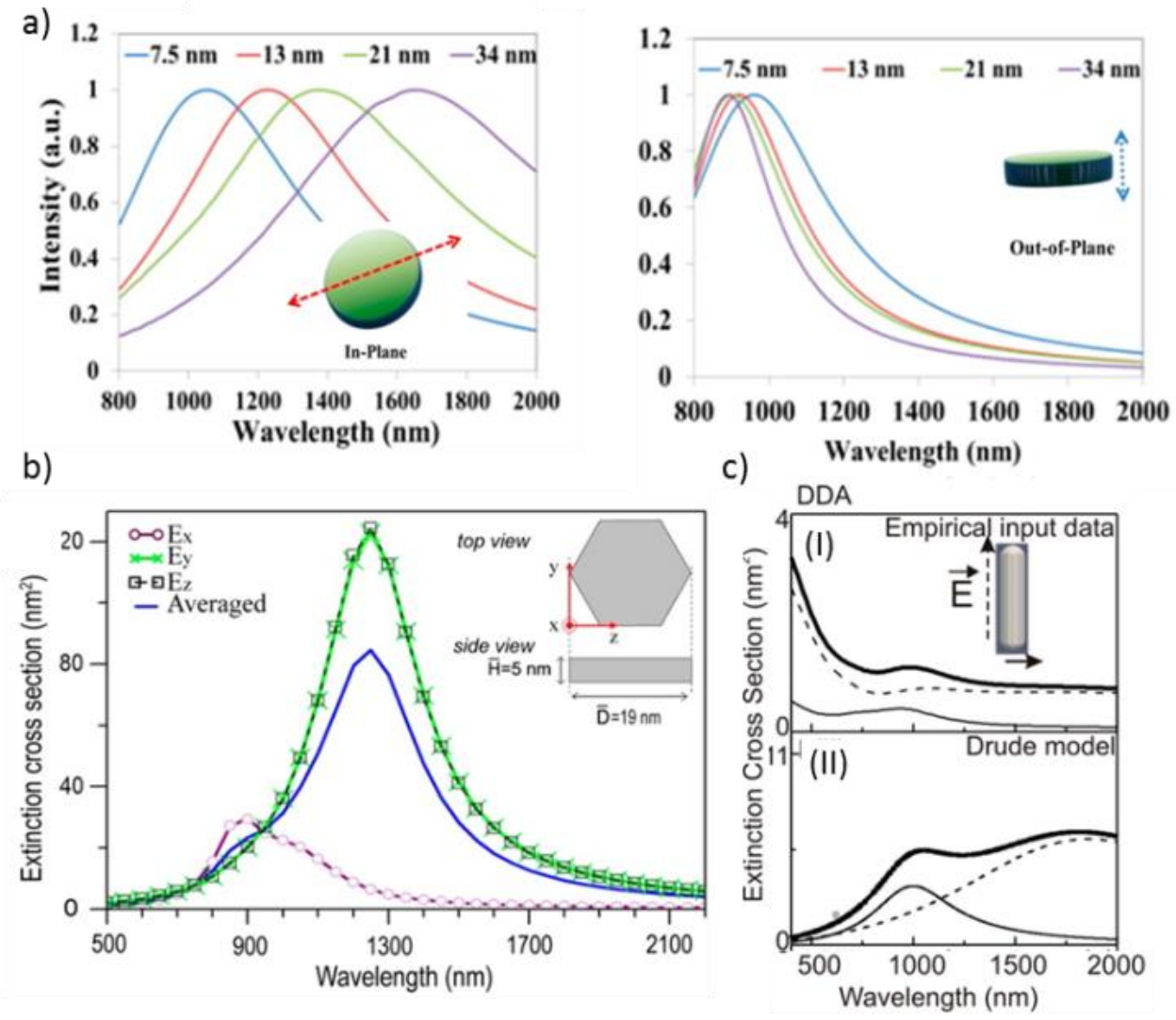

**Figure 23.** a) Simulated in-plane and out-of-plane spectra for four covellite CuS nanoplatelet sizes with thickness of 4 nm calculated with the full-wave time-harmonic field theory and the Drude model. Reprinted with permission from ref [203] Copyright 2015 American Chemical Society. b) Computed DDA extinction scattering cross sections of a CuS covellite nanodisk of 5 nm in thickness and 9 nm in width calculated with the Drude parameters for $\omega_p$ = 4.5 eV, $\varepsilon_\infty$ = 9 and $\gamma$ = 0.25 eV, for incident light polarized along the transverse and longitudinal-axis directions and with orientational averaging. Reprinted with permission from ref [156] Copyright 2013 American Chemical Society. c) Calculated DDA extinction spectra for a $Cu_{2-x}Te$ nanorod

(length × width, 21 × 5 nm), (I) calculated using an empirical dielectric function for $Cu_{2-x}$Te (ref [206]) and (II) using the Drude model; the black thin solid curves represent the T-bands; the dashed thin curves represent the L-bands; and the black thick solid curves represent the averaged extinction spectra of the nanorods. While the empirical dielectric function represents well the experimental data, the Drude model overestimates in the range of the longitudinal plasmon resonance. Adapted with permission from ref [184] Copyright 2013 American Chemical Society.

Kriegel *et al.*[184] demonstrated on the other hand that the Drude model does not sufficiently describe the experimental extinction spectra of copper chalcogenide NCs governed by copper vacancies. For three different shapes of $Cu_{2-x}$Te NCs (rods, dots and tetrapods) with $x > 0$ the authors were not able to extract one common dielectric function based on the Drude model that is able to reproduce all three different shapes. In particular, in elongated structures the longitudinal plasmon resonance was overestimated with the Drude model (**Figure 23**c). Indeed such resonance is heavily damped in the experimental data. The authors explained this by partial carrier localization with respect to free carriers in the Drude model. This finding appears to be valid for the entire family of copper chalcogenides of type $Cu_{2-x}$E (E=S, Se, Te) and demonstrates that the role of vacancies in the structure results in a deviation of the purely metallic character, i.e. that such carriers cannot be treated as fully free.[184]

### 5.2. The case of metal oxide NCs

As discussed earlier in this paper, also metal oxide NCs display a plasmonic response when doped heavily. Their dielectric function is influenced by the choice of material,[9,196] the mechanism of doping[9,196] and even by the synthesis protocol.[34] In the case of ICO (indium doped cadmium oxide) for example, the optical response of the free carriers is well described by the Drude theory with a constant damping term $\Gamma$, leading to a symmetric line shape of the plasmon resonance peak.[38,207] However, in metal oxide NCs the incorporation of dopants might lead to additional scattering centers in the crystal. In ITO, for example, the presence of ionized impurities was shown to result in a frequency dependence of the damping term $\Gamma(\omega)$ described by the extended Drude function.[9,208] Here, the damping constant can be divided into two regimes: the low-frequency regime, where the frequency of the incoming light wave is smaller than the plasmon resonance $\omega_p$. The real part takes a constant value due to screening of the ionized impurities by the electron cloud. The screening effect breaks down for higher frequencies, when the response of the electron cloud is no longer coherent with the changes in the local electric field at the frequency of oscillation. As a consequence of this, the ionized impurities behave as bare Coulomb scattering centers, and the dynamical resistivity drops off.[9,209–211] The frequency dependent damping was deduced both empirically and semi-empirically for bulk ITO films and is com-posed of a low and a high frequency damping terms, with a cross over region.[9,34,209–211] This leads to an asymmetric line shape of the LSPR, being broader on the red side of the resonance.[9] The spectra of ITO NCs can be well reproduced when implementing the frequency dependent damping term *Γ(ω)*,[34,208] and the optically extracted carrier densities of photodoped ITO NCs are comparable to those counted by oxidative titration.[100,115,117] Notably, titrations determine carrier densities independent of the model chosen and therefore serve well for a critical evaluation of the applied theory.[117]

Lounis *et al.*[34] showed that depending on the synthesis protocol the spectral shape of spherical ITO NCs changes from asymmetric to symmetric. Elemental depth-profiling by X-ray photoelectron spectroscopy (XPS) unveiled that in the asymmetric line shaped samples the ionized impurities, i.e. dopant atoms, are distributed throughout the NC volume. In the other sample, the dopant atoms were found to be located at the NC surface only.[34] The authors performed a core-shell based modelling where the frequency dependent scattering is only valid for a layer at the NC surface, while the core part encounters a frequency independent damping term, similar to the Drude formulation used for noble metals. This result implies that the carriers move freely through the NC volume, without encountering the ionized impurities introduced through doping. The description of the core therefore simplifies back to the simple Drude model with a static damping term, while in the asymmetric line shaped sample the scattering centers in the entire NC volume require the frequency dependent damping.[34] This example shows how careful optical modelling supported by structural analysis can help to unravel the chemistry of doping and the role of scattering centers on the plasmonic response.

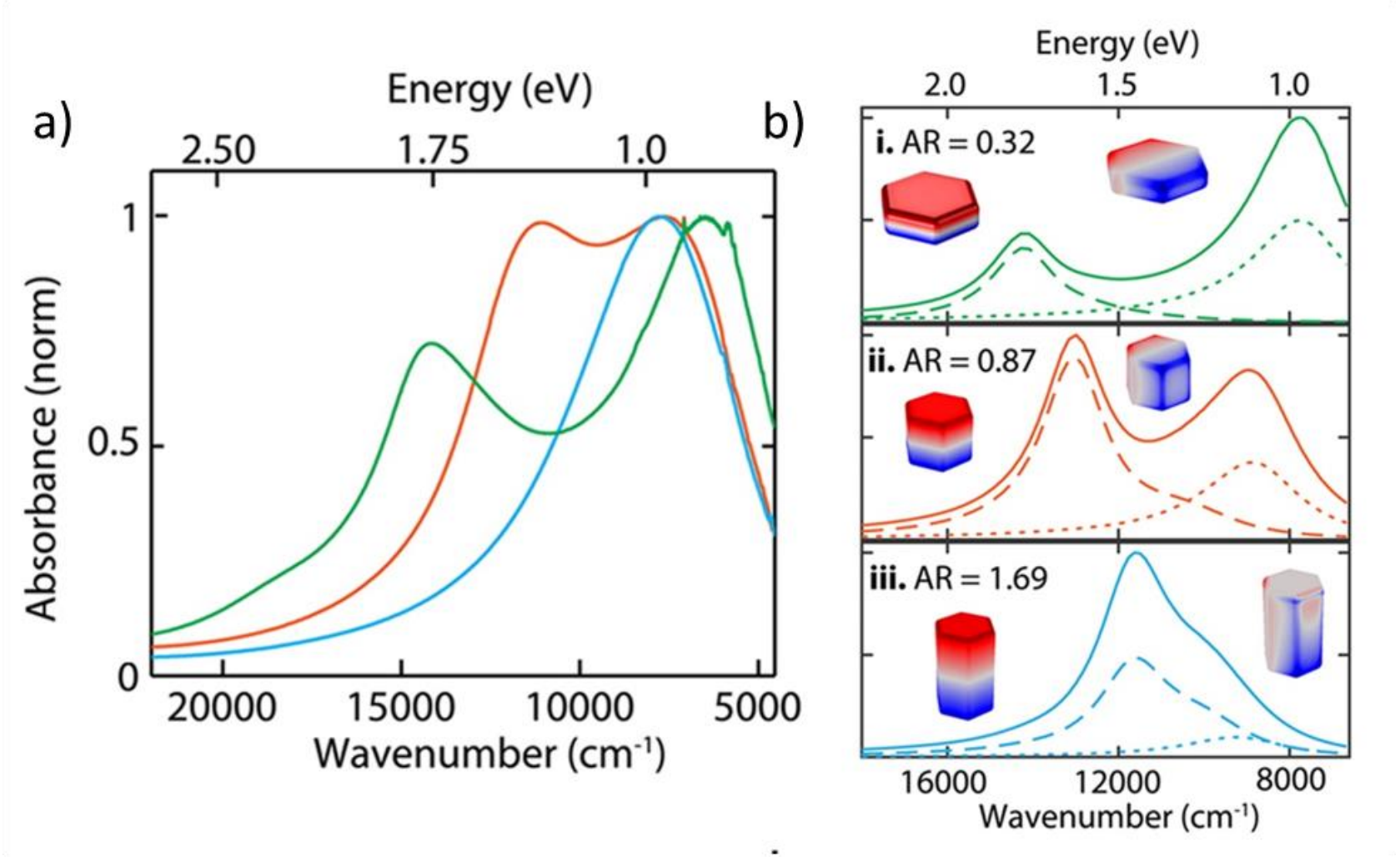

**Figure 24.** a) Experimental optical absorption spectra of $CsWO_3$ platelets (green), prisms (orange), and rods (blue) showing the shape dependent optical response. (b) Theoretically simulated

LSPR spectra for the same shapes as in (a) employing the anisotropic dielectric function of $CsWO_3$. Solid lines represent the overall LSPR response by summing the longitudinal mode (dashed lines) and two transverse modes (dotted lines). The insets give the simulation of the dipolar surface charge distributions calculated at longitudinal and transverse peak wavelengths. Adapted with permission from ref [36] Copyright 2016 American Chemical Society.

In another work by Kim *et al.*, the influence of anisotropic NC shape was further correlated to the intrinsic crystal structure and their collaborative effects on the LSPR explored.[36] By synthesizing $CsWO_3$ nanorods of various aspect ratios, it was concluded that the anisotropy of the underlying crystal strongly influences the location of the LSPR modes. The Drude parameters effective mass $m^*$, the high frequency dielectric constant $\varepsilon_\infty$ and the damping $\Gamma$ describing the dielectric response of the free carriers in the NIR are known to be strongly dependent on the crystallographic axis. And indeed, only by taking into account the anisotropy of the dielectric function in the different crystallographic directions were the authors able to reproduce the experimental spectra (compare **Figure 24**a and b).[36] This finding demonstrates that the conventional description of the LSPR that assumes isotropic dielectric constants is insufficient to describe the plasmonic response of doped metal oxide NCs with anisotropic crystal structures. The main message from this work is that synthetic control of shape and crystalline anisotropies of semiconductor nanocrystals provides an additional means to rationally tune LSPR line shape and near-field enhancement.[36]

In Table 1 we summarize the Drude parameters of copper chalcogenide NCs and some of the most common metal oxide NCs obtained by fitting the LSPRs of NC dispersions or films. For comparison also the parameters of noble metals such as Au and Ag are given.

**Table 1. Summary of Drude parameters obtained by fitting the LSPRs of NC dispersions or films of copper chalcogenide NCs and some of the most common metal oxide NCs. For comparison also the parameters of Au and Ag are given.**

| | $\varepsilon_\infty$ | $\omega_p$ (eV) | n ($cm^{-3}$) | $m^*/m_0$ (kg) | $\Gamma$ (eV) | |
|---|---|---|---|---|---|---|
| $Cu_{1.85}S$ | 10 | 1.78-2.08 | 0.9.2-1.26 × $10^{21}$ | 0.21 | | [196] |
| $Cu_{1.8-1.4}S$ | 2.0-2.6 | 2.83-3.19 | 4.7-5.99 × $10^{21}$ | 0.8 | | [150] |
| CuS | 3.8-4.4 | 2.94-3.21 | 4.56-6.08 × $10^{21}$ | 0.8 | | [150] |
| CuS[1] | 9 | 4.5 | ~$10^{22}$ | | 0.25-0.5 | [156] |
| CuS[2] | 10.1 | 4.02 | 0.9 × $10^{22}$ | | 0.82 | [203] |
| $Cu_{1.96}S$<br>$Cu_{7.2}S_4$<br>CuS | | | 1.0 × $10^{21}$<br>3 × $10^{21}$<br>1.6 × $10^{21}$ | 0.8<br>0.78<br>0.55 | | [205]<br>[205]<br>[205] |
| $Cu_{2-x}S$<br>$Cu_{2-x}Se$ | 11.56<br>11 | | | 0.8<br>0.5 | | [46]<br>[46] |
| $Cu_{2-x}Se$ | | | | | | [13,202] |
| $Cu_{2-x}Se$ | 10 | 4.45 | $3 \times 10^{21}$ | | 0.43 | [212] |
| $Cu_{2-x}Te$ | 8.4, | 4.1 | | | 0.6 | [184] |
| ITO | 4.26 | 1.91 | $8.3 \times 10^{20}$ | - | 0.37 ($\Gamma_L$)<br>*0* ($\Gamma_H$)<br>*0.67* ($\Gamma_X$)<br>*0.21* ($\Gamma_W$)[4] | [208] |
| ITO | 4 | 4.45-4.86 | 3.03-3.50 $\times 10^{21}$ | 0.4 | 0.124 | [196] |
| ITO | | 0.72-0.93 | 1.5-2.49 $\times 10^{20}$ | - | 0.1132 | [34] |
| ICO | 5.3 (CdO) [213] | - | 0.81-1.33 $\times 10^{21}$ | 0.42[5] | 0.034-0.138 | [38,207,213] |
| $Ce:In_2O_3$ | | 0.71-0.93 | 1.51-2.55 $\times 10^{20}$ | 0.4 | 0.86-0.13 | [94] |
| FICO | 5.6 | 2.32–2.44 | 1.68–1.86 $\times 10^{21}$ | 0.43[5] | 0.07 | [112] |
| $Cs_xWO_3$:<br>c-axis<br>a-axis | <br>6.3<br>5.8 | <br>4.66<br>3.18 | | <br>0.75<br>1.62 | <br>0.22<br>0.35 | [36] |
| AZO | 3.7 | | 2.2-3.6 $\times 10^{20}$ | 0.6 | 0.029 | [214] |
| Ag | 3.7 | 9.1 | | | 0.018 | [215,216] |
| Au | 9.84 | 9 | | | 0.067 | [215,216] |

[1]Similar accuracy of the fit was achieved by choosing different sets of parameters ($\varepsilon_\infty$=7,$\omega_p$=4.3eV or $\varepsilon_\infty$=11,$\omega_p$=4.7eV).[156]
[2]Similar accuracy of the fit was achieved by choosing different sets of parameters ($\varepsilon_\infty$=14.1,$\omega_p$=4.54eV or $\varepsilon_\infty$=12,$\omega_p$=4.23eV).[203]
[3]The authors found that the same dielectric function does not reflect the other shapes synthesized due to localization effects on the carriers.
[4]Frequency dependent damping.
[5]Non-parabolic effective mass (for ICO given by the equation $m^* = m_0^* \sqrt{1 + 2C \frac{\hbar^2}{m_0^*} (3\pi^2 n)^{2/3}}$ with $m_0^*$=0.17$m_e$ and the non-parabolicity factor C=0.5$eV^{-1}$).[112,207,217]

# 6. Response of degenerately doped semiconductor NCs to light

The unique characteristic of LSPR modification in degenerate semiconductor nanostructures brings along new means to control the light matter interaction in these structures. The LSPR tunability can be exploited to control the overlap between two resonances for the study of fundamental physical properties, such as plasmon-plasmon coupling or exciton-plasmon interaction. The control over the carrier density to an ultralow doping level opens the route to quantum plasmonics. In the following section, we review on recent studies regarding the response of degenerately doped semiconductor NCs to light, covering fundamental plasmonic characteristics in the pristine structures to complex hybrid interactions.

## 6.1. Fundamental optical phenomena in degenerate semiconductor nanostructures or heterostructures thereof

In two independent studies on degenerately doped metal oxide NCs, the influence of sample inhomogeneity was addressed.[20,214] The most common reasons for resonance linewidth broadening stem from inhomogeneous size and shape distributions. However, in degenerately doped semiconductors, an additional, strong influence from different levels of doping varying from NC to NC is expected. Indeed, Johns *et al.*[214] found that the ensemble LSPR of MO NCs was substantially broadened with respect to spectra obtained from single NCs. The latter were assessed by synchrotron infrared nano-spectroscopy, where LSPRs of single AZO and ITO NCs were directly recorded. NC linewidths down to $600\,cm^{-1}$ were measured, which is less than half of the ones of the ensemble. Moreover, such LSPR linewidths are narrower than those of noble metal nanoparticles, such as gold, holding promise for a wide range of applications where intense local fields in the NIR are required. This analysis suggests that dopant incorporation is heterogeneous, thus, resulting in broad ensemble linewidths of the LSPR, instead of large damping due to lossy plasmonic materials.[214] Similarly, Machida *et al.*[20] found that significant differences in the dielectric properties of each NC are responsible for the observed LSPR broadening. The authors developed a model which goes beyond size and shape inhomogeneity with the aim to address variations that are primarily ascribed to dopant desorption on the NC surface and decrease in oxygen vacancies in each NC.[20]

In plasmonic nanostructures the study of the quantum plasmon has attracted attention in recent years, addressing the discretization of the energy levels for very small sizes, opposed to the band type character usually observed.[218] In degenerately doped nanostructures that are small enough to support quantum confinement (i.e. with discrete energy levels) the low number of carriers can exhibit multiple transitions of individual carriers between quantized states. Thus, the control over size and carrier density allows addressing questions such as the difference between intraband excitations of carriers and the excitation of the LSPR, or what is the number of carriers required to support an LSPR.[46–48,219,220] Schimpf *et al.*[48] showed that experimental plasmon resonance energies of photodoped ZnO nanocrystals with controlled sizes and carrier densities diverge from the classical Drude model predictions. This result was assigned to the existence of quantum plasmons in these NCs. It was found that the Lorentz oscillator model more adequately describes the data, illustrating a bridge between single-electron and LSPR regimes, from quantum confinement in small NCs to Drude plasmons in large NCs.[48] Similarly, theoretical calculations by Zhang *et al.*[46] demonstrated that a small number of carriers within a confined nanostructure will exhibit single-carrier excitations between quantized energy levels, followed by a transition from the size-quantization regime to the classical regime of plasmon oscillations as carriers are progressively added.[46] Jain *et al.*[47] established a plasmon-in-a-box model to describe such limiting cases and found that intraband transitions of single carriers between levels dominate the optical response, while collective excitations begin when the carrier density reaches critical values, because Coulomb interactions between carriers begin to overcome the confinement of Fermi level carriers to the lattice. Notably, the authors found a transition regime between quantum plasmons and classical plasmons, where both excitations coexist in an intermediate regime.[47] These results together highlight the fundamental difference of this new class of plasmonic material with respect to noble metals, as the number of carriers can be precisely controlled to limiting cases, having important consequences for both the understanding and future applications of degenerately doped semiconductor NCs.[46–48]

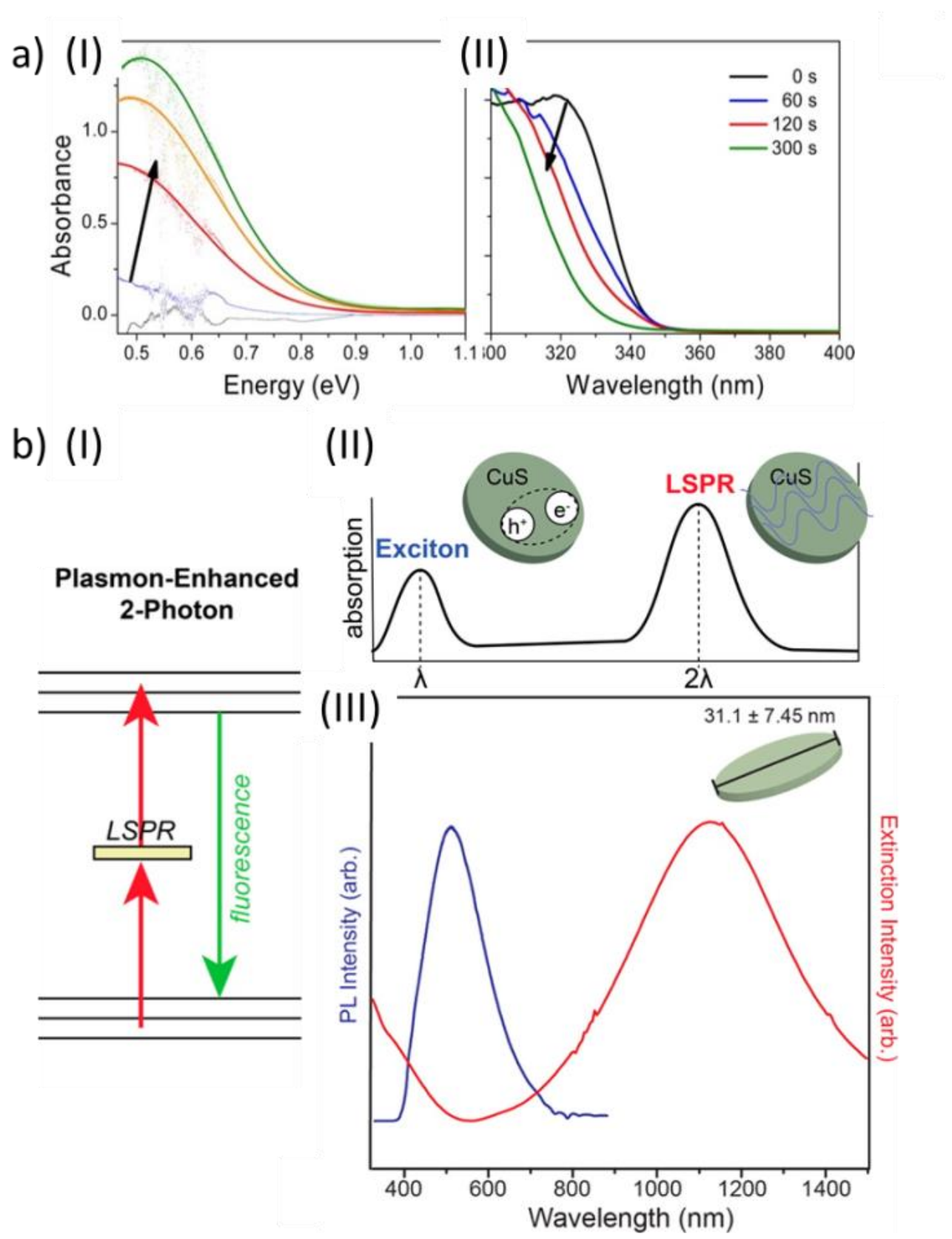

**Figure 25.** (a) Photoinduced LSPR absorption of ZnO NCs. UV irradiation of ZnO NCs in de-aerated methanol results in the accumulation of electrons in the conduction band (I) inducing the onset of an LSPR mode and a (II) concomitant bleaching of the excitonic absorption followed by a Moss–Burstein shift of the absorption band edge. Reprinted with permission from ref [122] Copyright 2013 American Chemical Society. b) I) Schematic of the electronic transitions for plasmon-enhanced two-photon processes. (II) Schematic of the absorbance profile for Plasmon-enhanced two-photon absorption. (III) Photoluminescence emission spectra and extinction spectra for CuS nanodisks. Adapted with permission from ref [45] Copyright 2016 American Chemical Society.

The increase of the carrier density in plasmonic semiconductors coupled to the modification of the LSPR, additionally induces an effective increase of the band gap, known as the Moss-Burstein effect for degenerately doped semiconductors.[11,12,38,100] As shown by Faucheaux *et al.*,[122] the introduction of free carriers into MO NCs performed via photodoping leads to the appearance of an LSPR in the NIR, which increases with increasing doping. Simultaneously the excitonic resonance is gradually weakened followed by a shift of the absorption band edge due to the Moss-Burstein effect (**Figure 25**a (I) and (II)).[122] In $Cu_{2-x}S$ NC similar findings were reported, where LSPR evolution leads to the loss of the excitonic transition and the decrease of PL intensity. Indeed, the high level of doping and the excess carriers in the band structure weaken the excitonic resonance and introduce nonradiative relaxation pathways that shorten the exciton recombination time[221] and decrease the photoluminescence quantum yield due to Auger recombination.[100]

The study of exciton-plasmon coupling in the so called plexcitonic systems[222–224] is of high interest for the exploration of electron transfer e.g. for photocatalysis,[225,226] emission enhancement or suppression,[223] or in the study of Fano resonances.[227–229] Targeted fabrication e.g. *via* cation exchange enables the design of nanostructures with energetic match of excitonic and plasmonic transitions.[221] The controlled switching can facilitate tuning of the exciton-plasmon interaction from an ON to an OFF state and vice versa. In early studies on $Cu_{2-x}Te$-CdTe nano-heterostructures, only a weak exciton plasmon interaction was found, while the faster decay of the excitonic transition was primarily ascribed to Auger processes.[221] In a recent study by Marin *et al.*[45] a novel platform for engineering plasmon–exciton coupling within one material - covellite (CuS) - was shown and exploited for photon upconversion processes (in which the absorption of two or more photons leads to higher energy emission). More specifically, a two-photon absorption process, assisted by the existence of an LSPR at twice the wavelength of the emission, was observed in covellite nanodisks (see sketches in **Figure 25**b (I) and (II)).[45] By substituting S with Se the LSPR was fine-tuned, such that it better overlapped with the two-photon absorption edge, a 3-fold enhancement of the two-photon fluorescence emission was observed (**Figure 25**b (III)). These results are unique, as the doped semiconductor NCs serve as both the plasmonic and photoluminescent components. This study is the first to show an “all-in-one” platform for engineering plasmon–exciton coupling in the absence of a physical or chemical interface and is highly interesting for applications were multiphoton absorbance is required, such as bioimaging and photodynamic therapy.[45] In another work, Zhou *et al.*[230] exploited plasmonic $Cu_{2-x}S$ NCs as near-infrared antennas to enhance the upconversion luminescence of nanoparticles. These were made of an inorganic crystalline host, such as $NaYF_4$:$Yb^{3+}$,$Er^{3+}$ ($NaYF_4$), embedding transition metal, lanthanide or actinide dopant ions.[230] In hybrid films of $Cu_{2–x}S$-$MoO_3$-$NaYF_4$ the upconversion luminescence was highly improved, with a notable power-dependent behavior: at lower power, the interaction of the LSPR with the upconverting NCs plays the major role. In the higher excitation power regime instead, electron diffusion *via* a two-photon interband transition of $Cu_{2–x}S$ was proposed. The latter is unique to the ener-gy band structure of semiconducting $Cu_{2-x}S$ NCs that is characterized by a lower indirect valley and a higher direct valley. This unique band structure is completely different from that of noble metals and highlights interesting opportunities of this new type of plasmonic materials.[230]

Besides exciton-plasmon coupling, also the study of fundamental interparticle plasmon coupling becomes accessible in nano heterostructures containing degenerately doped semiconductors with tunable LSPRs. In structures composed of a noble metal and the inherently plasmon tunable degenerate semiconductor, such as Au-$Cu_{2-x}Se$ nanorods[44] or Au-$In_2O_3$ nanodimers,[231] the overlap between the two plasmon resonances can be modified by adjusting the position and the intensity of the degenerate semiconductor LSPR to enable strong, intermediate or weak plasmon coupling. Thus, the cross talk between plasmon resonances can be tuned.[44] Muhammed *et al.*[44] demonstrated that the NIR plasmonic behavior in such bi-plasmonic Au-$Cu_{2-x}Se$ nanostructures is reversibly switched under oxidative and reductive conditions so that the metal core alone, the metal core and the semiconductor shell, or the shell alone dominate the NIR plasmonic response of the assembly (**Figure 26**). The light-absorbing properties spanning both the first and the second NIR windows opens interesting scenarios for photothermal therapy or theranostics.[44]

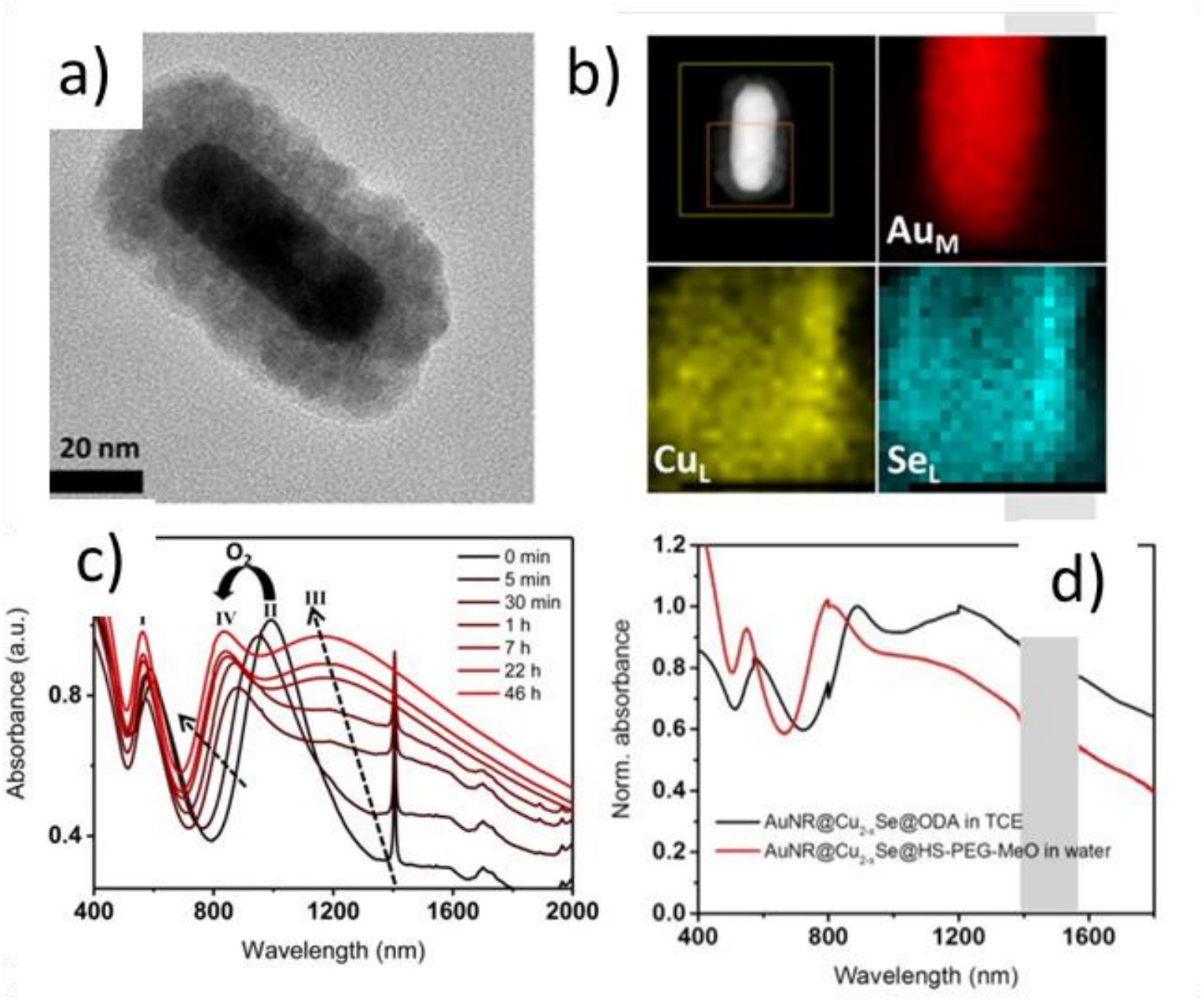

**Figure 26.** (a) HRTEM image of metal/semiconductor plasmonic core–shell nanoparticle cluster (AuNR@$Cu_{2-x}Se$). (b) Elemental maps corresponding to the most intense lines of each element: Au, Cu and Se. (c) Time-dependent evolution of the vis–NIR extinction spectra of AuNR@$Cu_2Se$ NPCs in TCE during oxidation upon air exposure (the total exposure time from the black to the red curve is 46 h). Plasmon peaks are labeled as I, II, III, and IV. (IV) Vis–NIR

extinction spectra of AuNR@$Cu_{2-x}Se$ NPCs as oxidized in TCE (black curve) and after ligand exchange and redispersion in water (red curve) for their utilization in the bio-medical field. Adapted with permission from ref [44] Copyright 2015 American Chemical Society.

Also the interparticle coupling and the formation of plasmonic hot spots is of high interest in degenerately doped semiconductor NCs, with the advantage that the LSPR frequency is in the NIR, highlighting their use for NIR enhancing spectroscopies, such as Raman or IR spectroscopy.[205,232,233] Near-field interactions are expected to be greatly influenced by NC size, shape, and composition, but in particular also carrier density and carrier mobility should play a great role. Although field enhancement factors in degenerately doped metal oxide NCs are generally order of magnitude smaller than those of noble metals (due to lower carrier densities), the advantage of tuning the LSPR wavelength to specific ranges is interesting. Moreover, for plasmonic applications in the NIR similar enhancement factors can be reached for much smaller nanoparticle sizes.[234] Additionally, synthesis protocols of compound semiconductors provide a wide range of shapes, including anisotropic nanostructures.[235,236] As an example, octahedron shaped indium doped cadmium oxide (ICO) NCs were shown to deliver high field enhancement factors in the NIR, between 1600 and 2200 nm.[38] Instead, only weak electric field enhancement was reported for the $Cu_{2-x}Te$ nanorods, most probably due to the partial localization of carriers in copper chalcogenides, as discussed above (section 5 Optical modelling of degenerately doped semiconductor nanostructures.).[184]

In compound semiconductors, crystal anisotropy should play an important role in determining near-field interactions. To investigate this peculiarity, Hsu *et al.*[150] formed close-packed films of nanodisks via compression at an air–water interface, which were then transferred to a solid support by dip-coating. Two different inter-particle orientations were obtained, either orthogonal or parallel with respect to the air–water interface. Interparticle plasmon coupling was found to be most prominent when Cu sulfide nanodisks are aligned in a side-by-side arrangement, leading to dipole-dipole coupling between the in-plane LSPR modes, with red shifts ($\Delta\lambda$) reaching up to 419 ± 8.2 nm for the closest-packed side-by-side arrangement The strength of coupling increased with decreasing distance.[150] A much weaker coupling was found by Chen *et al.*[232] in face-to-face assembled $Cu_7S_4$ nanodisks. The poor plasmonic coupling of stacks of nanodisks suggests that the LSPR decay length for Cu sulfide nanodisks is anisotropic, due to the inherent anisotropy of the electrical conductivity, which is much greater along one specific crystallographic direction,[237] and of the hole mobility as well. Such peculiarities in some doped semiconductors need to be taken into account in strongly anisotropic nanostructures in order to correctly understand their optical properties, as discussed previously for in $CsWO_3$[36] (details in the section 5 Optical modelling of degenerately doped semiconductor nanostructures.). On the other hand, they can also represent an extra handle in designing electromagnetic hot spots or coupled plasmonic architectures.

### 6.2. Ultrafast optical response of degenerately doped semiconductor nanostructures

The photo-induced signal manipulation of plasmonic nanomaterials represents a step forward towards the all-optically controlled signal processing.[68,111,238] On ultrafast timescales, applications in active plasmonics and optical switching come in reach.[111,238,239] The ultrafast signal modulation is particularly interesting in the NIR operating at telecommunication wavelengths,[68,111,238,239] where degenerately doped semiconductor nanostructures show their plasmonic response, tunable to specific wavelength ranges of interest.[9,11–13,22,100] Moreover, specific properties of their band structure in some of the cases, such as the non-parabolicity of the conduction band,[239] the presence of the bandgap[109,111] or the anisotropy of the crystal structure,[36] deliver new options to manipulate their plasmonic response. Large signal response combined with ultrafast recovery times make them attractive for the design of ultrafast nanodevices.[68,109,111,239] Indeed, several studies on doped semiconductors that appeared in recent years present interesting and promising results.[3,12,24,109,111,156,239–244]

The temporal evolution of the non-linear optical response of a plasmonic material follows a specific trend, observed in two well separated temporal dynamics: a short-time dynamic of several picoseconds, and a longer decay dynamic of several hundreds of picoseconds. The strong perturbation of the free carriers after excitation with an intense laser pulse (pump pulse) in the region of the LSPR creates a non-thermal distribution of the Fermi gas. Strong *carrier-carrier* scattering leads to the establishment of a new Fermi distribution within several hundreds of femtoseconds, resulting in a thermalized free carrier gas with higher temperature than the lattice (‘hot’ carriers). Following this, the carriers cool down via interaction with the lattice through *carrier-phonon* coupling (initial fast decay) and the subsequent release of energy to the environment (*phonon-phonon* coupling) (**Figure 27** a and b).[245,246] The variation of the carrier and lattice temperatures after excitation induces a temporal modification of the dielectric function of the material, which can be probed at controlled time delays after excitation with a weak probe pulse.[247,248] Indeed, the non-linearities observed in the region of the LSPR in degenerately doped semiconductor NCs show a very similar modulation.[3,12,24,109,156,239–244] Xie *et al.*[156] studied CuS nanodisks of varying sizes with similar height and varying widths with transient absorption spectroscopy. The temporal evolution exhibited an initial fast decay within a few picoseconds, ascribed to the initial phonon emission (*carrier–phonon*). Subsequently, a slower decay (of several hundreds of picoseconds) was observed and ascribed to *phonon–phonon* scattering and heat dissipation into the solvent.[156] A change in the equilibrium volume of the NC is triggered as a result of the ultrafast laser pulse induced heating of the lattice (due to *carrier–phonon* coupling), which is much shorter than the lattice heat dissipation. This ultimately results in the excitation of coherent vibrational (breathing) modes.[156,249,250] The breathing mode induced volume change of the NC in turn affects the surface plasmon resonance condition and is observed as oscillations overlaying the second slower decay to the red and the blue of the maximum bleach non-linearity. By fitting the period of the oscillation with a damped cosine function, Xie *et al.*[156] found that the vibrational motion of their CuS nanodisks is confined along the radius of the nanodisk, mostly propagating radially along the longitudinal-axis direction. The period of the modulated oscillation was observed to increase linearly with increasing diameter unambiguously assigning the symmetric radial breathing mode of the nanodisks that produce small periodic changes to the nanodisk volume along their major axis.[156]

The laser induced heating of $Cu_{2-x}Se$ NCs was further evaluated by Scotognella *et al.*[242] by modeling the temporal dynamics with the so-called Two Temperature Model (TTM) employed for noble-metal nanoparticles. It quantitatively accounts for the picosecond dynamics related to the cooling of the 'hot' carrier gas, $T_C$, generated after pump absorption, and the subsequent heat transfer to the lattice, $T_L$. The following coupled equations [247]:

$$\begin{cases} \gamma T_c \frac{dT_C}{dt} = -G(T_C - T_L) + P_A(t) \\ C_L \frac{dT_L}{dt} = G(T_C - T_L) - G_L(T_L - T_0) \end{cases} \quad \textbf{(8)}$$

take into account the heat capacity of the carrier gas ($\gamma T_C$), with $\gamma$ the so-called *carrier heat capacity constant*. $C_L$ is the heat capacity of the lattice, $G$ and $G_L$ are the *carrier-lattice* coupling factor and the lattice-environment coupling factor respectively. $T_0$ is the environmental temperature and $P_A(t)$ is the pump power density absorbed in the volume of the sample.[239] The initial heating of the carrier gas results in a smearing of the Fermi distribution, and gives rise to a modulation of the inter-band transition probability for the probe in both real and imaginary parts of the dielectric function. This gives rise to the modulation $\Delta\varepsilon_\infty$ of the $\varepsilon_\infty$ that scales quadratically with the carrier temperature $T_C$. Instead the heating of the lattice results in a modulation of the $\Gamma$ parameter due to the increase of the free carrier scattering with increasing lattice temperature.[239] The temperature-dependent dielectric function of $Cu_{2-x}Se$ is thus given by:

$$\varepsilon(T_C, T_L) = \varepsilon_\infty + \Delta\varepsilon_\infty(T_C) + \frac{\omega_p}{\omega^2+\Gamma^2(T_L)} + i\frac{\omega_p\Gamma(T_L)}{\omega(\omega^2+\Gamma^2(T_L))} \quad (9)$$

The theoretical prediction of this semi-empirical model turns out to be in excellent agreement with the experiments on $Cu_{2-x}Se$ nanoparticles, and allows the extraction of the phonon coupling constant, which compares reasonably well with those found in noble metals after accounting for the lower free carrier density. Moreover, it was found that the lower carrier density results in a smaller carrier heat capacity and a much higher effective carrier temperature than in noble metal nanoparticles at comparable fluencies and is responsible for a sub-picosecond kinetics that is much faster than in noble metals (**Figure 27** a and b).[239]

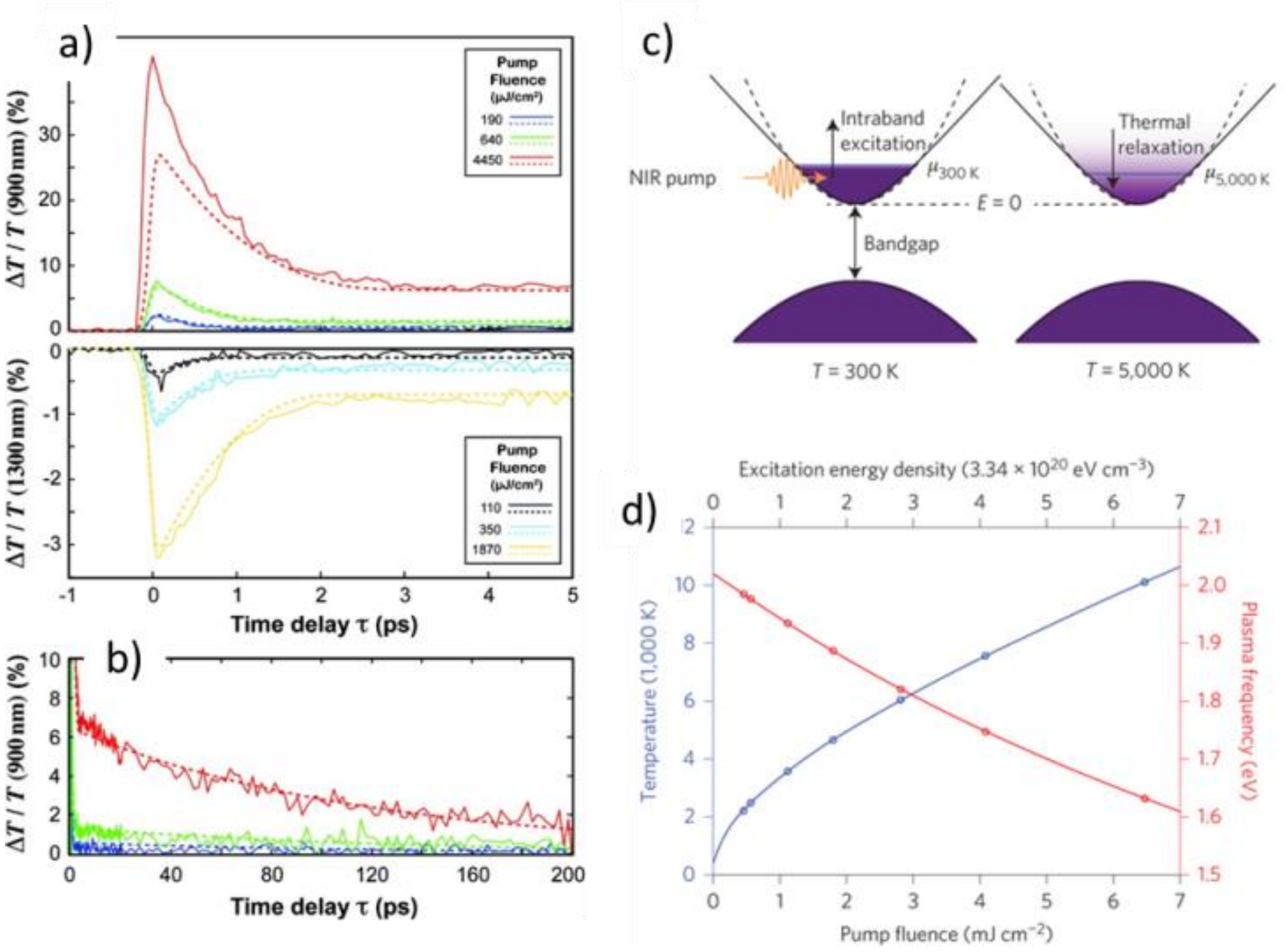

**Figure 27.** a) Differential transmission dynamics from a solution of $Cu_{1.85}Se$ NCs at 900 nm and at 1300 nm probe wavelength and (b) long-time dynamics at 900 nm illustrating the typical two step decay of the LSPR response (experimental versus the numerical results - solid lines versus dashed lines, respectively). Adapted with permission from ref [242] Copyright 2011 American Chemical Society. c) Schematic diagram depicting the electron configurations and electronic processes involved in the intraband pumping of the FICO NCs' LSPR. Solid lines and dashed lines sketch the non-parabolic and parabolic dispersions, respectively. d) Calculated plasma frequency and electron temperature depending on the calculated excitation energy density and the corresponding experimental pump fluencies. Reprinted by permission from Macmillan Publishers Ltd: Nature Photonics (ref. [239]), copyright 2016.

Similarly, Guo *et al.*[239] reported the ultrafast plasmon modulation of the near- and mid-IR plasmon resonances of ITO nanorod arrays (ITO-NRAs) by intraband pumping. Much higher carrier tem-peratures with respect to noble metal nanoparticles by similar fluencies were reached also in this work, due to the lower electron density and the concomitant smaller electron heat capacity of ITO (**Figure 27** c and d).[239] Notably, the authors observed that the non-linear optical response of the LSPRs in ITO-NRAs is not a result of the Fermi-surface smearing and the related change in $\varepsilon_\infty$. Instead, it was found that the observed pump-induced redshifts of the LSPRs arose from a change of the plasma frequency of ITO, which depends on the carrier density *n* and inversely on the carrier effective mass *m** (see equation 2). Since intraband pumping does not result in a variation of the carrier density *n*, the reduction of $\omega_p$ was ascribed to a change in the effective mass of the carriers *m**, due to the non-parabolicity of the conduction band (an effect observed in several highly-doped metal oxides[207,239]). Differently from a parabolic band dispersion, to which a constant effect mass is associated, a non-parabolic band structure implies that the effective mass depends on the wavevector (hence on the energy) of the carriers and *m** increases with increasing energy. The resulting significant plasma frequency modulation leads to large transient optical changes up to the near and mid IR range, peculiar to this type of plasmonic material (**Figure 27** c and d).[239] These materials are also called epsilon-near-zero materials. They display a large magnitude of switching of the NIR LSPRs at the sub-picosecond timescale. In another example, intraband photoexcitation of fluorine and indium codoped CdO (FICO) NCs results in a red-shift of the plasmon feature, consequence of the low electron heat capacity and the conduction band non-parabolicity of such oxide. Already at low excitation fluencies (<5 $mJ/cm^2$) large changes in light transmission (+86%) and index of refraction (+85%) were observed.[217] This suggest that these materials are very promising for all-optical switching, wavefront engineering, and ultrafast switching at terahertz frequencies.[112,217]

A modulation of the plasma frequency $\omega_p$ can also be achieved by changing the carrier density *n*. Photodoping is a tool to optically control the carrier density in MOs (see chapter 3.2.2 Photodoping of MO NCs) by prompting interband transitions. As the free carrier density of the doped semiconductor is much lower than in metals, the LSPRs of these materials are more sensitive than metals to absolute changes in the carrier concentration. This enhanced sensitivity to the carrier density makes doped semiconductors appealing as active optical components in devices where the reversible modulation of either the intensity or the frequency of the plasmon resonance can be exploited for ultrafast switching. Indeed, in a similar sample of ITO NRAs Tice *et al.*[109] showed that band gap excitation modulates the plasma frequency of ITO on the ultrafast time scale by the injection of electrons into, and their subsequent decay from, the conduction band of the rods. Carrier density increases of about ~13% were achieved.[109] Similar results were shown for FICO NCs.[112,217] Kriegel *et al.*[112] showed that the temporarily increased carrier density (*n*) after the ultrafast photodoping decays monoexponentially resembling elec-

tron−hole recombination at the bandgap. Pump-power-dependent measurements display an additional red-shift of the derivative shaped signal with increasing pump power in correspondence with a further increased carrier density. *Via* optical modeling of the transient signal an increase of the carrier density by approximately 7.5% was extracted from those measurements.[112] Notably, the transient optical response of FICO NCs when pumped into the LSPR was assigned to the non-parabolic conduction band and the thus induced changes in the effective mass of the carriers, as mentioned earlier.[112,217] Also in a work by Kinsey *et al.*[111] the transmissivity of ITO and AZO films showed signal modulation of up to 40% with an ultrafast (<1 ps) response in both transmission and reflection was observed for all applied fluencies. The excess carrier density was extracted to be roughly 10 % together with the recombination times that enable switching speeds exceeding the terahertz regime. This ultrafast recovery time is believed to be due to additional defects present in the crystal, delivering an additional tool to control the speed of carrier recombination, thus, optical signal manipulation.[111]

# 7. Applications of degenerately doped semiconductor NCs

## 7.1. Exploiting the characteristics of degenerately doped semiconductor NCs for sensing applications

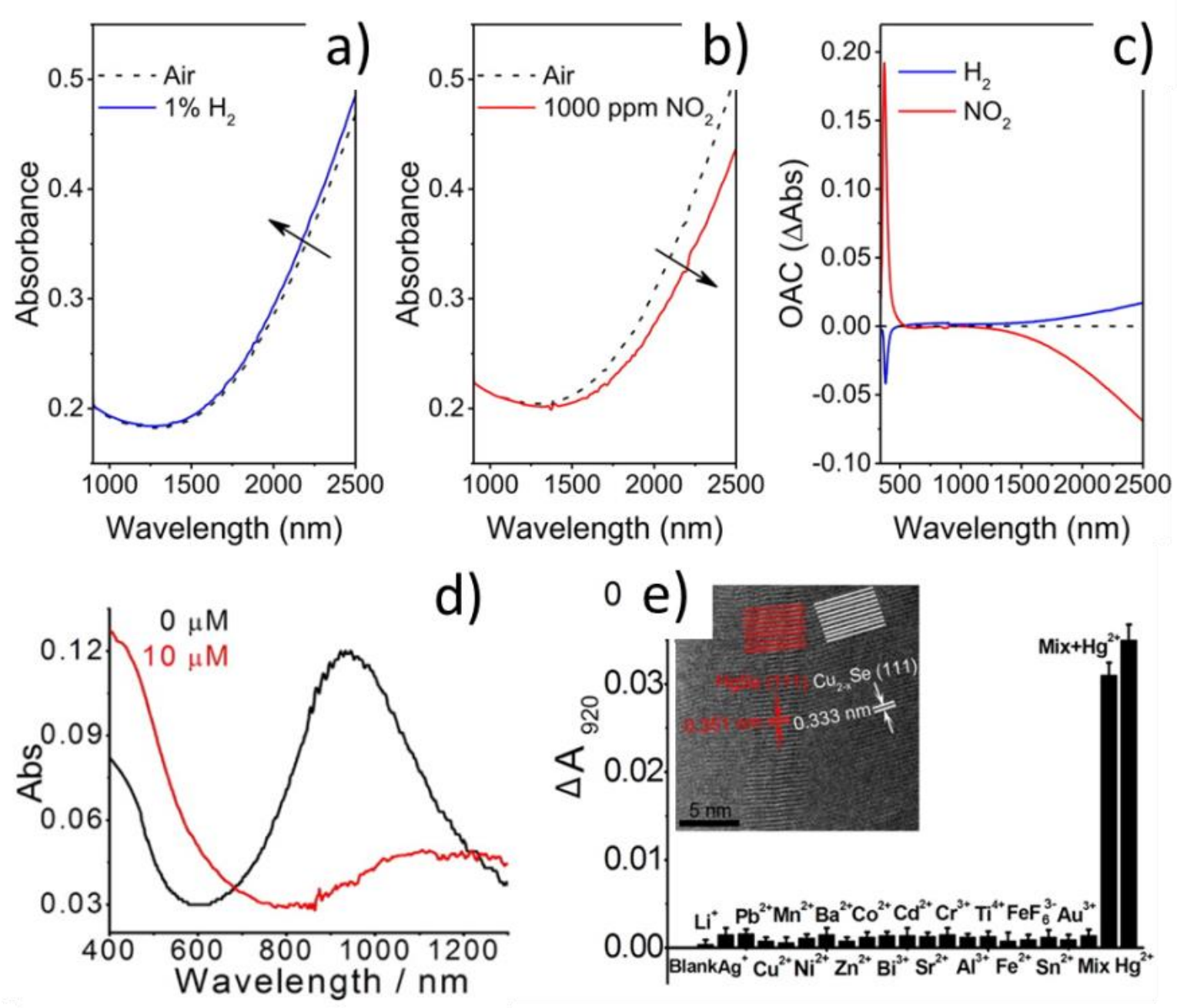

**Figure 28.** a-c) Optical absorption spectra for GZO10 sample when exposed to (a) 1% vol $H_2$ and (b) 1000 ppm of $NO_2$ at 150 °C operating temperature. A blue-shift (a) and red-shift (b) of the LSPR leads to (c) an overall optical absorbance change ($Abs_{Gas} – Abs_{Air}$) at 2000 nm to positive values when exposed to $H_2$ and to negative values when exposed to $NO_2$. Adapted with permission from ref. [252] Copyright 2016 American Chemical Society. d) Absorption spectra

of the $Cu_{2-x}Se$ NCs in the absence and presence of 10 μM $Hg^{2+}$. e) Absorption responses of the $Cu_{2-x}Se$ NCs to some transition/heavy metal ions in environmental water samples showing the high affinity to $Hg^{2+}$. Inset: HRTEM image of the $Cu_{2-x}Se$ NCs in the presence of 10 μM $Hg^{2+}$ illustrating the HgSe shell. Adapted with permission from ref. [251] Copyright 2016 American Chemical Society.

Many sensor applications involving plasmonic semiconductor NCs were demonstrated that are based on similar approaches as for conventional plasmonic nanoparticles, such as NC aggregation and the thus following LSPR shift[51,253] or refractive index changes.[254] Nevertheless, there are several recent reports that rely on the specific characteristics coming along with degenerately doped semiconductor NCs. For example, in MO NCs the capability to uptake and release capacitive carriers from their conduction band has attracted attention for their use as dispersible plasmonic NC sensors that optically track redox reactions.[31] Mendelsberg *et al.*[31] demonstrate that electron transfer in aqueous systems are remotely tracked when plasmonic ITO NCs are introduced to a system out of redox equilibrium. Electron transfer events occur until equilibrium is reached, leaving the ITO NCs behind with a modulated free carrier concentration tracked as LSPR shift. The authors showed that it is possible to monitor such processes with single-electron sensitivity by fitting the LSPR with the Mie theory and Drude model, thus, noninvasively and quantitatively tracking redox chemical events for organic, inorganic, biogenic, and even living cells.[31] Similarly, copper chalcogenide NCs were suggested as plasmonic probes of redox chemistry as the LSPR of vacancy doped semiconductor NCs is sensitive to redox chemical events.[11–13,30,244] The electron transfer events were tracked *via* changes of the LSPR, through the perturbation of their carrier density. Such plasmonic probing was shown for redox reactions, electrochemical charging/discharging, or ligand binding.[30]

In a recent work gallium doped zinc oxide (GZO) NCs were presented as novel plasmonic sensors for the detection of hazardous gases. The authors exploit the strong sensitivity of the LSPR in GZO NCs to chemical and electrical changes at the NCs' surface to optically detect both, oxidizing and reducing gases. GZO NCs were deposited into uniform and homogeneous films with visible transparency. The changes in the NIR plasmonic absorption was exploited to selectively sense hydrogen ($H_2$) and nitrogen dioxide ($NO_2$): probing the intensity at proper wavelengths (e.g. 2000 nm) results in a positive variation of the signal for $H_2$ and in a negative variation for $NO_2$. The detected shifts arise from the variation in electron density when the metal oxide interacts with the gases, causing a blue-shift of the LSPR peak and increase in carrier density when interacting with $H_2$, and a red-shift of the plasmon peak due to the removal of electrons when interacting with oxidizing gases like $NO_2$ (**Figure 28**a-c). Chemiresistive measurements confirm these findings and suggest highly doped semiconductor NCs as both NIR plasmonic and chemiresistive sensors acting towards highly sensitive and selective devices with low operation temperatures.[252]

Another interesting approach to sensing applications is to exploit plasmon coupling in nano-heterostructures composed of two different plasmonic sub-moieties, one of which is a degenerately doped semiconductor. For example in a heterostructure of copper chalcogenide and gold, a minor change in the copper chalcogenide LSPR due to the uptake or release of carriers leads to a detectable shift of the gold plasmon resonance due to the altered plasmonic coupling.[44] Thus, the sensing though a modulation of the copper chalcogenide LSPR in the NIR is translated into the visible spectral range. Along these lines ultrathin copper sulphide films were produced by chemical bath deposition. The films were made of CuS (covellite) nanoclusters, with grain sizes of approximately 50 nm. The CuS films were then covered with Pd nanoparticles of approximately 4–5 nm by electron beam deposition. The resulting optical hydrogen sensor changes its color from dark green to brown, which makes the proposed CuS–Pd an eye-readable sensor. The color change is ascribed to a decrease in the LSPR of CuS in the NIR due to a reduction of the free carrier density upon exposure to $H_2$ gas. The proposed CuS–Pd optical sensor can efficiently detect the presence of up to 0.8% of $H_2$ gas in air.[251] A promising direction that takes advantage of the specific affinity of copper chalcogenide NCs to metal ions was recently highlighted by *Xie et al.* and is based on the sequestration of heavy metals.[32] The uptake of heavy metals, such as $Cd^{2+}$ and $Hg^{2+}$, into the crystal of $Cu_{2-x}S$ NCs leads to a quenching of the plasmon resonance, which enables a simultaneous tracking of the process together with the purification of the solution. The guest metal cations are thereby incorporated by breaking up the S−S bonds (operated by the presence of a reducing agent). The released Cu species are mostly recaptured by the not exchanged CuS domains or by the remaining CuS NCs leading to the LSPR quenching.[32] Along similar lines, Zhang *et al.*[255] have reported a new colorimetric platform for sensitive and selective sensing of $Hg^{2+}$ by using NIR plasmonic $Cu_{2-x}Se$ NCs which exploits the strong tendency of $Hg^{2+}$ to exchange the copper cations in $Cu_{2-x}Se$ NCs, yielding a HgSe layer around the host NCs. The absorbance in the visible is thereby red shifting, with a simultaneous dramatic decrease of the NIR plasmon resonance. Tracking of such shifts reveals $Hg^{2+}$ quantification down to about 10 nM as shown in various water samples, combined with a high selectivity. Due to favorable analytical performance, the proposed $Cu_{2-x}Se$ NPs based system has potential applications in monitoring trace amounts of $Hg^{2+}$ ions in various real samples, even in drinking water (**Figure 28**).[255] The interesting part of these approaches to sensing is that the sensitivity is based on real chemical reactions, differently from the refractive-index based probing exploited in noble metal nanoparticle sensors.

### 7.2. Near field enhancing spectroscopy with NIR plasmonic semiconductor nanostructures

Light-matter interaction at the nanoscale can be strongly enhanced in the regions where the electronic light field is localized. This can occur at so called hot spots, where noticeable field intensity enhancement occurs within a small volume. Traditionally, noble metal substrates, such as gold, are employed for this purpose, bringing along high field enhancement factors. But also plasmonic semiconductor NCs were shown to provide high enhancement factors. In both cases the hot spots decay exponentially in about 10 nm.[234] This scale defines the length-scale of surface sensitive techniques relying on near field enhancement, such as surface enhanced infrared absorption spectroscopy (SEIRA) or surface enhanced Raman scattering (SERS). The electric field enhancement can lead to an increase in the signal intensity by orders of magnitude for molecules adsorbed on a specific surface. This is exploited as a powerful analytical tool for the identification of chemical and biological analytes adsorbed to the substrate surface.[42,256,257] Cong *et al.*[258] showed that the use of vacancy-containing $W_{18}O_{49}$ nanostructures as the SERS substrate yielded SERS enhancement factors as high as $3.4 \times 10^5$, comparable to noble metal SERS substrates (without the presence of hot spots). The importance of oxygen vacancies and thus the plasmonic response was evidenced by deliberately controlling the vacancy density upon substrate annealing in $WO_{3-x}$ nanostructures. The authors assigned the Raman enhancement to the formation of a charge-transfer state between the adsorbate and the semiconductor substrate, strongly involving the vacancy states within the bandgap of tungsten oxide.[258] Similarly CuTe nanorods with a very weak plasmonic peak did not yield any SERS signal, while other shapes with pronounced LSPR band did, highlighting the importance of the NIR LSPR in the enhancement process of the SERS signal (**Figure 29**a and b).[42] In a different example molybdenum oxide was investigated as a SERS substrate and surprisingly, initially the SERS activity of the vacancy doped, plasmonic $MoO_{3-x}$ was found to be lower than the one of non-plasmonic commercial $MoO_3$.[259] This trend was explained by considering that plasmonic $MoO_{3-x}$ was also photocatalytically active in degrading the adsorbed chemical species on the SERS substrate, and this was linked to the presence of oxygen vacancies on the surface of the $MoO_{3-x}$ substrate. To track the distribution of oxygen vacancies with the SERS activity, the authors subjected the $MoO_{3-x}$ surface to mild annealing. This lead to the elimination of surface $Mo^{5+}$ species and oxygen vacancies, resulting in $MoO_{3-x}/MoO_3$ core–shell structures, which possess a plasmonic $MoO_{3-x}$ core that preserves the initial free carrier concentration (~ $10^{21}$ $cm^{-3}$) and a non-plasmonic $MoO_3$ shell. The elimination of $Mo^{5+}$ centers and oxygen vacancies at the nanostructure surface prevent the photocatalytic degradation of the analyte, but instead support the electromagnetic SERS enhancement due to the plasmonic $MoO_{3-x}$ core leading to SERS signals of the order of $10^5$, comparable to noble metals (**Figure 29**c and d).[259]

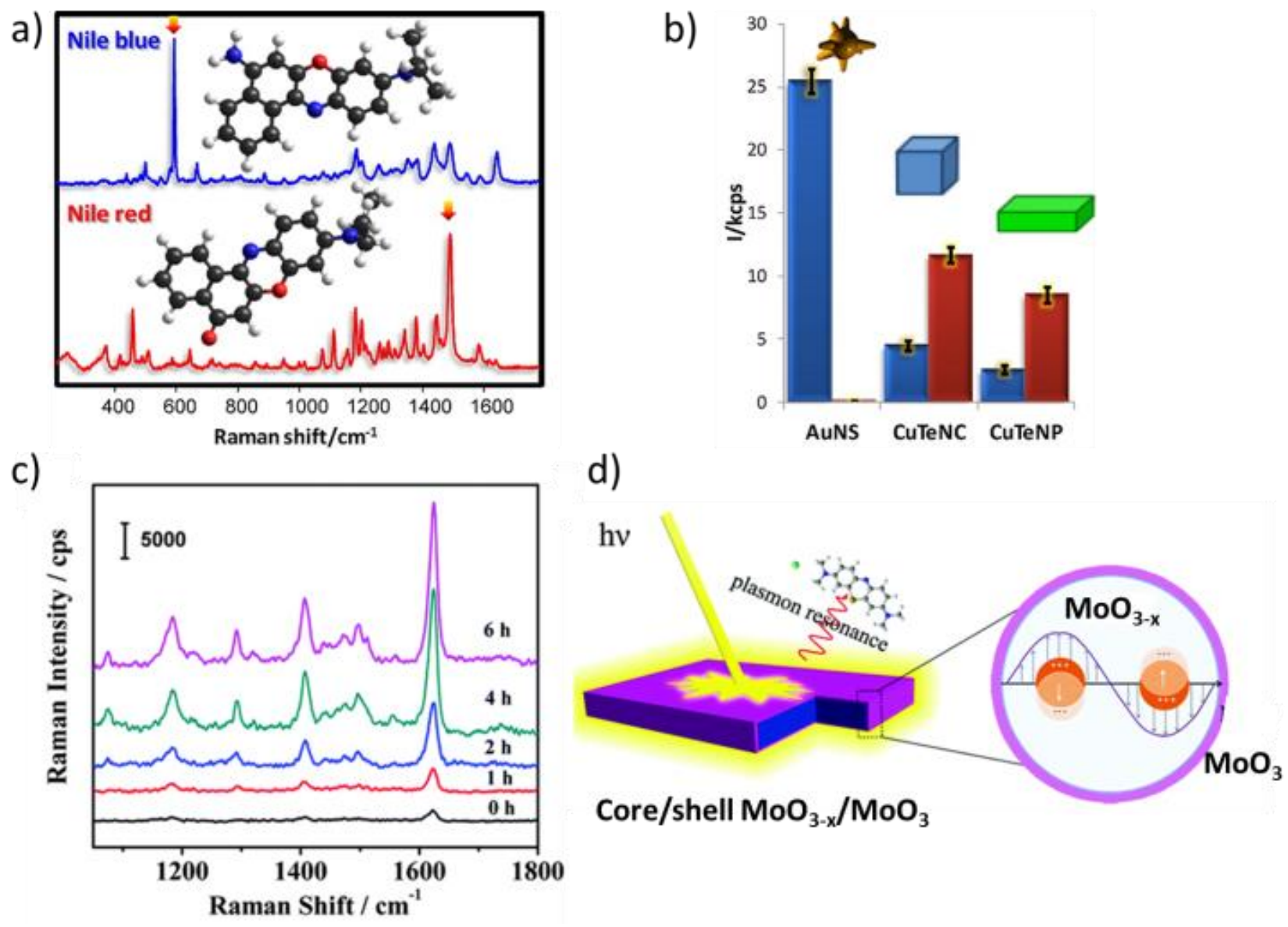

**Figure 29.** a) SERS spectra of Nile blue and Nile red. b) Signal enhancement of Nile blue (blue pillars) and Nile red (red pillars) when using gold nanostars, CuTe nanocubes, and CuTe nanoplates. c) SERS spectra of methylene blue on plasmonic $MoO_{3-x}$ nanosheets for different annealing times. Adapted with permission from ref [42] Copyright 2013 American Chemical Society. d) Schematic of the core/shell $MoO_{3-x}/MoO_3$ nanosheets depicting the proposed mechanism for SERS enhancement that occurs due to the nanoshell-isolated electromagnetic enhancing mechanism for plasmonic $MoO_{3-x/}MoO_3$ nanosheets. Adapted from ref [259] with permission of The Royal Society of Chemistry.

Shape tailoring enhances the signal intensity for hot spots. As such CuTe cubes demonstrate slightly higher signal enhancement compared to plates, and indium doped cadmium oxide (ICO) octahedra show extremely enhanced signals with respect to their spherical counterparts.[38] The field enhancement decreases significantly when the corners are smoother,[38,42] as demonstrated by Agrawal *at al.*[38] by studying the field enhancement with increasing rounding of the corners. While this is well known also from noble metals nanostructure, degenerately doped semiconductor NCs additionally provide the advantage to shift the plasmonic response to the NIR, an important spectral range for biological applications.[42,256,257] Noble metal LSPRs are mostly located in the visible spectral range but can be extended to the NIR if the particles are grown bigger. Larger sizes however make their applicability in living organisms less viable. Comparison of calculated extinction spectra of noble metal and doped MO nanoantenna arrays shows that the same spectral position of the LSPR in the NIR corresponds to a four times shorter MO nanoantenna with respect to the noble metal.[234] For sensing applications, the density of hot spots on a substrate influences dramatically the overall efficiency of a sensor. Thus, the smaller size of the nanoantenna based on metal oxides results in a 30 times higher number of hot spots on a substrate with respect to the noble metal. As such, a large area sensor based on a dense array of doped MO nanoantennas has the potential to outperform an array of noble metal nanoantennas, even though each individual MO nanoantenna shows an order of magnitude lower hot spot field intensity enhancement, as a result of the order of magnitude lower carrier density.[234]

Degenerately doped semiconductor nanostructures provide the option to extend the collection of functional groups that can be adsorbed to their surface.[42,260] For example the efficiency of CuTe NCs as SERS sensors was tested by investigating two different analytes.[42] The enhancement factors were compared with gold nanostars, which are among the most efficient SERS structures (**Figure 29**a).[261–263] SERS spectra of Nile blue and Nile red in contact with the CuTe NCs show enhancement factors up to around $10^6$. Although the amino-containing Nile blue gave higher SERS factors when investigated with the gold nanostars, the Nile red did not show any signal enhancement in the presence of gold. This was explained by the low affinity of ketones for gold compared to CuTe and demonstrates the possibility to extend the library of analytes that can be detected by plasmonic semiconductor NCs.[42] In a different work, magnetite ($Fe_3O_4$) NCs were found to provide SERS signal from oxalic acid and cysteine *via* an electric field enhancement. This study highlights magnetite as a possible oxide substrate for SERS study of adsorbed organic molecule with in principle fewer restrictions on the binding mechanisms.[260]

A great advantage provided by doped semiconductors is the lower production of heat with respect to their noble metal counterparts. It is well known that noble metals suffer from high losses and absorption resulting in local heating due to the transduction of the absorbed radiation into thermal energy. However, the increase of the local temperature around the nanoparticle can influence the electromagnetic response of the system.[264] As such, novel heavily doped semiconductor nanostructures were suggested, such as GaP or phosphorous doped Ge, which show ultralow optical losses in the investigated NIR regime, representing a novel approach to applications that require field enhancement, but in the same time omit local heating. Indeed, it was shown that their mid- to NIR plasmonic response was exploited for the resonant detection of molecular vibrational fingerprints.[264,265] One can envisage the *in situ* tunability of the carrier density as an additional tool for resonance-specific sensing in the near to mid-IR, by switching on/off and by fine-tuning the LSPR to specific vibrational signatures.[38]

### 7.3. Photon to energy conversion with heterostructures based on degenerately doped semiconductor NCs

Nano-heterostructures based on plasmonic semiconductor nanostructures are interesting for photocatalytic application,[33,83,99,143,195,266–283] as for example in combination with carbon-based materials[268–270,284,285] or metal co-catalysts.[33,83,99,143,195,266,267,271–283] The decoration of semiconductor nanocrystals with metal co-catalysts, such as platinum or gold, represents a promising approach for catalytic hydrogen production. The co-catalyst acts as an electron sink for the photogenerated carriers, thereby improving the charge separation, which is beneficial for triggering photocatalytic reactions.[225,226,286–289] A recent study by Manzi *et al.*[144] demonstrated that Pt decorated $Cu_{2-x}S$ nanorods serve as efficient photocatalysts for reducing $CO_2$ with excellent selectivity over hydrogen production. The materials were prepared starting from Pt-decorated CdS nanorods (in an aqueous environment in aerobic conditions), by using the photoexcited electrons in the conduction band of the initial nanorods to reduce $Cu^{2+}$ ions to $Cu^{+}$. The latter were then engaged in cation exchange reactions with the $Cd^{2+}$ ions of the nanorods. Carbon monoxide evolution measurements showed an activation time for the photocatalytic production, which matches the time required to complete the cation exchange reaction from CdS to $Cu_{2-x}S$. This indicates that during the exchange reaction the photogenerated electrons reduce $Cu^{2+}$ instead of $CO_2$, which leads to the formation of $Cu_{2-x}S$. Once the cation exchange is completed, the reduction of $CO_2$ is triggered. With this method the authors presented a way to circumvent the need for protective atmosphere and non-aqueous medium during the cation exchange synthesis of $Cu_{2-x}S$, a prerequisite for their use in photocatalysis. Moreover, the work demonstrated that the $Cu_{2-x}S$ nanorods are efficient and selective catalysts for $CO_2$ reduction to both carbon monoxide and methane under visible light illumination (**Figure 30**a).[144]

Notably, a correlation between the hole carrier density, i.e. plasmonic response and the photocatalytic activity was demonstrated in some cases.[272,290] In a particularly interesting example involving Pd decorated $Cu_7S_4$ nanocrystals, the authors showed that the NIR LSPR of $Cu_7S_4$ was exploited for light-harvesting, while the catalytic features of the Pd co-catalyst helps to drive solar photocatalysis of organic synthesis reactions. Wavelength selective excitation confirmed the involvement of the LSPR in the photocatalytic activity. This effect was ascribed to the transfer of hot holes generated by plasmon excitation in the NIR, which were then injected into the Pd co-catalyst. The hole-rich Pd co-catalyst then serves as effective catalytic site to promote the chemical reactions (**Figure 30**b).[271]

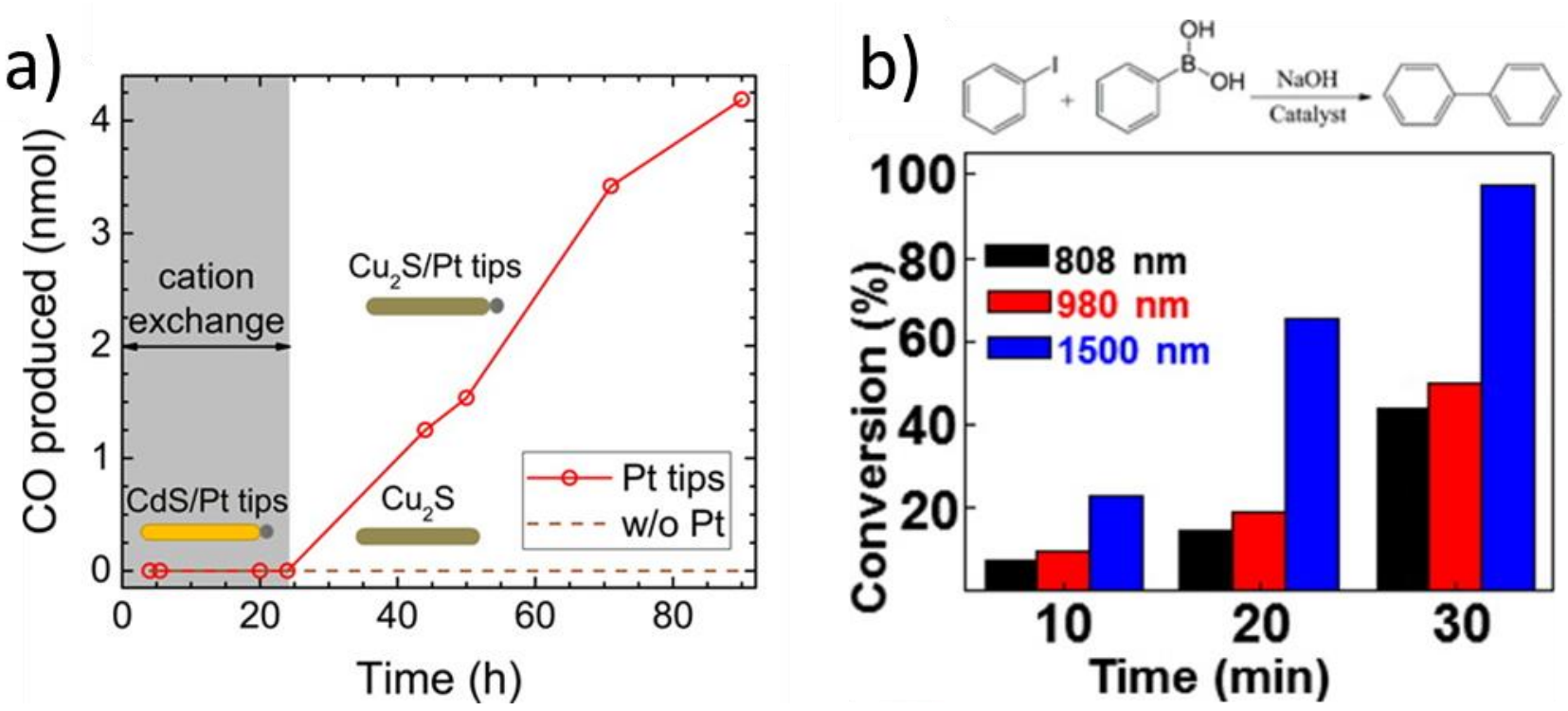

**Figure 30.** a) CO evolution from Pt decorated CdS/$Cu_2S$ NRs in the presence of Cu(II) after illumination with a 447 nm laser. Adapted with permission from ref [144] Copyright 2015 American Chemical Society. Photocatalytic activity of $Cu_7S_4$-Pd nanostructure for Suzuki coupling reactions of iodobenzene with phenylboronic acid under 808, 980, and 1500 nm laser irradiation for 30 min. Adapted with permission from ref [271] Copyright 2015 American Chemical Society.

A recent work exploited the NIR absorption from degenerately doped semiconductor NCs to design an efficient near-infrared nanophotodetector:[291] the authors fabricated germanium nanoneedle arrays coated with a single layer graphene on which ITO NCs were spin coated. High sensitivity of the device structure was found to 1500-nm illumination with a high response speed, excellent spectral selectivity and good reproducibility in a wide range of switching frequencies. The involvement of ITO NCs resulted in improved performance with respect to the device without ITO decoration. The main effect responsible for the improved performance was related to plasmon induced hot electron injection from the ITO NCs to the single layer graphene in combination with the light confinement effect of the supporting Ge nanoneedle array.[291] The works highlighted above highlight the possibility to exploit the NIR LSPR for light harvesting upon plasmon induced hot carrier extraction in the NIR.[33,271,292] As also shown for example, in $Cu_{2-x}S$ nanoplate/reduced graphene oxide electrodes serving as active electrocatalysts for the oxygen reduction reaction. [143] Three different doping levels of the copper chalcogenide component were tested, where the highest electrocatalytic activity was found for the most heavily doped sample. The increase of the catalytic activity was attributed to its significantly higher concentration of free holes.[143] Also in quantum dot sensitized solar cells the potential benefits of the doped versus the undoped semiconductor NC was unraveled.[293,294] Indeed, in $Cu_xIn_yS_2$ NCs showing plasmonic response in the NIR improved performance was shown with respect to the non-plasmonic device.[294] This enhanced photovoltaic response was assigned to augmented charge excitation due to the near-field antenna effects in the plasmonic NCs.[294]

Copper chalcogenide NCs were also used as high conductive electrodes and in solar cells to improve the charge transport.[138,138,295–303] In a recent work, a layer of CuS NCs was used as an inorganic hole-selective layer in inverted planar perovskite solar cells.[304] In the inverted device usually two planar charge transport layers selectively pass electrons and holes created in the perovskite light-absorbing layer to the corresponding electrodes. Due to high hole mobility, excellent chemical stability and low cost, CuS NCs were suggested as hole-selective contacts and displayed superior performance with respect to the device without the CuS NCs layer. This indicates that CuS NCs serve as potential in-organic hole-selective contacts for perovskite solar cells.[304] Films of 2D nanosheets of copper sele-nide outperform films of spherical NCs of the same materials under mechanical stress. For example copper selenide nanosheets with lateral dimensions of up to 3 μm were used to prepare flexible conduc-tive films by simple drop-casting:[305] the electrical conductivity of the films was almost fully recov-ered after bending, compared to a much severe drop in conductivity from films made of spherical NCs. This makes nanosheets better suited than spheres in flexible electronics applications.[305]

### 7.4. Tunable optoelectronic devices involving tunable plasmonic NCs

The dynamic control of the LSPR of degenerately doped semiconductor NCs presents an exceptional opportunity to develop tunable optoelectronic devices. Important functionalities can be achieved when the NCs are assembled in a controlled fashion. For example in photonic crystal structures of alternating layers of plasmonic semiconductor NCs and dielectrics, selective tuning of the NIR LSPR and the pho-tonic band can be achieved by controlling the carrier density of the doped NC films. This highlights their use for sensing, tunable light filtering or electrochromic devices.[254,306,307] Varying doping levels and capacitive charging are possible ways to induce NIR LSPR tunability that can be interesting for effective NIR absorbing solar heat-shielding[78,80,308,309] or for electrochromic devices.[10,310]

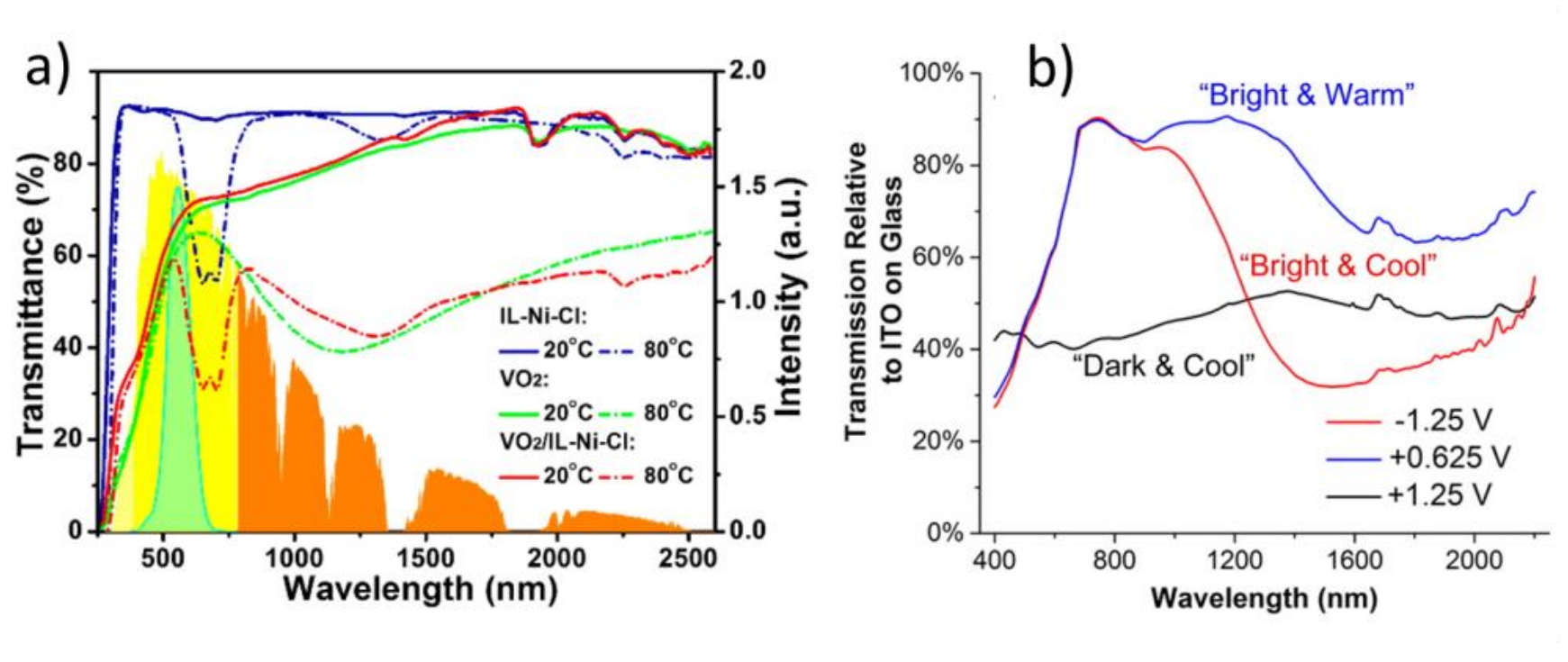

**Figure 31.** a) Transmission spectra at two different temperatures: 20 °C and 80 °C of pure liq-uid–nickel–chlorine (IL–Ni–Cl) complexes, pure $VO_2$ film and $VO_2$/IL-Ni-Cl hybrid film. The yellow-orange area depicts the visible and NIR range, while the cyan area indicates the region of highest eye sensitivity. Adapted with permission from ref [311] Copyright 2016 American Chemical Society. b) Transmission spectra of the poly(3,3'-dimethyl-2,2'-bithiophenyl) ($PMe_2T_2$)–ITO composite film at −1.25 V (red), + 0.625 V (blue), and +1.25 V (black) versus ferrocenium ion/ferrocene ($Fc^+$/Fc) redox couple. Adapted with permission from ref [312] Copyright 2016 American Chemical Society.

Matsui *et al.*[308] showed that the three dimensional assembly of surface-modified ITO NCs leads to plasmon coupling induced high reflectance performance in the near- and mid-IR range due to interparticle plasmon coupling on large-size flexible substrates.[308] In a different example, vanadium oxide was used as thermochromic smart coatings, switchable with temperature (**Figure 31**a).[79,311,313–315] Thus, such design assembly appears as a powerful tool to integrate the properties of the NCs to macroscale functionality.[10] The application of an electric field to such a functional structure delivers a way to control the spectral properties of the device. Indeed, in the latter example vanadium oxide was also shown to undergo transformation induced color changes by applying an electrochemical potential.[313,316] Reversible LSPR tuning is also possible by manipulation with an applied potential by charging and discharging of the degenerately doped NC assemblies.[10,22,310,317] Also selective NIR and visible light switching was shown in doped anatase titania NCs. The selective modification was induced by controlling capacitive charging for NIR LSPR manipulation, and lithium intercalation for visible light absorption. The latter induces structural phase transitions that result in visible light absorption variation of the NCs. This delivers a way for multimodal electrochromic applications in smart windows.[87] Particular interesting switching properties arise when composites are formed between the plasmonic NCs and other materials. Such composite electrochromic materials enable to independently tune the transmission of visible and NIR across several modes of distinct operation, such as "bright and warm", "bright and cool", and "dark and cool" (**Figure 31**b). [29,129,312,318,319] Also nanoscale structuring plays a very important role, leading to enhanced performances and faster switching times due to increased surface area and improved stability.[29,125,129,319,320] Such tunable devices are useful for making efficient spectrally-selective smart windows, tunable filtering devices, sensors or other devices such as dynamic real-time plasmonic sensors.[10,254,306,310,321]

### 7.5. Biomedical applications of degenerate semiconductor NCs

Since the discovery of the plasmonic properties of copper chalcogenide NCs, these materials have been suggested for photothermal therapy (**Figure 32**a).[27,28,42,43,322–334] The plasmon resonance of copper chalcogenides in the NIR is optimally located in the biological window (700–1100 nm), where the human tissue is partly transparent. This enables light irradiation to enter the tissue with minimal ab-sorption and excite the plasmon resonance. This energy is then released in form of heat, leading to a temperature increase and possible cell death when the therapeutic agent is located in close vicinity to the tumor site (**Figure 32**).[27,28,42,43,322–331] For example $Cu_{2-x}Se$ NCs with strong NIR plasmonic absorption peaking around 980 nm produce significant photothermal heating when excited with 800 nm light.[27] Their transduction efficiency was around 22%, comparable to nanorods and nanoshells of gold. Moreover, it was found that the release of toxic ions or the generation of high levels of reactive oxygen species from CuTe and $Cu_{2-x}S$ NCs result in additional photodynamic activity.[28,42] Even higher photothermal transduction efficiency was shown for nano-heterostructures combining copper chalcogenides with gold.[43,322,335–337] For example, deep tissue photothermal therapy was performed with Au-$Cu_9S_5$ core-shell nanoparticles by exciting into the second NIR window with a 1064 nm laser.[43]

Combining such therapeutic agents with non-invasive, sensitive and accurate molecular imaging techniques for early diagnosis of diseases represents a novel and powerful approach. Indeed, the X-ray computer tomography images were enhanced after injection of Au-$Cu_9S_5$ core-shell nanoparticles, due to the presence of the higher atomic number element Au, which has excellent X-ray attenuation ability.[43] Another promising direction in this field is represented by photoacoustic imaging, a technique in which a transient thermal expansion in the biological tissue, induced by a laser pulse, is detected as ultrasonic emission.[338–343] It was found for example that Au-$Cu_{2-x}Se$ NCs represent efficient tools for in vivo non-invasive photoacoustic imaging in deep tissue with imaging depths of 17 mm using a 1064 nm laser source.[343] In yet another approach, the functionalized ligand shell of the $Cu_4S_7$ NCs was loaded with $^{19}F$ to achieve simultaneous $^{19}F$ magnetic resonance imaging and photothermal therapy.[344,345] Magnetic resonance imaging with $^{19}F$ is a powerful molecular imaging technique, due to its high sensitivity and negligible background contribution.[344,345] Along similar lines, the introduction of $^{64}Cu$ ions to the photothermal agent serves well for imaging based on positron emission tomography.[52,346,347] An example in this direction is represented by the approach developed by Riedinger *et al*. to introduce $^{64}Cu$ in CuS NCs:[52] by taking advantage of the unique ability of copper chalcogenides to uptake metal cations into their crystal, $^{64}Cu$ was post-synthetically incorporated in CuS NCs as radiolabel for photothermal probing. Moreover, the simplicity of the preparation scheme, which involves the use of radioactive species only as a last step, makes the protocol easily transferable to the clinical practice.[52]

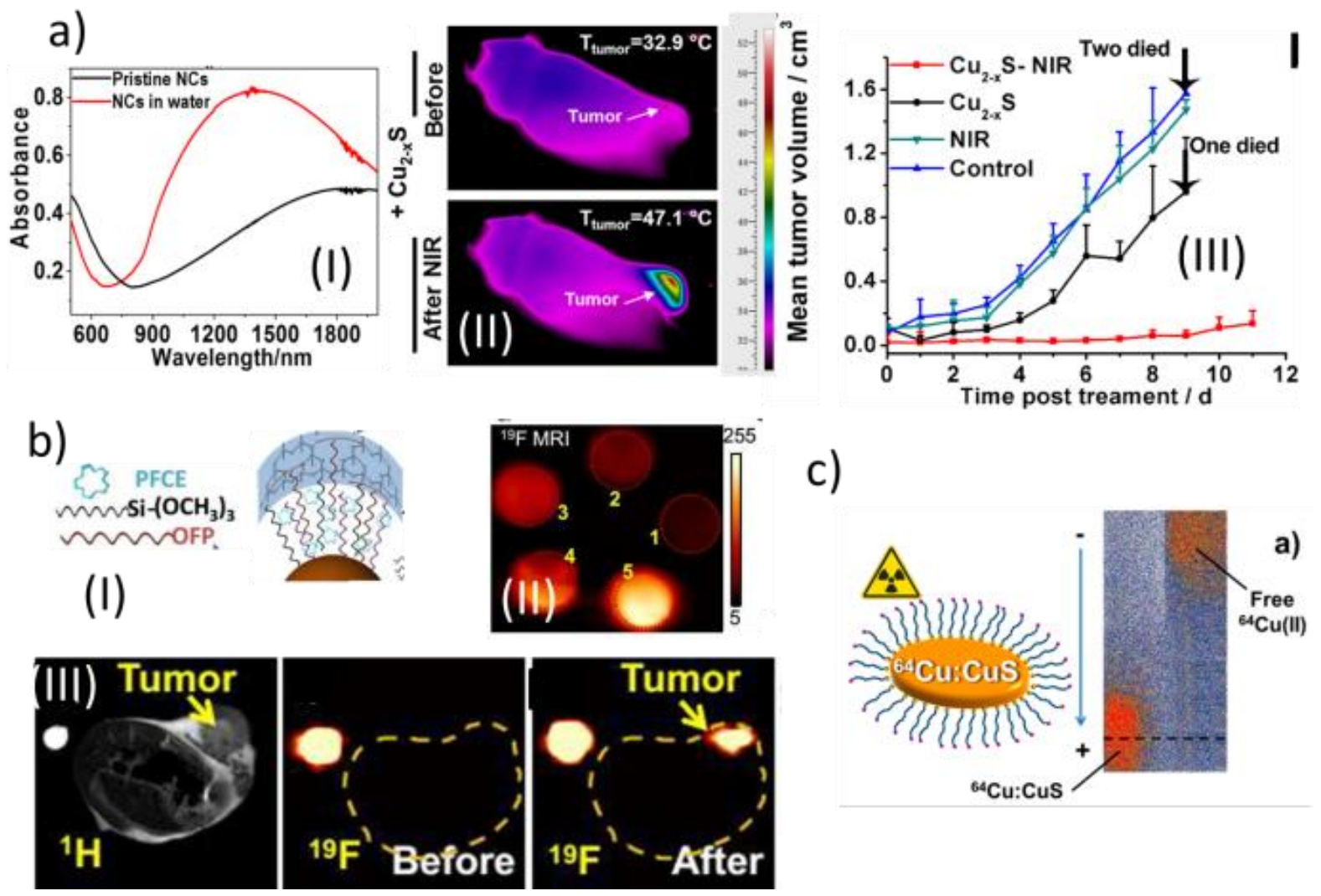

**Figure 32.** a) (I) Optical absorbance of the pristine $Cu_{2-x}S$ NCs in toluene (black curve) and after ligand exchange and water transfer (red curve). (II) Infrared thermal images of the tumor site post injection of $Cu_{2-x}S$ NCs before and after NIR laser light irradiation. (III) Mean tumor volumes were measured in the control group, NIR group, $Cu_{2-x}S$ group, and $Cu_{2-x}S$-NIR group. Adapted with permission from ref. [28]; copyright 2015 American Chemical Society. b) (I) Illustration of the multifunctional $^{19}F$ loaded $Cu_{1.75}S$ nanoprobe; (II) $^{19}F$ MR images of $^{19}F$ loaded $Cu_{1.75}S$ nanoprobe with various $^{19}F$ concentrations. (III) From left to right: *In vivo* $^{1}H$ and $^{19}F$ MRI of the tumor-bearing mice before and after the injection of $^{19}F$ loaded $Cu_{1.75}S$ nanoprobe. Adapted with permission from ref. [344]; copyright 2016 American Chemical Society. c) Sketch of the radioactive $^{64}Cu$ incorporated CuS for radiolabeling and the agarose gel electrophoresis analysis of free $^{64}Cu(II)$ and $^{64}Cu$:CuS NCs demonstrating the loading of $^{64}Cu$ into the NCs. Adapted with permission from ref. [52]; copyright 2016 American Chemical Society.

Drug delivery systems are receiving increasing attention in the field of cancer therapy, with the possibility of remote and noninvasive control over drug release for improved treatment efficacy and reduced side effects. Promising results were reported by loading nanoparticles, such as gold nanorod and hollow $Cu_7S_4$ yolk-shell NCs, with specific drugs.[336] Thermosensitive drug delivery systems were also developed, which exploit the photothermal properties of the NIR plasmonic copper chalcogenide NCs leading to the photothermal release of the pre-loaded drugs. For example, $Cu_{1.75}S$ NCs coated with thermosensitive polymers (prepared by *in situ* polymerization or incapsulation in thermosensitive microgels) can act as smart nanoplatforms for synergistic, hyperthermia-based, *in vitro* and *in vivo* cancer therapies.[348,349] A novel platform for the use of plasmonic $Cu_{2-x}S$ NCs in biomedicine as intelligent light-driven sterilants was reported by Liu *et al.*.[350] Upon 980 nm laser irradiation, the NIR plasmonic

NCs acted as photothermal agents. Additionally, the generation of high levels of reactive oxygen species was observed, which lead to cellular injury due to apoptosis-related death and excess expression of caspase proteins in Sertoli cells. The NIR-mediated toxicity towards Sertoli cells, both *in vitro* and *in vivo*, was induced through the cooperation of photothermal and photodynamic effects, with an extremely high bio-compatibility.[350] Along a similar line, Wang *et al.*[28] observed that $Cu_{2-x}S$ NCs, upon NIR irradiation, not only acted as photothermal agents, but additionally as photodynamic agents, by generating high levels of reactive oxygen species, which the authors linked to the local release of Cu(I) species.[28]

### 7.6. Interaction of copper chalcogenide NCs with cations for tailored material synthesis

A particularly interesting characteristic of copper chalcogenide NCs is their ability to engage in cation exchange reactions, making them an attractive tool to synthesize novel materials, such as alloys based on copper chalcogenides[85,155,351] or to create more complex heterostructures composed of more than one material.[44,85,352] As an example, De Trizio *et al.*[85] showed that, starting from $Cu_{2-x}Se$ NCs and depending on the valency of the exchanging cation (Sn in their case), either alloy (when using $Sn^{4+}$ ions) or heterostructured NCs (when using $Sn^{2+}$ ions) were formed. In another example, dual interface epitaxial heterostructured NCs of $Cu_{2-x}S$ and ZnS were formed, through cation exchange with $Zn^{2+}$, whereby the thickness of the inner-disk layer could be tuned to form two-dimensional, single atomic layers (<1 nm) (see **Figure 33**a).[353] Notably, the reaction of these heterostructured NCs with trialkylphosphines, which are usually serving as surfactans for NCs leads to the selective chemical etching of the $Cu_{2-x}S$ part in the presence of oxygen creating unique morphologies (**Figure 33**b). In a multistep process S is removed from the NCs followed by diffusion of excess Cu ions into the remaining particle. This ultimately results in the quenching of the LSPR due to the reduction of carrier concentration. This work presents an attractive approach integrating top-down methods, such as chemical etching, with the bottom-up size and shape control of colloidal nanoparticle synthesis.[353] Tu *et al.*[352] demonstrated instead that the coordination number of the entering cations dictates the reaction pathways in cation exchange reactions of $Cu_{2-x}Te$ nanocubes. As such, cations with coordination number 4, such as $Cd^{2+}$ or $Hg^{2+}$, result in Janus-like heterostructures of $Cu_{2-x}Te$/CdTe and $Cu_{2-x}Te$/HgTe, as shown in the HRTEM image of **Figure 33**c. Instead, cations with the coordination number 6, such as $Pb^{2+}$ or $Sn^{2+}$ react with the NCs forming $Cu_{2-x}Te$/PbTe (**Figure 33**d) or $Cu_{2-x}Te$/SnTe core/shell nano heterostructures. The differences in the final structures are attributed to the size of the cations, as follows: the relatively small tetrahedrally coordinated cations can easily diffuse and nucleate in the crystal at a preferred region. This ultimately leads to the formation of Janus-like architectures. The octahedrally coordinated cations instead are probably slower diffusers, due to their larger size, and this limits their reaction to the surface of the NCs, thereby forming core/shell nanostructures. Notably, an annealing process of these core/shells led to the transformation into Janus like structures, as obtained by cation exchange with the tetrahedrally coordinated cations (**Figure 33**c and d).[352]

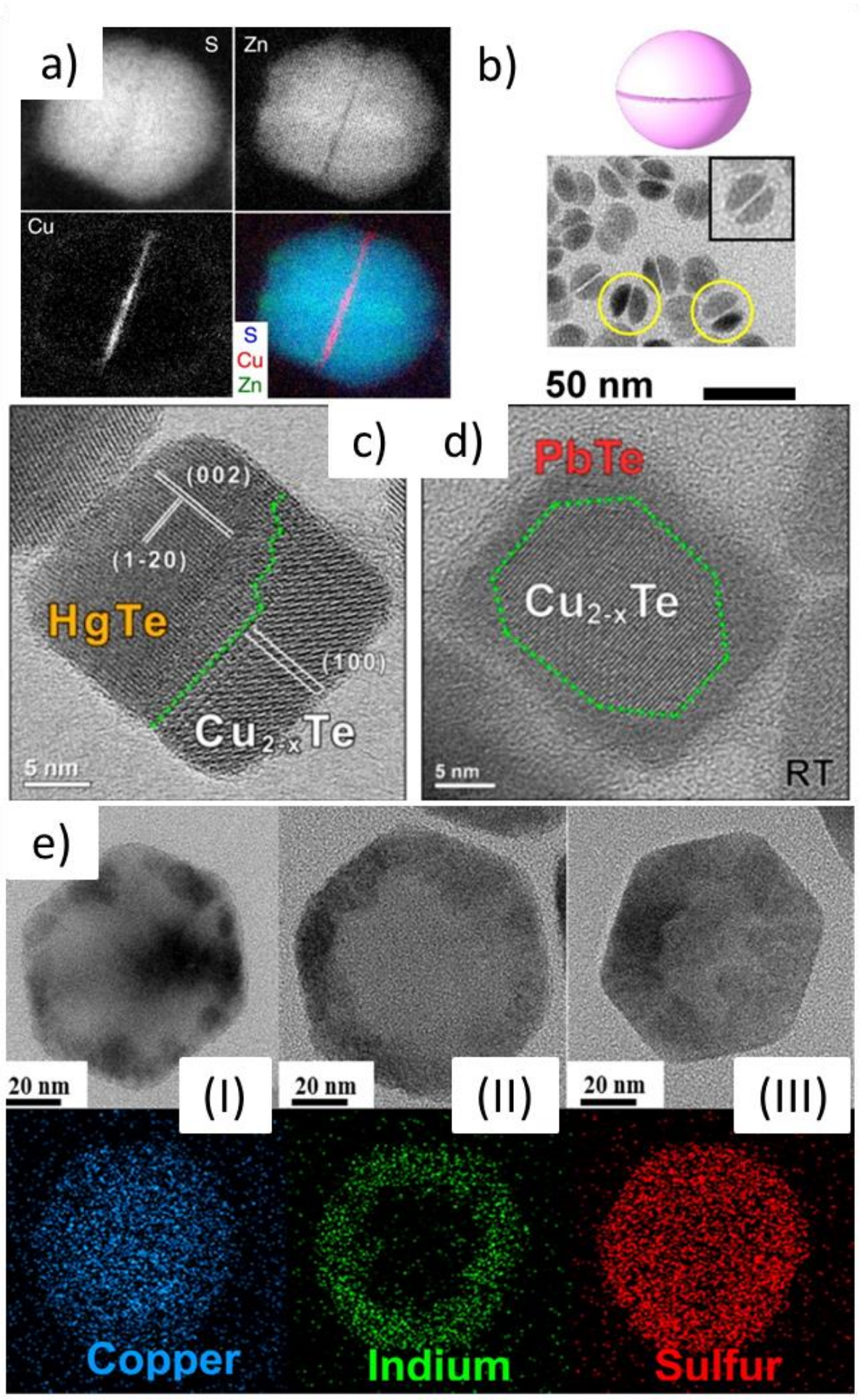

**Figure 33.** a) Epitaxial heterostructured NCs of $Cu_{2-x}S$ and ZnS with a $Cu_{2-x}S$ two-dimensional inner-disk layer tunable in thickness down to a mono layer. EELS mapping verifies the chemical composition of Cu and S in the internal disk region and Zn and S in the surrounding regions. Adapted with permission from ref [353] Copyright 2014 American Chemical Society. b) Schematic and TEM image of the etched $Cu_{2-x}S$–ZnS heterostructures (by reacting them with trialkylphosphines for 1 h). It appears that $Cu_{2-x}S$ is completely removed, leaving apparently connected ZnS caps. Adapted with permission from ref [354] Copyright 2014 American Chemical Society. c) and d) c) HRTEM image shows Janus-like NCs are made of a monocrystalline $Cu_{2-x}Te$ domain and a monocrystalline HgTe domain, with a sharp interface between the two domains indicated by a green dashed line. d) nano heterostructures prepared by partial cation exchange of $Cu_{2-x}Te$ NCs with $Hg^{2+}$ and $Pb^{2+}$ cations, respectively. Adapted with permission from ref [352] Copyright 2016 American Chemical Society. HRTEM images of $Cu_{2-x}Te$/PbTe

core/shell NCs before the in situ annealing treatment at 200 °C, which causes the transition from a core/shell architecture to a more stable Janus-like morphology, as a consequence of the minimization of the interfacial energy. e) TEM and HAADF-STEM images of $Cu_{2-x}S/CuInS_2$ core/shell structure indicating the growth of the outer shell $CuInS_2$ (from I to III). Each color spot represents an element of blue - copper, green - indium, red - sulfur. Adapted with permission from ref [355] Copyright 2016 American Chemical Society.

The role of copper vacancies on the cation exchange reaction was investigated by Lesnyak *et al.*,[356] in which the reactivity of stoichiometric and non-stoichiometric copper selenide NC with two different divalent cations ($Zn^{2+}$ and $Cd^{2+}$) was studied. Both cases delivered partially exchanged Janus-like heterostructures in which the two domains shared an epitaxial interface. An acceleration of the exchange process was found in the non-stoichiometric compounds, likely due to the presence of a large density of copper vacancies, which facilitated cation diffusion.[356] Similarly, it was found that the cation-exchange reaction between Cu and In atoms of $Cu_{2-x}S$ hexagons is more favorable at crystal planes where Cu vacancies are more easily formed. This was determined by density functional theory calculations, which showed that the lateral surface of the hexagonally shaped NC is more reactive towards vacancy formation and thus exchange occurs preferably from the lateral plane. This then leads to the formation of heteronanostructured $Cu_{2-x}S/CuInS_2$ nanodisks, with topology of chemical composition as shown in **Figure 33**e.[355]

In a different approach, $CuInSe_2$ nanoparticles were synthesized by a redox induced phase transformation of the copper selenide nanoparticle template upon addition of elemental indium.[351] The elegance of this strategy stands on the use of a sacrificial template of copper diselenide $CuSe_2$ that has a chemical and structural similarity to that of $CuInSe_2$. The pyrite form of $CuSe_2$ is a high-pressure phase, but was successfully synthesized under ambient conditions by employing high-energy shaker ball milling.[351] The solid-state transformation of $CuSe_2$ to $CuInSe_2$ is believed to occur via the concurrent oxidation of In to $In^{3+}$ and the reduction of $Cu^{2+}$ and $Se_2^{2-}$ to $Cu^+$ and $Se^{3-}$. The altering oxidation states of the Cu and Se atoms induce a reorganization of the atoms within the $CuSe_2$ crystal that together with the diffusion of $In^{3+}$ ions result in the high temperature cubic structure (sphalerite) of the $Cu^{1+}(In^{3+})\ (Se^{2-})_2$ phase.[351] Also copper indium sulfide (CIS) and quaternary copper indium zinc sulfide (CIZS) NCs were produced by means of cation exchange with $In^{3+}$ and $Zn^{2+}$.[357] Following the synthesis of $Cu_2S$ NCs, $In^{3+}$ was incorporated by a cation exchange reaction, and the resulting NCs were then further reacted with $Zn^{2+}$ to prepare the quarternary compound. In this last step the $Zn^{2+}$ partially replaces the $Cu^+$ and $In^{3+}$ cations located at the NC surface, facilitating the formation of a ZnS-rich shell.[357] $Cu_3SbSe_3$ nanorods were produced by a diffusion-controlled chemical transformation of binary anisotropic 1D $Sb_2Se_3$ nanostructures. Notably, also intermediate core/shell $Sb_2Se_3$/$Cu_3SbSe_3$ nanorods could be isolated.[170]

# 8. Outlook and perspectives

Research on plasmonic semiconductor NCs has advanced considerably in recent years, leading to good control over NC shapes and sizes, doping levels, materials, and the fabrication of hybrid structures. The two most investigated types of degenerately doped semiconductor NCs to date, copper chalcogenides and metal oxides, show very similar properties in terms of their optical response and plasmon tunability, although the two classes are based on fundamentally different doping (and thus tuning) mechanisms. While in the copper chalcogenides it is mostly the ratio of Cu to chalcogenide that determines the carrier density and the plasmonic response in the NIR, in doped metal oxides it is the level of doping. Active tuning in the latter is driven by the introduction of additional charge carriers to the conduction band of the material, as opposed to structural modifications in the copper chalcogenides. With the help of advanced characterization tools and elemental analysis, a good understanding of the doping mechanisms was brought together, unraveling structural changes upon controlled vacancy formation and LSPR tuning in copper chalcogenides, or the role of doping sites, vacancies, impurities, and their location throughout the crystal in metal oxides.

Although a considerable progress was made on understanding the dielectric response and the plasmonic properties of these materials, we remain still with numerous open questions. In copper chalcogenides it is of major importance to study in depth the structural changes upon redox-chemical treatment, which largely influences all parameters related to their dielectric function. This is particularly important for a more reliable extraction of the carrier density from the optical spectra of copper chalcogenides NCs. For example, how does the carrier damping change when moving from the copper vacant copper chalcogenides ($Cu_{2-x}E$, with E=S, Se, Te) to the limit case of CuE composition? Initial works attested a purely metallic character to CuE , while the copper vacant one appears to suffer from additional damping or carrier localization due to vacancy formation [156,184,305]. The resulting dielectric functions obtained by band structure calculations of covellite CuS shows remarkable agreement with the experimental optical spectra [358]. However, questions remain as to the changes to the band structure upon vacancy formation. In metal oxide NCs, the attested differences in the chemical reactivity of carriers added via capacitive charging and chemical doping [66] should lead to differences in their Drude dielectric response, an aspect that to date has not been studied in detail. This however would be especially critical in applications (for example in sensing) where one needs to estimate the number of charge carri-ers within the nanostructure [31]. Do capacitive charge carriers versus those introduced via chemical doping display a different response to light reflected in the Drude parameters? What is the role of the charge balancing counter ions and how much do they influence the plasmonic response? What is the role of the injection depth within the NC volume and how does this affect the exact position of the LSPR [126,130]?

The affinity of copper chalcogenide NC to uptake metal cations has been proposed as a way to extract heavy metals from solution [32,255]. The resulting LSPR shifts and the changes in optical properties were proposed as an associated sensing tool. Initial steps in this direction have shown very promising results, although more work must be done to improve the sensitivity and the detection threshold. For example, it is now well ascertained that the NIR LSPR of CuS NCs shows a response to the uptake of such cations, which manifests itself in a quenching of the NIR LSPR, but the extent of such quenching is not correlated to the type of cation that is up taken [32,255]. It would be therefore extremely useful to develop a cation-specific read-out mechanism. Also, the signal sensitivity might be increased when translating the LSPR signal changes from the NIR to the visible, ultimately facilitating detection. This could be achieved in a heterostructure type of architecture, in which for example the NIR plasmonic domain is coupled to a fluorescent domain or to a noble metal plasmonic domain. In both cases, the coupling between the two domains would be modified by the changes in the NIR response, and would translate to a change in the signal in the visible range. Doped MO NCs, photodoped MO NCs as well as copper chalcogenide NCs are sensitive to a manifold of oxidation and reduction reagents (including oxygen), and the LSPR by itself is sensitive to ambient conditions. For a reasonable and selective sensing of specific chemical reactions, the NCs need to be protected from their susceptibility to other redox chemical reactions and the LSPR needs to be stabilized. Stabilization and control over the LSPR could then also be exploited in for surface enhanced infrared absorption spectroscopy (SEIRA) or surface enhanced Raman scattering (SERS). Indeed, it would be of major interest to extend this control to *in-situ* tunability to overlap with specific resonances. The designated enhancement of specific vibrational signatures might enable selective sensing.

The plasmonic properties of copper chalcogenide NCs were proposed for photothermal therapy in various independent works, including their use for theranostics serving as both radiotracers and as photothermal probes in tumor ablation treatments [27,28,42,43,322–331] (see **Figure 32**a). Metal oxide NCs have also been suggested for photothermal therapy, although to a lesser extent [359–363]. Biological applications require a tight control over the chemistry involved, as for example the release of toxic ions from copper chalcogenide NCs is potentially an unwanted side effect [28,42]. Possibly a properly chosen doped metal oxide can help to overcome those problems. On the other hand, it has been shown recently that the release of Cu(I) ions from $Cu_{2-x}S$ NCs, when they are heated with a NIR laser light, induces the formation of reactive oxygen species that trigger cell apoptosis, and therefore the NCs can serve the dual purpose of photothermal and photodynamic agents.[28]

As amply discussed in this review, metal oxide NCs have shown capability to uptake and store extra charge carriers. This is particularly useful for electrochromic devices, where a controlled application of an electric field enables selective darkening of the window in specific wavelength regions. Depending on whether the UV needs to be blocked or heating has to be minimized, either the UV-vis or the NIR region can be addressed. A manifold of different materials combinations with specific structures might enhance the ability to selectively switch the absorption properties of degenerately doped semiconductor NCs and to protect them from degradative side effects [319,364,365].

A promising direction to go is the direct light to energy conversion of plasmonic semiconductor NCs by exciting the NIR LSPR. This allows extension of the light absorption to the NIR, where a large portion of the incoming sunlight is located, but also for tunable NIR detectors [271,291]. Initial efforts have been undertaken in this direction, although a more systematic approach would be required. In this regard, a variety of different heterostructures can be screened as photocatalysts, by studying for example the role of the shape, size, materials, and doping levels on their catalytic activity. One would need to ascertain to what extent the NIR plasmon resonance is involved in such activity, and the limiting factors, as to the exact band alignment or the interfaces in heterostructures need to be unraveled. Size, shape and dielectric constant of the medium will together determine whether light is more effectively absorbed or scattered, and also the directional dependence of the plasmon oscillation plays a major role. The targeted extraction of capacitively introduced charges (either through optical processes, such as photodoping, or through electrochemical doping) by an optically triggered dis-charging event, such as plasmon induced hot electron extraction, could be an interesting direction to go for solar light conversion, NIR light detection, or to trigger chemical reactions such as photocatalysis, as well.

Initial steps to other potential directions employing degenerately doped semiconductor NCs have already been taken, such as their use as piezoelectric sensors [366], for enhanced microwave absorption [367], for efficient light-induced water evaporation [368], as photothermal fungicide agents for sunlight driven disinfection [369], as non-linear optical materials operating in the telecommunication region [370–372], or for the study of artificial polariton bandgaps at infrared frequencies [373]. Many applications exploiting the new properties of these materials are still at a very early stage and demand for further systematic development. The fundamentally different doping mechanism in vacancy doped copper chalcogenide NCs with respect to doped metal oxide NCs also results in profoundly different targeted applications. Nevertheless, this summary shows that degenerately doped semiconductor NCs hold strong potential for a manifold of different technologies.

## Acknowledgements

This work was supported by the Global Fellowship MOPTOPus (Marie Curie Actions) of the European Union's Horizon 2020 [Grant number 705444].